\documentclass{elsarticle}
\pdfoutput=1 

\usepackage[a4paper, top=1.0in, bottom=1.2in, left=1.0in, right=1.0in]{geometry}
\usepackage[utf8]{inputenc}
\usepackage[tbtags]{amsmath} 

\usepackage{stix2} 

\usepackage{placeins} 
\usepackage{xcolor}
\usepackage{booktabs} 
\usepackage{mathtools} 
\usepackage{stmaryrd} 

\usepackage{hyperref} 
\hypersetup{pdfauthor={Buist et al.},pdftitle={Energy-stable two-fluid model}}

\usepackage{tikz} 
\usetikzlibrary{shapes,arrows,arrows.meta,matrix,decorations.pathmorphing,shapes.geometric,positioning,calc}  
\def\centerarc[#1](#2)(#3:#4:#5){ \draw[#1] ($(#2)+({#5*cos(#3)},{#5*sin(#3)})$) arc (#3:#4:#5); } 

\renewcommand{\v}[1]{\ensuremath{\mathbf{#1}}} 
\newcommand{\gv}[1]{\ensuremath{\mbox{\boldmath$ #1 $}}} 
\newcommand{\abs}[1]{\left| #1 \right|} 

\renewcommand{\d}[2]{\frac{\mathrm{d} #1}{\mathrm{d} #2}} 
\newcommand{\pd}[2]{\frac{\partial #1}{\partial #2}} 
\newcommand{\pdd}[2]{\frac{\partial^2 #1}{\partial #2^2}} 
\newcommand{\pddd}[2]{\frac{\partial^3 #1}{\partial #2^3}}  
\newcommand{\dotp}[2]{\langle #1 \, {,} \, #2 \rangle} 

\usepackage[english]{babel} 
\usepackage[titletoc,title]{appendix}

\usepackage[shortcuts]{extdash} 

\begin{document}

\title{Energy-stable discretization of the one-dimensional two-fluid model} 
\author[add1,add2]{J.F.H. Buist}
\ead{jurriaan.buist@gmail.com, corresponding author}
\author[add1]{B. Sanderse}
\author[add3]{S. Dubinkina}
\author[add4]{C.W. Oosterlee}
\author[add5]{R.A.W.M. Henkes}

\address[add1]{Scientific Computing Group, Centrum Wiskunde \& Informatica (CWI), Amsterdam, The Netherlands}
\address[add2]{Department of Process \& Energy, Delft University of Technology, Delft, The Netherlands}
\address[add3]{Department of Mathematics, Vrije Universiteit Amsterdam, Amsterdam, The Netherlands}
\address[add4]{Mathematical Institute, Utrecht University, Utrecht, The Netherlands}
\address[add5]{J.M. Burgerscentrum, Delft University of Technology, Delft, The Netherlands}

\begin{abstract}

In this paper we present a complete framework for the energy-stable simulation of stratified incompressible flow in channels, using the one-dimensional two-fluid model.
Building on earlier energy-conserving work on the basic two-fluid model, our new framework includes diffusion, friction, and surface tension. 
We show that surface tension can be added in an energy-conserving manner, and that diffusion and friction have a strictly dissipative effect on the energy. 

We then propose spatial discretizations for these terms such that a semi-discrete model is obtained that has the same conservation properties as the continuous model.
Additionally, we propose a new energy-stable advective flux scheme that is energy-conserving in smooth regions of the flow and strictly dissipative where sharp gradients appear. 
This is obtained by combining, using flux limiters, a previously developed energy-conserving advective flux with a novel first-order upwind scheme that is shown to be strictly dissipative.

The complete framework, with diffusion, surface tension, and a bounded energy, is linearly stable to short wavelength perturbations, and exhibits nonlinear damping near shocks.
The model yields smoothly converging numerical solutions, even under conditions for which the basic two-fluid model is ill-posed.
With our explicit expressions for the dissipation rates, we are able to attribute the nonlinear damping to the different dissipation mechanisms, and compare their effects.

\end{abstract}



\maketitle

\section{Introduction}

The one-dimensional two-fluid model (TFM) is a cross-sectionally averaged model for two-phase flow in pipes and channels.
Velocities and phase fractions are resolved only along the main direction of flow, for each fluid separately.
This yields an efficient model that is useful when calculations are needed quickly, or when many calculations need to be made.
It is most commonly used for flow assurance in oil and gas or $\mathrm{CO}_2$ transport \cite{AursandHammerMunkejordEtAl2013, GoldszalDanielsonBansalEtAl2007}, and for safety analysis of steam-water flows in nuclear reactors \cite{BerryZouZhaoEtAl2014}.

The TFM possesses the ability to dynamically simulate the Kelvin-Helmholtz instability which arises at the interface between two fluids flowing at different velocities.
This is a valuable property since it is essential in predicting the transition from stratified flow to slug flow, a type of flow which is typically unwanted due to the large loads it places on the pipe \cite{Fabre2003}.
However, for the basic TFM, when the difference between the two fluids' velocities is large, the instability is unphysically severe.
Linear stability analysis shows an unbounded growth rate at short wavelengths, leading to the conclusion that the model is ill-posed \cite{DinhNourgalievTheofanous2003, DrewPassman1999, Montini2011}. 
For the basic model, with only first-order terms, the results of the linear stability analysis can be compared to those of a characteristic analysis: short wavelength unbounded instability implies complex eigenvalues \cite{RamshawTrapp1978}. 

The stability issue is intertwined with a modeling issue.
Due to the averaged one-dimensional nature of the TFM, not all small-scale dynamics of the instability can be resolved, and there is uncertainty on how to model their effect on the averaged flow.
The TFM implicitly carries the \textit{long wavelength assumption}, implying that the TFM can only accurately model perturbations with a wavelength longer than the fluid depth \cite{HolmasSiraNordsveenEtAl2008, Montini2011}.
It is precisely at the poorly modeled short wavelengths that the catastrophic instability takes place.

The issue has led some researchers to use regularizing terms such as an artificial interfacial pressure force which completely eliminates the instability, for both long and short wavelengths \cite{Bestion1990, EvjeFlatten2003, LiouNguyenChangEtAl2006}.
Others have proposed regularizing terms which only eliminate instability below a desired cut-off wavelength, in the form of artificial diffusion, added both to mass and momentum equations \cite{BonzaniniPicchiPoesio2017, HolmasSiraNordsveenEtAl2008}.
Finally, researchers strive for stabilization through the systematic inclusion of missing physics \cite{DrewPassman1999, LopezdeBertodanoFullmerClausseEtAl2017}.
Small-scale stabilizing effects include molecular and turbulent diffusion (in axial direction) \cite{FullmerLopezdeBertodanoRansom2011}, surface tension \cite{RamshawTrapp1978}, and mixing \cite{Castro-DiazFernandez-NietoGonzalez-VidaEtAl2011}.

Beyond the question of the growth of small perturbations, which is answered by linear stability analysis, lies the question of the growth of large perturbations, for which the full nonlinear behavior of the model must be taken into account \cite{LopezdeBertodanoFullmerClausseEtAl2017}.
For related models, namely the single-layer and two-layer shallow water equations, the mechanical energy acts as an entropy function, and as a nonlinear bound on the solution \cite{BouchutMoralesdeLuna2008, FjordholmMishraTadmor2009, vanReeuwijk2011}.
An energy conservation equation can be derived from the governing equations, leading to the conclusion that energy is a secondary conserved quantity of the model, following the terminology of \cite{Veldman2021a}.
Energy-conserving discretization schemes, in which the energy conservation property of the continuous equations is retained, have been designed in order to prevent numerical instability \cite{GassnerWintersKopriva2016, vantHofVeldman2012}.
In \cite{BuistSanderseDubinkinaEtAl2022} we showed that the basic TFM satisfies an energy conservation equation like the shallow water equations, and developed an energy-conserving finite volume scheme which satisfies a semi-discrete energy conservation equation.

However, in the presence of shocks, the derivation of the energy conservation equation for the continuous model no longer holds, and energy needs to be dissipated \cite{Jameson2008a}.
Energy-conserving schemes without dissipation will produce numerical oscillations in the presence of shocks. 
Therefore energy-stable schemes are designed, by taking an energy-conserving scheme as a baseline, and adding strictly dissipative terms, which can only cause a decrease of the energy \cite{CastroFjordholmMishraEtAl2013, FjordholmMishraTadmor2011}.
These dissipative terms typically take the form of numerical diffusion which is proportional to grid cell size, and preferably dissipate the minimum required amount of energy, and only in the vicinity of shocks, where it is needed.

The TFM requires mechanisms both for dissipation in shocks, and for suppression of the unbounded linear instability. 
We follow the approach of \cite{FullmerRansomLopezdeBertodano2014} and stabilize short wavelength perturbations through the inclusion of axial (momentum) diffusion and surface tension.
In this work, we fit these effects, along with wall and interface friction, into our energy-consistent framework  \cite{BuistSanderseDubinkinaEtAl2022}.
Diffusion and friction are shown to be strictly dissipative, surface tension is shown to be energy-conserving, and we present a spatial discretization of these terms that retains these properties. 
Importantly, we propose a novel discretization of the advective flux that is energy stable, with numerical dissipation acting near discontinuities in the solution. 

The extended framework possesses bounded linear growth rates (with damping at short wavelengths), and possesses a nonlinear bound on the energy.
It possesses multiple mechanisms for dissipation, which can be quantified using explicit expressions for the various dissipation rates. 
The energy-stable nature of the semi-discrete model, consistent with its continuous counterpart, provides additional fidelity in the accuracy of the numerical solution.
The framework yields grid-converged numerical solutions, with well-resolved shocks, for flow states for which the basic TFM is linearly ill-posed.

The analysis of the continuous model is given in \autoref{sec:continuous}, starting with a review of the basic model and its energy behavior, followed by the results for the extended model, and then a detailed analysis of each term separately. 
In \autoref{sec:semi-discrete}, these steps are repeated for the semi-discrete model, with the addition of an analysis of the newly proposed advective flux discretization, showing that it is energy stable. 
The stability of the TFM is discussed in detail in \autoref{sec:stability}, in order to motivate the additions to the basic model.
In \autoref{sec:numerical_experiments}, the energy and stability properties predicted by analysis are verified using numerical experiments.
We test the capability to model a traveling wave, and a growing wave which develops into a shock, and take a detailed look at the different components of the dissipation near the shock.   
Our conclusions are given in \autoref{sec:conclusion}.

\section{Energy conservation and the continuous two-fluid model}
\label{sec:continuous}

\subsection{Governing equations for the basic model}
\label{ssec:continuous/model}

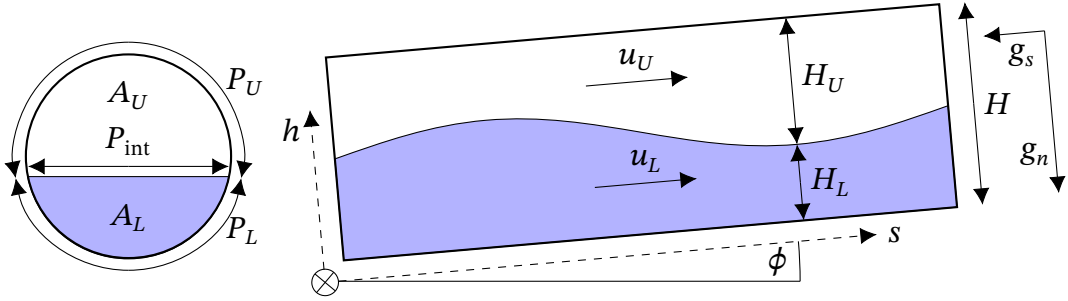
\begin{figure}[!htb]
\large \centering
\begin{tikzpicture}[scale=1.35,baseline=(current bounding box.north)]

\begin{scope}[shift={(0,0.235)},rotate=5]

\filldraw[fill=blue!30!white, draw=black] (-3,0) sin (-1.5,0.25) cos (0,0) sin (1.5,-0.25) cos (3,0) -- (3,-1) -- (-3,-1) -- cycle;

\draw[-{Latex[width = 2.2mm, length = 2.2mm]},dashed] (-3.2, -1.075)--(-3.2,0.5);
\node [left] at (-3.2,0.3) {$h$};
\draw[-{Latex[width = 2.2mm, length = 2.2mm]}] (3.25, 0)--(3.25,-1);
\draw[-{Latex[width = 2.2mm, length = 2.2mm]}] (3.25, 0)--(3.25,1);
\node [right] at (3.25,0) {$H$};

\draw[-{Latex[width = 2.2mm, length = 2.2mm]},dashed] (-3.07, -1.2)--(2.2,-1.2);
\node [right] at (2.2,-1.2) {$s$};
\centerarc[](-3.07,-1.2)(0:-5:4.5)
\draw[] (-3.07, -1.2)--(1.413,-1.592); 
\node [left] at (1.413,-1.4) {$\phi$};

\draw[] (-3.2,-1.2) circle (0.13);
\draw[] (-3.292,-1.108) -- (-3.108,-1.292);
\draw[] (-3.108,-1.108) -- (-3.292,-1.292);

\draw[thick] (-3,-1) rectangle (3,1);

\draw[-{Latex[width = 2.2mm, length = 2.2mm]}] (-0.5, -0.5)--(0.5,-0.5);
\node [above] at (0,-0.5) {$u_L$};

\draw[-{Latex[width = 2.2mm, length = 2.2mm]}] (-0.5, 0.5)--(0.5,0.5);
\node [above] at (0,0.5) {$u_U$};




\draw[-{Latex[width = 2.2mm, length = 2.2mm]}] (1.5, -0.625)--(1.5,-0.25);
\draw[-{Latex[width = 2.2mm, length = 2.2mm]}] (1.5, -0.625)--(1.5,-1);
\node [right] at (1.5,-0.625) {$H_L$};

\draw[-{Latex[width = 2.2mm, length = 2.2mm]}] (1.5, 0.375)--(1.5,-0.25);
\draw[-{Latex[width = 2.2mm, length = 2.2mm]}] (1.5, 0.375)--(1.5,1);
\node [right] at (1.5, 0.4) {$H_U$};

\draw[-{Latex[width = 2.2mm, length = 2.2mm]}] (4.0, 0.65)--(4.0,-0.95);
\node [left] at (4.06,-0.57) {$g_n$};
\draw[-{Latex[width = 2.2mm, length = 2.2mm]}] (4.0, 0.65)--(3.4,0.65);
\node [below] at (3.78,0.65) {$g_s$};

\end{scope}



\filldraw[fill=blue!30!white, draw=black] (-5.9798,-0.2) -- (-4.0202,-0.2) arc (-11.087:-168.913:1) -- cycle;

\draw[thick] (-5,0) circle (1);

\node at (-5,-0.6) {$A_L$};
\node at (-5,0.6) {$A_U$};

\centerarc[-{Latex[width = 2.2mm, length = 2.2mm]}](-5,0)(-90:-168.913:1.12);
\centerarc[-{Latex[width = 2.2mm, length = 2.2mm]}](-5,0)(-90:-11.087:1.12);
\node[right] at (-4.15,-0.73) {$P_L$};

\centerarc[-{Latex[width = 2.2mm, length = 2.2mm]}](-5,0)(90:191.087:1.12);
\centerarc[-{Latex[width = 2.2mm, length = 2.2mm]}](-5,0)(90:-11.087:1.12);
\node[right] at (-4.15,0.73) {$P_U$};

\draw[-{Latex[width = 2.2mm, length = 2.2mm]}](-5,-0.1) -- (-4.0202,-0.1);
\draw[-{Latex[width = 2.2mm, length = 2.2mm]}](-5,-0.1) -- (-5.9798,-0.1) ;
\node[above] at (-5,-0.1) {$P_\mathrm{int}$};
\end{tikzpicture}%
\normalsize
\caption{A schematic of stratified two-fluid flow in ducts (a circular pipe segment is shown as an example) described by the one-dimensional TFM.}
\label{fig:continuous/model/two-fluid_schematic}
\end{figure}

The one-dimensional two-fluid model (TFM) is a cross-sectionally averaged model for two-phase flow in a closed conduit \cite{IshiiMishima1984, StewartWendroff1984}. 
The conduit can take different forms, such as a pipe with a circular cross section, as depicted in \autoref{fig:continuous/model/two-fluid_schematic}, a duct with a rectangular cross section, or (more abstractly) a two-dimensional channel with a cross section of zero width.  
In all cases, the model can be obtained by defining control volumes for the two fluids separately, which are assumed to be stratified with a sharp interface between them, and setting up integral mass and momentum balances for these control volumes. 
No energy balance is needed, since the flow is assumed to be isothermal \cite{Munkejord2006}.
The mass and momentum balances are divided by their length $\Delta s$, the limit $\Delta s \rightarrow 0$ is taken, and the resulting equations are written in terms of cross-sectionally averaged variables, which are functions only of the streamwise coordinate $s$ and time $t$. 
Important assumptions made in this process include that the magnitude of the velocity normal to the direction of flow is small, and that the derivative of the velocity along the direction of flow is small; this is known as the long wavelength assumption \cite{HolmasSiraNordsveenEtAl2008, Montini2011}.  
The flow is also assumed to be incompressible, and along the normal direction it is in hydrostatic balance. 

The model, in conservative form, is given by \cite{BuistSanderseDubinkinaEtAl2022, SanderseSmithHendrix2017}:
\begin{equation}
\pd{\v{q}}{t}  + \pd{\v{f}( \v{q} )}{s}  +  \v{j}(\v{q}) \pd{p}{s} = \v{0},
\label{eq:continuous/model/conservative_v1}
\end{equation}
with the conservative variables $\v{q}(s,t)$ representing a mass per unit length or momentum per unit length:
\begin{equation*}
\v{q}^T = 
\begin{bmatrix}
q_1 &
\quad q_2 &
\quad q_3 &
\quad q_4
\end{bmatrix}
=
\begin{bmatrix}
\rho_U A_U &
\quad \rho_L A_L &
\quad \rho_U u_U A_U &
\quad \rho_L u_L A_L
\end{bmatrix}.
\end{equation*}
The conservative variables can be written in terms of the primitive variables, namely the cross-sections $A_U$ and $A_L$ (related to the heights $H_U$ and $H_L$) which are occupied by the upper and lower fluids respectively, the densities $\rho_U$ and $\rho_L$ of each fluid, and the streamwise (averaged) velocities $u_U$ and $u_L$. 
In \eqref{eq:continuous/model/conservative_v1}, the fluxes are given by
\begin{equation*}
\v{f}(\v{q})^T =
\begin{bmatrix}
f_1(\v{q}) &
\quad f_2(\v{q}) &
\quad f_3(\v{q}) &
\quad f_4(\v{q})
\end{bmatrix}
=
\begin{bmatrix}
q_3 &
\quad q_4 &
\quad \frac{q_3^2}{q_1} -  \rho_U g_n  \widehat{H}_U &
\quad \frac{q_4^2}{q_2} - \rho_L g_n  \widehat{H}_L
\end{bmatrix}.
\end{equation*}
Here $\widehat{H}_U = \widehat{H}_U(\v{q})$ and $\widehat{H}_L = \widehat{H}_L(\v{q})$ are geometric quantities, defined in \autoref{sec:geometric_relations}, which are part of the terms known as the level gradients, which describe the effect of the variation of the hydrostatic pressure along $s$. 
The symbol $g_n = g \cos{(\phi)}$ represents the normal component of gravity, $\phi$ being the pipe inclination angle. 

The pressure $p$ appearing in the equations denotes the pressure at the interface between the two fluids.
Its derivative is weighted by the vector $\v{j}$, which is given by
\begin{equation*}
\v{j}(\v{q})^T = 
\begin{bmatrix}
0 &
\quad 0 &
\quad \frac{q_1}{\rho_U}  &
\quad \frac{q_2}{\rho_L} 
\end{bmatrix}.
\end{equation*}
Since the upper and lower fluid together fill the pipe with cross-section $A$, the system is subject to the volume constraint 
\begin{equation*}
\frac{q_1}{\rho_U} + \frac{q_2}{\rho_L} = A,
\label{eq:continuous/model/volume_constraint}
\end{equation*}
which implies the volumetric flow constraint \cite{BuistSanderseDubinkinaEtAl2023}
\begin{equation}
    \pd{Q}{s} = 0, \quad \text{with} \quad  Q(\v{q}) = \frac{q_3}{\rho_U} + \frac{q_4}{\rho_L} = u_U A_U + u_L A_L.
 \label{eq:continuous/model/volumetric_flow_constraint}
\end{equation}

\subsection{Energy conservation for the basic model}
\label{sec:continuous/basic}

The basic TFM has been shown in \cite{BuistSanderseDubinkinaEtAl2022} to conserve the following mechanical energy:
\begin{equation}
e_b(\v{q})=\rho_U g_n \widetilde{H}_U +  \rho_L g_n \widetilde{H}_L + \frac{1}{2}  \frac{q_3^2}{q_1} + \frac{1}{2}  \frac{q_4^2}{q_2}.
\label{eq:continuous/basic/e_definition_1} 
\end{equation}
Here $ \widetilde{H}_U = \widetilde{H}_U(A_U(q_1,\rho_U))$ and $\widetilde{H}_L = \widetilde{H}_L (A_L(q_2,\rho_L))$ are geometric terms representing the centers of mass of the upper and lower fluids respectively (see \autoref{sec:geometric_relations}). 
For the 2D channel geometry, \eqref{eq:continuous/basic/e_definition_1} reduces to the following primitive form:
\begin{equation*}
e_{b,\mathrm{ch}} = \frac{1}{2} \rho_U g_n H_U^2 + \rho_U g_n  H_U H_L +  \frac{1}{2} \rho_L g_n H_L^2  + \frac{1}{2}  \rho_U  u_U^2 H_U + \frac{1}{2}\rho_L u_L^2 H_L.
\end{equation*}
Given a mechanical energy $e_b(\v{q})$, we can define the following vector:
\begin{equation}
\v{v}_b(\v{q})^T \coloneqq \left[\pd{e_b}{\v{q}}\right]
=
\begin{bmatrix}
- \frac{1}{2}  \frac{q_3^2}{q_1^2} +  g_n \d{\widetilde{H}_U}{A_U} &
\quad - \frac{1}{2} \frac{q_4^2}{q_2^2} +  g_n \d{\widetilde{H}_L}{A_L} &
\quad  \frac{q_3}{q_1} &
\quad \frac{q_4}{q_2}
\end{bmatrix}.
\label{eq:continuous/basic/v_definition} 
\end{equation}
Taking the dot product of this vector with the governing equations given by \eqref{eq:continuous/model/conservative_v1} yields
\begin{equation}
\dotp{\v{v}_b}{\pd{\v{q}}{t} } + \dotp{\v{v}_b}{ \pd{\v{f} }{s} } + \dotp{\v{v}_b}{ \v{j} \pd{p}{s}}  = 0.
\label{eq:continuous/basic/continuous_energy_equation_v1}
\end{equation}
Using the geometric relations \eqref{eq:geometric_relations/essential_geometric_relations}, the volumetric flow constraint \eqref{eq:continuous/model/volumetric_flow_constraint}, and assuming $g_n$ to be constant along $s$, it can be shown that these terms can be written in conservative form \cite{BuistSanderseDubinkinaEtAl2022}:
\begin{equation*}
\dotp{ \v{v}_b}{ \pd{\v{q}}{t}  } = \pd{e_b}{t}, \quad \dotp{\v{v}_b}{\pd{\v{f}}{s}} = \pd{h_f}{s}, \quad \dotp{\v{v}_b}{ \v{j} \pd{p}{s}} = \pd{h_p}{s},
\end{equation*}
with
 \begin{equation}
h_f = g_n   q_3 \d{\widetilde{H}_U}{A_U} + g_n  q_4  \d{\widetilde{H}_L}{A_L}   + \frac{1}{2}   \frac{q_3^3}{q_1^2} + \frac{1}{2} \frac{q_4^3}{q_2^2},
\label{eq:continuous/basic/hf_definition} 
\end{equation}
and
\begin{equation}
h_p = Qp.
\label{eq:continuous/basic/hp_definition} 
\end{equation}

Therefore, \eqref{eq:continuous/basic/continuous_energy_equation_v1} reduces to the local energy conservation equation
\begin{equation}
\pd{e_b}{t} + \pd{h_b}{s}  = 0,
\label{eq:continuous/basic/continuous_energy_equation_v3}
\end{equation}
with $h_b = h_f + h_p$, 
which describes how the energy $e_b(s,t)$ at a specific point in space changes due to an inflow or outflow.
In case of periodic or closed boundaries, integrating this equation over a section of pipe yields the global energy conservation equation 
\begin{equation}
\d{E_b}{t} = -\left[ h_b \right]_{s_1}^{s_2}  = 0, \quad \text{with} \quad E_b(t) = \int_{s_1}^{s_2} e_b \, \mathrm{d} s.
\label{eq:continuous/basic/global_energy_equation}
\end{equation}
This shows that the mechanical energy is a secondary conserved quantity of the TFM (in contrast to the primary conserved quantities of mass and momentum).

\subsection{Energy equation for the extended model}
\label{ssec:continuous/summary}

Having set up the basic TFM and its energy conservation equation, we will extend it in an energy\-/consistent manner, with three additions that make it linearly well-posed and energy stable.
We will show, in the following subsections, that friction and diffusion have a strictly dissipative effect, while surface tension can be added in an energy-conserving manner. 
In previous work \cite{BuistSanderseDubinkinaEtAl2023}, the energy-conserving nature of streamwise gravity and the energy input due to a driving pressure gradient have been demonstrated.

The model, extended with all the additional terms, is given by 
\begin{equation}
\boxed{\pd{\v{q}}{t}  + \pd{\v{f}}{s}  +  \v{j} \pd{p}{s} =  \pd{\v{d}}{s} + \v{s} + \v{c}_{g} + \v{c}_f + \v{c}_p}
\label{eq:continuous/summary/extended_governing_equations}
\end{equation}
with $\partial \v{d}/\partial s$ representing diffusion, $\v{c}_{f}$ representing friction, and $\v{s}$ representing surface tension.
The expressions for these terms will be given in \eqref{eq:continuous/physical_diffusion/model_addition_d}, \eqref{eq:continuous/friction/model_addition_cf}, and \eqref{eq:continuous/surface_tension/general_model_addition_s}, respectively.
The extended model includes the following contributions from streamwise gravity, indicated in \autoref{fig:continuous/model/two-fluid_schematic} with $g_s = g \sin{(\phi)}$:
\begin{equation*}
\v{c}_g^T = 
\begin{bmatrix}
0 &
\quad 0 &
\quad -  g \sin{(\phi)} q_1  &
\quad -  g \sin{(\phi)} q_2
\end{bmatrix},
\end{equation*} 
and from a constant driving pressure gradient, which can be applied in cases with periodic boundary conditions in order to balance against streamwise gravity and friction:
\begin{equation*}
\v{c}_p^T = 
\begin{bmatrix}
0 &
\quad 0 &
\quad -  \frac{q_1}{\rho_U} \pd{p_0}{s}  &
\quad -  \frac{q_2}{\rho_L} \pd{p_0}{s} 
\end{bmatrix}.
\end{equation*} 

For the extended model given by \eqref{eq:continuous/summary/extended_governing_equations}, the following energy conservation equation will be derived:
\begin{equation}
\boxed{\pd{e}{t} + \pd{h}{s} = -\epsilon + c_p}
\label{eq:continuous/summary/local_energy_equation}
\end{equation}
with 
\begin{gather}
e = e_b + e_g + e_\sigma, \label{eq:continuous/summary/extended_energy} \\
h = h_b + h_g + h_d + h_\sigma, \\
\epsilon = \epsilon_d + \epsilon_f.
\end{gather}
Equation \eqref{eq:continuous/summary/local_energy_equation} is the first main novel result of this work.
It shows that the mechanical energy $e$, which consists of kinetic, potential, and surface energy, is locally conserved except for the dissipating effects of diffusion and friction.
The upcoming subsections will give the expressions \eqref{eq:continuous/physical_diffusion/h_d} for $h_d$, \eqref{eq:continuous/physical_diffusion/epsilon_d} for $\epsilon_d$, \eqref{eq:continuous/friction/epsilon_f} for $\epsilon_f$, and \eqref{eq:continuous/surface_tension/final_results} for $e_\sigma$ and $h_\sigma$.
Contributions from streamwise gravity are present in the energy and the energy flux \cite{BuistSanderseDubinkinaEtAl2023}:
\begin{equation*}
e_g = g y \left( q_1 + q_2 \right), \quad h_g = g y \left( q_3 + q_4 \right), \quad \text{with} \quad \d{y}{s} =  \sin(\phi(s)),
\end{equation*}
while the driving pressure gradient adds a source term:
\begin{equation*}
c_p = - Q \pd{p_0}{s},
\end{equation*}
which is strictly positive in a flow which is aligned with its driving pressure gradient, e.g.\ $Q>0$ and $\partial p_0/\partial s <0$.
This term differs from the others in that it represents an externally applied force, and therefore does not adhere to the strictly dissipative behavior of the flow itself. 

Upon integrating the local energy equation over a periodic domain, the conservative term $\partial h/\partial s$ in \eqref{eq:continuous/summary/local_energy_equation} vanishes. 
Besides conservative terms, the new energy equation has an explicit sink term $-\epsilon$ which remains present in the global energy equation:
\begin{equation}
\boxed{\d{E}{t}  = -\mathcal{E} + C_p} \quad \text{with} \quad E(t) = \int_{s_1}^{s_2} e \, \mathrm{d} s, \quad \mathcal{E} = \int_{s_1}^{s_2} \epsilon \, \mathrm{d} s, 
\label{eq:continuous/summary/global_energy_equation}
\end{equation}
\begin{equation*}
C_p = \int_{s_1}^{s_2} c_p  \, \mathrm{d} s = -  Q \pd{p_0}{s} L,
\end{equation*}
with $L=s_2-s_1$ the length of the domain. 
Disregarding the (optional) externally supplied energy source, the energy-conserving basic model has been supplemented with a sink term  which will shown to be strictly negative, leading to the dissipation of energy, and an energy-stable model. 

Each addition to the model independently results in additional terms in the energy equation.
The combined result of all these additions was given here. 
In the following subsections, the novel terms in \eqref{eq:continuous/summary/local_energy_equation} will be derived separately. 

\subsection{Physical diffusion}
\label{ssec:continuous/physical_diffusion}

Our first novel contribution in the continuous setting is that we show that adding viscous diffusion terms to the TFM has a strictly dissipative effect, which can be quantified using an expression for the dissipation rate.
We refer to these viscous terms as ``physical diffusion'' in contrast to the artificial diffusion of \cite{BonzaniniPicchiPoesio2017, FullmerLeeLopezdeBertodano2014, HolmasSiraNordsveenEtAl2008}, and the numerical diffusion which will be discussed in \autoref{sec:semi-discrete}. 
The physical diffusion terms naturally appear in the derivation of the model, but are typically neglected due to the long wavelength assumption, with the argument that the TFM cannot accurately resolve the scale at which these terms act.
However, they are important in bounding the linear instability of short wavelength perturbations (see \autoref{sec:stability}), and in bounding nonlinear shocks through dissipation (see \autoref{ssec:numerical_experiments/double_wave}). 

In the TFM, physical diffusion takes the form of the term $\partial \v{d}/\partial \v{s}$ as included in \eqref{eq:continuous/summary/extended_governing_equations}, with $\v{d}$ given by \cite{FullmerRansomLopezdeBertodano2014, Montini2011}
\begin{equation}
\v{d}^T =
\begin{bmatrix}
0 &
\quad 0 &
\quad \nu_{\mathrm{eff},U} q_1 \pd{}{s} \frac{q_3}{q_1} &
\quad \nu_{\mathrm{eff},L} q_2 \pd{}{s} \frac{q_4}{q_2}
\end{bmatrix}.
\label{eq:continuous/physical_diffusion/model_addition_d}
\end{equation}
We use the effective viscosity model of \cite{FullmerLopezdeBertodanoRansom2011}, which combines the molecular viscosity $\nu_m$ with a turbulent viscosity $\nu_t$.
This serves as a closure term for small scale fluctuations that are not resolved by the model:
\begin{equation*}
\nu_{\mathrm{eff}} = C_\epsilon \left( \nu_m + \nu_t \right),
\end{equation*}
with $C_\epsilon$ an adjustment factor.
The parameters $\nu_t$ and $C_\epsilon$ are empirical: they can be based on fully resolved (higher dimensional) simulations, specific to a given test case.
Physical diffusion conserves momentum, since it can be written in conservative form.

We now consider the effect of physical diffusion on the energy.
Unlike the addition of streamwise gravity, the addition of diffusion does not change the energy definition. 
There is a contribution of the extra terms to the left hand side (LHS) of the energy equation, which is given by
\begin{equation*}
-\dotp{\v{v}_b}{\pd{\v{d}}{s}},
\end{equation*}
with $\v{v}_b$ given by \eqref{eq:continuous/basic/v_definition}. 
Some manipulation yields (for smooth solutions)
\begin{equation}
-\dotp{\v{v}_b}{\pd{\v{d}}{s}} = -\frac{q_3}{q_1} \pd{}{s} \left( \nu_{\mathrm{eff},U} q_1 \pd{}{s} \frac{q_3}{q_1} \right) -\frac{q_4}{q_2} \pd{}{s} \left( \nu_{\mathrm{eff},L} q_2 \pd{}{s} \frac{q_4}{q_2} \right) 
= \pd{h_d}{s} + \epsilon_d,
\label{eq:continuous/physical_diffusion/derived_energy_equation_contribution}
\end{equation}
with
\begin{gather}
h_d = -  \nu_{\mathrm{eff},U} q_1 \frac{1}{2}  \pd{}{s} \frac{q_3^2}{q_1^2}  -   \nu_{\mathrm{eff},L} q_2 \frac{1}{2}  \pd{}{s} \frac{q_4^2}{q_2^2},
\label{eq:continuous/physical_diffusion/h_d} \\
\epsilon_d = \nu_{\mathrm{eff},U} q_1 \left( \pd{}{s} \frac{q_3}{q_1} \right)^2 + \nu_{\mathrm{eff},L} q_2 \left( \pd{}{s} \frac{q_4}{q_2} \right)^2.
\label{eq:continuous/physical_diffusion/epsilon_d}
\end{gather}

The terms included in the energy flux $h_d$ are energy-conserving, since they can be written in conservative form. 
The remaining terms, collected in $\epsilon_d$, are not conservative.
They are strictly positive, since $\nu_{\mathrm{eff},U}$, $q_1$, $\nu_{\mathrm{eff},L}$, and $q_2$ must be positive, and the square of the differential terms must be positive.
Therefore, when moved to the right hand side (RHS), it becomes clear that $-\epsilon_d$ is a strictly negative sink term.
In conclusion, we have proven analytically that physical diffusion leads to dissipation of the energy given by \eqref{eq:continuous/summary/extended_energy}, with dissipation rate $\epsilon_d$.

\subsection{Friction terms}
\label{ssec:continous/friction}

Our second novel contribution in the continuous setting is that we prove that wall and interface friction add a strictly dissipative sink term to the energy equation.
The friction term $\v{c}_f$ can be added to the model as in \eqref{eq:continuous/summary/extended_governing_equations}, with
\begin{equation}
\v{c}_f^T = 
\begin{bmatrix}
0 &
\quad 0 &
\quad \tau_U P_U + \tau_\mathrm{int} P_\mathrm{int} &
\quad \tau_L P_L - \tau_\mathrm{int} P_\mathrm{int}
\end{bmatrix}.
\label{eq:continuous/friction/model_addition_cf}
\end{equation}
The wall stresses $\tau_U$ and $\tau_L$ represent the shear stresses acting at the pipe perimeters $P_U$ and $P_L$, that are in contact with the upper and lower fluids, respectively. 
The interface stress $\tau_\mathrm{int}$ represents the shear stress at the interface $P_\mathrm{int}$ between the two fluids.
The stress terms in the model are the averaged effect of local stresses on the averaged flow, and in order to express these in terms of the averaged variables, closure relations are required.
These typically take the following form \cite{TaitelDukler1976}:
\begin{equation}
\tau_{L} = -\frac{1}{2} f_L \rho_L u_L |u_L|, \quad \tau_{U} = -\frac{1}{2} f_U \rho_U u_U |u_U|, \quad  \tau_{\mathrm{int}} = -\frac{1}{2} f_{\mathrm{int}} \rho_U \left(u_U - u_L \right) |u_U - u_L|,
\label{eq:continuous/friction/coventional_closure_terms}
\end{equation}
in which $f_L$, $f_U$, and $f_{\mathrm{int}}$ are friction factors that require further closure relations, which are functions of the solution $\v{q}$ (see \autoref{sec:friction_closure_relations}).

We now consider the effect of wall and interface friction on the energy.
The contribution of the extra terms to the RHS of the energy equation is
\begin{equation*}
+\dotp{\v{v}_b}{\v{c}_f},
\end{equation*}
with $\v{v}_b$ given by \eqref{eq:continuous/basic/v_definition}.
Carrying out the multiplication, substituting \eqref{eq:continuous/friction/coventional_closure_terms}, and some rewriting yields
\begin{equation}
\dotp{\v{v}_b}{\v{c}_{f}} = \frac{q_3}{q_1} \left(  \tau_U P_U + \tau_\mathrm{int} P_\mathrm{int} \right) +  \frac{q_4}{q_2} \left(  \tau_L P_L - \tau_\mathrm{int} P_\mathrm{int} \right)  
 = -\epsilon_f,
\end{equation}
with
\begin{equation}
\epsilon_f =  \frac{1}{2} f_U \rho_U \left( \frac{q_3}{q_1} \right)^2 \abs{ \frac{q_3}{q_1} } P_U  + \frac{1}{2} f_L \rho_L \left( \frac{q_4}{q_2} \right)^2 \abs{ \frac{q_4}{q_2} } P_L 
  + \frac{1}{2}  f_{\mathrm{int}}  \rho_U  \left( \frac{q_3}{q_1} - \frac{q_4}{q_2} \right)^2 \abs{\frac{q_3}{q_1} - \frac{q_4}{q_2} }P_\mathrm{int}.
\label{eq:continuous/friction/epsilon_f}
\end{equation}
Since $f_U$, $f_L$, $f_\mathrm{int}$, $\rho_U$, $\rho_L$, $P_U$, $P_L$, and $P_\mathrm{int}$ must be positive, and the rest of the terms are either quadratic or absolute, all three terms in \eqref{eq:continuous/friction/epsilon_f} must be positive.
Therefore,  $-\epsilon_f$ will act as a sink in the energy equation, which represents the dissipation of energy due to friction.
In conclusion, we have proven analytically that wall and interface friction have a strictly dissipative effect on the energy given by \eqref{eq:continuous/summary/extended_energy}.

\subsection{Surface tension}
\label{ssec:continuous/surface_tension}

Our third novel contribution in the continuous setting is that we show that surface tension can be added to TFM in such a way that the total energy is conserved.
Surface tension is an important addition since it makes the model linearly well-posed (see \autoref{sec:stability}).
However, if surface tension were to be added in a non-conservative manner, it would spoil the energy-stable nature of the model.
Therefore, it is key to find an energy-conserving form of the surface tension. 

The effect of surface tension in the TFM is typically modeled through its effect on the pressure.
This effect is to introduce a discontinuity in the pressure at the interface.
The pressure difference is given by \cite{Montini2011, RamshawTrapp1978}
\begin{equation}
\Delta p = - \sigma \kappa = \sigma \pdd{H_L}{s} \left[ 1 + \left(\pd{H_L}{s}\right)^2 \right]^{-3/2},
\label{eq:continuous/surface_tension/unsimplified_pressure_difference}
\end{equation}
with $\sigma$ the surface tension and $\kappa$ the streamwise curvature of the interface, with the interface assumed flat along the other direction.
This is the Young-Laplace equation for the TFM.

Similar to \cite{FullmerRansomLopezdeBertodano2014}, we include the effect of this pressure difference through the term $\v{s}$ in \eqref{eq:continuous/summary/extended_governing_equations}, with
\begin{equation}
\v{s}^T = 
\begin{bmatrix}
0 &
\quad 0 &
\quad 0 &
\quad \frac{q_2}{\rho_L} \pd{\Delta p}{s}
\end{bmatrix}.
\label{eq:continuous/surface_tension/general_model_addition_s}
\end{equation}
This is a general way to write the surface tension.
Typically in literature \cite{BarneaTaitel1994, FullmerRansomLopezdeBertodano2014, Montini2011, RamshawTrapp1978}, the assumption $\left(\partial H_L/\partial s\right)^2 \ll 1$ will be made to approximate \eqref{eq:continuous/surface_tension/unsimplified_pressure_difference} as:
\begin{equation}
\Delta p \approx \sigma \pdd{H_L}{s} \approx \frac{\sigma}{P_\mathrm{int}} \pdd{A_L}{s}.
\label{eq:continuous/surface_tension/dp_approximation}
\end{equation}
Note that for the specific case of a 2D channel geometry ($P_\mathrm{int}=1$, $A_L=H_L$), the two approximations in \eqref{eq:continuous/surface_tension/dp_approximation} are equivalent.

We note that, unlike the basic model, surface tension of the form given by \eqref{eq:continuous/surface_tension/general_model_addition_s} is not momentum\-/conserving, as it cannot be written in conservative form. 
This is caused by the one-dimensional nature of the model and stands in contrast to higher-dimensional, unaveraged models, where surface tension does conserve momentum \cite{Remmerswaal2023}.

Though \eqref{eq:continuous/surface_tension/general_model_addition_s} is not momentum-conserving, it can still be energy-conserving, and we aim to find a set of expressions for $\Delta p$ and the surface energy such that the contribution of surface tension to the energy conservation equation is of conservative form. 
Physically, the surface tension is associated with an energy, proportional to the surface area, that is conserved in combination with the mechanical energy \cite{Remmerswaal2023}. 
The surface area in the one-dimensional two-fluid model (see \autoref{fig:continuous/model/two-fluid_schematic}) will depend on $\partial H_L/\partial s = P_\mathrm{int}^{-1} \partial A_L/\partial s$, and $P_\mathrm{int}$. 
Therefore, we introduce the following general form for the surface energy:
\begin{equation}
e_\sigma = e_\sigma(S_\mathrm{int}, P_\mathrm{int}), \quad \text{with} \quad S_\mathrm{int} = \pd{A_L}{s},
 \label{eq:continuous/surface_tension/general_energy}
\end{equation}
with the functional dependencies specified as
\begin{equation*}
q_2 = q_2(A_L),  \quad P_\mathrm{int} = P_\mathrm{int}(A_L), \quad S_\mathrm{int} = S_\mathrm{int}(s,t), \quad A_L = A_L(s,t).
\end{equation*}

The additional term on the LHS of the energy equation due to surface tension as given by \eqref{eq:continuous/surface_tension/general_model_addition_s} is
\begin{equation*}
 - \dotp{\v{v}_b}{ \v{s} } =   -  \frac{q_4}{\rho_L}  \pd{\Delta p}{s}.
\end{equation*}
In order for the model addition to be energy-conserving, the following condition must hold:
\begin{equation}
-  \frac{q_4}{\rho_L}  \pd{\Delta p}{s} = \pd{h_\sigma}{s} + \pd{e_\sigma}{t},
 \label{eq:continuous/surface_tension/energy-conserving_condition}
\end{equation}
for in this case the addition to the energy equation will be of conservative form. 
We will now derive a relation between $\Delta p$ and $e_\sigma$ such that \eqref{eq:continuous/surface_tension/energy-conserving_condition} holds. 

The time derivative of the energy given by \eqref{eq:continuous/surface_tension/general_energy} is defined by
\begin{equation*}
\pd{e_\sigma}{t} =  \pd{e_\sigma}{S_\mathrm{int}} \pd{S_\mathrm{int}}{t}  + \pd{e_\sigma}{P_\mathrm{int}} \pd{P_\mathrm{int}}{t},
\end{equation*}
and under the assumption of smooth solutions, and through substitution of the mass conservation equation for the lower fluid (the second equation of \eqref{eq:continuous/model/conservative_v1}), can be rewritten in the following manner:
\begin{equation*}
\begin{split}
\pd{e_\sigma}{t} 
&=  \pd{e_\sigma}{S_\mathrm{int}} \pd{}{s} \left(  \rho_L^{-1}  \pd{q_2}{t} \right) + \pd{e_\sigma}{P_\mathrm{int}} \d{P_\mathrm{int}}{A_L} \rho_L^{-1} \pd{q_2}{t}  \\
&=  -\pd{e_\sigma}{S_\mathrm{int}} \pd{}{s} \left(  \rho_L^{-1}  \pd{q_4}{s} \right) - \pd{e_\sigma}{P_\mathrm{int}} \d{P_\mathrm{int}}{A_L} \rho_L^{-1} \pd{q_4}{s}  \\
&= - \pd{h_\sigma}{s} - \frac{q_4}{\rho_L} \pd{\Delta p}{s},
\end{split}
\end{equation*}
with
\begin{equation*}
\begin{split}
h_\sigma 
&= \frac{1}{\rho_L} \left(  \pd{e_\sigma}{S_\mathrm{int}}  \pd{q_4}{s} - q_4 \pd{}{s} \left( \pd{e_\sigma}{S_\mathrm{int}}   \right)+ q_4  \pd{e_\sigma}{P_\mathrm{int}} \d{P_\mathrm{int}}{A_L}  \right), 
\end{split}
\end{equation*}
and
\begin{equation}
\begin{split}
\Delta p
&=  \pd{}{s} \left( \pd{e_\sigma}{S_\mathrm{int}} \right) -  \pd{e_\sigma}{P_\mathrm{int}} \d{P_\mathrm{int}}{A_L}.
\end{split}
\label{eq:continuous/surface_tension/general_relation_dp_e_sigma}
\end{equation}
For an energy of the general form \eqref{eq:continuous/surface_tension/general_energy}, and a surface tension of the general form \eqref{eq:continuous/surface_tension/general_model_addition_s}, \eqref{eq:continuous/surface_tension/general_relation_dp_e_sigma} is the relation between the specific forms of $\Delta p$ and $e_\sigma$, that needs to be satisfied in order to achieve energy conservation. 

We now must find a set of expressions for $\Delta p$ and $e_\sigma$, that -- first -- satisfies \eqref{eq:continuous/surface_tension/general_relation_dp_e_sigma} and -- second -- makes physical sense. 
The most straightforward way to do this is to propose an energy based on physical considerations, substitute this in \eqref{eq:continuous/surface_tension/general_relation_dp_e_sigma}, and check if the resulting expression for $\Delta p$ compares to our expectation, which is that it take a form similar to \eqref{eq:continuous/surface_tension/unsimplified_pressure_difference} or \eqref{eq:continuous/surface_tension/dp_approximation}.
From a physical point of view, the energy should be given by $\sigma$ times the surface area, which can be expressed as
\begin{equation*}
 e_\sigma \left( S_\mathrm{int}, P_\mathrm{int} \right)= \sigma P_\mathrm{int}  \sqrt{ 1 + \left( \pd{H_L}{s} \right)^2 } =P_\mathrm{int}  \sqrt{ 1 + \left(P_\mathrm{int}^{-1} \pd{A_L}{s} \right)^2 } =P_\mathrm{int}  \sqrt{ 1 + \left(P_\mathrm{int}^{-1} S_\mathrm{int} \right)^2 }.
\end{equation*}
However, substituting this in \eqref{eq:continuous/surface_tension/general_relation_dp_e_sigma} yields an expression for $\Delta p$ that does not relate to \eqref{eq:continuous/surface_tension/unsimplified_pressure_difference} or \eqref{eq:continuous/surface_tension/dp_approximation}, and therefore cannot be physically justified. 
Mimicking the conventional approach of taking approximations such as \eqref{eq:continuous/surface_tension/dp_approximation}, we take the second order Taylor expansion of this energy around $S_\mathrm{int} =0$:
\begin{equation*}
e_\sigma \left( S_\mathrm{int}, P_\mathrm{int} \right) \approx \sigma \left( P_\mathrm{int} + \frac{1}{2}  P_\mathrm{int}^{-1}S_\mathrm{int}^2 \right).
\end{equation*}
Substituting this energy in \eqref{eq:continuous/surface_tension/general_relation_dp_e_sigma} yields the following expression for $\Delta p$:
\begin{equation}
\Delta p_\mathrm{int} = \frac{\sigma}{P_\mathrm{int}}  \pd{S_\mathrm{int}}{s} - \sigma \left( 1 + \frac{1}{2} P_\mathrm{int}^{-2} S_\mathrm{int}^2 \right) \d{P_\mathrm{int}}{A_L},
 \label{eq:continuous/surface_tension/dp_for_pipe_Taylor_expanded_energy}
\end{equation}
of which the first term can be recognized in \eqref{eq:continuous/surface_tension/dp_approximation}, but the second term can not. 

When the scope is reduced from arbitrary geometries to the specific case of the 2D channel geometry, for which $P_\mathrm{int}=1$ and $A_L=H_L$, \eqref{eq:continuous/surface_tension/dp_for_pipe_Taylor_expanded_energy} does match \eqref{eq:continuous/surface_tension/dp_approximation} exactly. 
This means that for the channel geometry, the combination
\begin{equation}
e_\sigma = \sigma \left( 1 + \frac{1}{2} \left( \pd{H_L}{s}\right)^2 \right), \quad \Delta p = \sigma \pdd{H_L}{s}, \quad \text{with} \quad h_\sigma = \frac{\sigma}{\rho_L} \left(\pd{q_4}{s} \pd{H_L}{s} - q_4 \pdd{H_L}{s} \right), 
\label{eq:continuous/surface_tension/final_results}
\end{equation}
is energy-conserving, and it can be justified physically, since $e_\sigma$ is an approximation of $\sigma$ times the surface area, and $\Delta p$ is an approximation of the Young-Laplace equation. 
This expression for $\Delta p$ can be substituted in \eqref{eq:continuous/surface_tension/general_model_addition_s} to obtain an energy-conserving form of the surface tension. 

We have therefore found a form of the surface tension $\v{s}$, and an associated surface energy $e_\sigma$, with which the basic model can be extended, while retaining its energy-conserving behavior. 
For the 2D channel geometry, this turned out to be equivalent to a standard form, often used in literature.

\section{Energy conservation and the semi-discrete two-fluid model}
\label{sec:semi-discrete}

\subsection{Semi-discrete equations for the basic model}
\label{ssec:semi-discrete/model}

With the energy analysis for the continuous model complete, we will continue to propose a discretization that inherits the energy properties of the three additions to the model on the discrete level. 
In order to obtain the same conservation properties for the discrete model as for the continuous model, the model must be discretized in a specific manner.
Therefore, the energy analysis guides the discretization.
In this subsection, we will first summarize the energy-conserving discretization of the basic model \cite{BuistSanderseDubinkinaEtAl2022}.

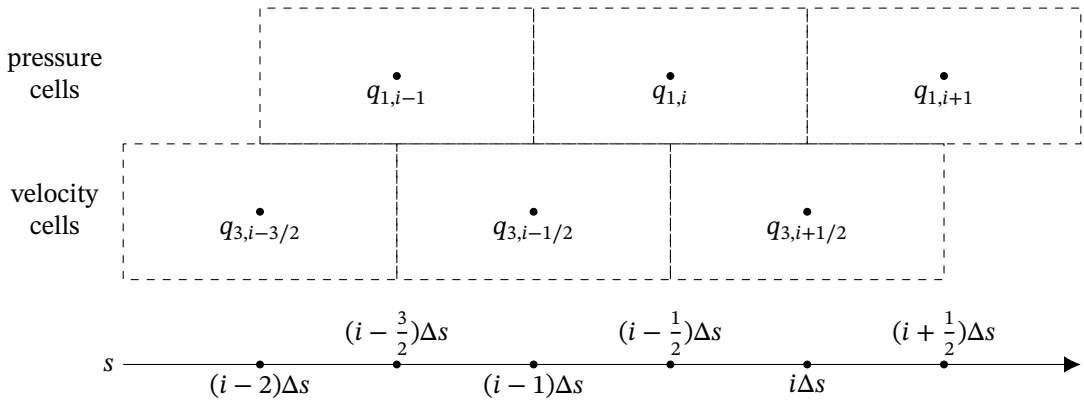
\begin{figure}[hbtp]
\centering%
\begin{tikzpicture}[scale=0.9]

\node at (-11,1.25) {\begin{tabular}{c} pressure \\ cells \end{tabular}};
\node at (-11,-0.75) {\begin{tabular}{c} velocity \\ cells \end{tabular}};

\draw[dashed] (-8,0.25) rectangle (-4,2.25);
\draw[fill=black] (-6,1.25) circle (0.05); 
\node [below] at (-6,1.25) {$q_{1,i-1}$};

\draw[dashed] (-4,0.25) rectangle (-0,2.25);
\draw[fill=black] (-2,1.25) circle (0.05); 
\node [below] at (-2,1.25) {$q_{1,i}$};

\draw[dashed] (-0,0.25) rectangle (4,2.25);
\draw[fill=black] (2,1.25) circle (0.05); 
\node [below] at (2,1.25) {$q_{1,i+1}$};

\draw[dashed] (-10,0.25) rectangle (-6,-1.75);
\draw[fill=black] (-8,-0.75) circle (0.05); 
\node [below] at (-8,-0.75) {$q_{3,i-3/2}$};

\draw[dashed] (-6,0.25) rectangle (-2,-1.75);
\draw[fill=black] (-4,-0.75) circle (0.05); 
\node [below] at (-4,-0.75) {$q_{3,i-1/2}$};

\draw[dashed] (-2,0.25) rectangle (2,-1.75);
\draw[fill=black] (-0,-0.75) circle (0.05); 
\node [below] at (0,-0.75) {$q_{3,i+1/2}$};

\node [left] at (-10,-3) {$s$};
\draw[-{Latex[width = 2.2mm, length = 2.2mm]}] (-10, -3)--(4,-3);

\draw[fill=black] (-8,-3) circle (0.05); 
\node [below] at (-8,-3) {$(i-2)\Delta s$};

\draw[fill=black] (-6,-3) circle (0.05); 
\node [above] at (-6,-3) {$(i-\frac{3}{2})\Delta s$};

\draw[fill=black] (-4,-3) circle (0.05); 
\node [below] at (-4,-3) {$(i-1)\Delta s$};

\draw[fill=black] (-2,-3) circle (0.05); 
\node [above] at (-2,-3) {$(i-\frac{1}{2})\Delta s$};

\draw[fill=black] (0,-3) circle (0.05); 
\node [below] at (0,-3) {$i\Delta s$};

\draw[fill=black] (2,-3) circle (0.05); 
\node [above] at (2,-3) {$(i+\frac{1}{2})\Delta s$};

\normalsize
\end{tikzpicture}
\caption{Staggered grid layout.\label{fig:staggered_grid}}
\end{figure}

We define the unknowns of the semi-discrete TFM on a staggered grid, depicted in \autoref{fig:staggered_grid}, in the following manner:
\begin{equation}
\v{q}_i(t) 
\coloneqq
\begin{bmatrix}
q_{1,i}(t) \\
q_{2,i}(t) \\
q_{3,i-1/2}(t) \\
q_{4,i-1/2}(t) 
\end{bmatrix}
=
\begin{bmatrix}
\left(\rho_U A_U \Delta s \right)_i \\
\left(\rho_L A_L \Delta s \right)_i \\
\left(\rho_U A_U u_U \Delta s \right)_{i-1/2} \\
\left(\rho_L A_L u_L \Delta s \right)_{i-1/2}
\end{bmatrix},
\label{eq:semi-discrete/model/discrete_variables_q}
\end{equation}
We introduce the following notation to denote central interpolation and jumps respectively:
\begin{gather}
\overline{a}_{i-1/2} \coloneqq \frac{1}{2} \left( a_{i-1} + a_{i} \right),  \quad \overline{a}_{i} \coloneqq \frac{1}{2} \left( a_{i-1/2} + a_{i+1/2} \right), 
\label{eq:semi-discrete/model/interpolation_definition} \\
\llbracket a_{i-1/2} \rrbracket  \coloneqq a_{i} - a_{i-1}, \quad \llbracket a_{i} \rrbracket  \coloneqq a_{i+1/2} - a_{i-1/2}.
\label{eq:semi-discrete/model/jump_definition}
\end{gather} 
The primitive variables can be extracted from \eqref{eq:semi-discrete/model/discrete_variables_q} through the following relations:
\begin{equation}
A_{U,i} = \frac{q_{1,i}}{\rho_U \Delta s}, \quad A_{L,i} = \frac{q_{2,i}}{\rho_L \Delta s}, \quad 
u_{U,i-1/2} = \frac{ q_{3,i-1/2}}{\overline{q}_{1,i-1/2}}, \quad u_{L,i-1/2} = \frac{ q_{4,i-1/2}}{\overline{q}_{2,i-1/2}}.
\label{eq:semi-discrete/model/primitive_translation}
\end{equation}

With this notation, the semi-discrete finite volume scheme can be written locally as
 \begin{equation}
\d{\v{q}_{i}}{t} + \llbracket \v{f}_{i} \rrbracket + \v{j}_i \left\llbracket p_{i-1/2} \right\rrbracket =  \v{0},
 \label{eq:semi-discrete/model/finite_volume_scheme}
\end{equation}
with
\begin{equation}
\v{f}_{i-1/2} \coloneqq 
\begin{bmatrix}
f_{1,i-1/2} \\
f_{2,i-1/2} \\
f_{3,i-1} \\
f_{4,i-1}
\end{bmatrix}
=
\begin{bmatrix}
\frac{q_{3,i-1/2}}{\Delta s} \\
 \frac{q_{4,i-1/2}}{\Delta s}  \\
  \overline{ \left( \frac{q_{3,i-1}}{  \overline{q}_{1,i-1} } \right) }  \frac{\overline{q}_{3,i-1}}{\Delta s}  - \rho_U g_n \widehat{H}_{U,i-1} \\
 \overline{ \left( \frac{q_{4,i-1}}{  \overline{q}_{2,i-1} } \right) }  \frac{\overline{q}_{4,i-1}}{\Delta s} - \rho_L g_n \widehat{H}_{L,i-1}
\end{bmatrix},
\quad \text{and} \quad
\v{j}_{i} \coloneqq 
\begin{bmatrix}
j_{1,i} \\
j_{2,i} \\
j_{3,i-1/2} \\
j_{4,i-1/2}
\end{bmatrix}
=
\begin{bmatrix}
0 \\
 0 \\
  \frac{\overline{q}_{1,i-1/2} }{\rho_U \Delta s} \\
 \frac{\overline{q}_{2,i-1/2} }{\rho_L \Delta s} 
\end{bmatrix}.
\label{eq:semi-discrete/model/conservative_numerical_flux_vector} 
\end{equation}
The semi-discrete version of the volume constraint is given by
\begin{equation*}
\frac{q_{1,i}}{\rho_U \Delta s} + \frac{q_{2,i}}{\rho_L \Delta s}  = A,
\end{equation*}
which implies the volumetric flow constraint \cite{BuistSanderseDubinkinaEtAl2023}
\begin{equation}
\llbracket Q_i \rrbracket = 0, \quad \text{with} \quad Q_{i-1/2}(\v{q}_{i}) \coloneqq \frac{q_{3,i-1/2}}{\rho_U \Delta s} + \frac{q_{4,i-1/2}}{\rho_L \Delta s}.
\label{eq:semi-discrete/model/volumetric_flow_constraint}
\end{equation}

\subsection{Energy conservation for the semi-discrete basic model}
\label{ssec:semi-discrete/basic}

The basic TFM, discretized as given above, has been shown in \cite{BuistSanderseDubinkinaEtAl2022} to conserve the following mechanical energy:
\begin{equation}
e_{b,i-1/2}  = \rho_U g_n \overline{\widetilde{H}}_{U,i-1/2} \Delta s +  \rho_L g_n \overline{\widetilde{H}}_{L,i-1/2} \Delta s + \frac{1}{2} \frac{q_{3,i-1/2}^2}{ \overline{q}_{1,i-1/2}  } + \frac{1}{2} \frac{q_{4,i-1/2}^2}{  \overline{q}_{2,i-1/2}  }.  
\label{eq:semi-discrete/basic/energy_definition}
\end{equation}
From this definition, the $\v{v}_b$ vectors can be calculated as
\begin{align*}
\v{v}_{b,i-1/2,i-1}  &\coloneqq \left[\pd{e_{b,i-1/2}}{\v{q}_{i-1}}\right]^T =
\begin{bmatrix}
- \frac{1}{4 } \frac{q_{3,i-1/2}^2}{  \overline{q}_{1,i-1/2}^2 } + \frac{1}{2} g_n \left(\d{\widetilde{H}_U}{A_U}\right)_{i-1}  \\
- \frac{1}{4} \frac{q_{4,i-1/2}^2}{  \overline{q}_{2,i-1/2}^2 } + \frac{1}{2} g_n \left(\d{\widetilde{H}_L}{A_L}\right)_{i-1} \\
0\\
0
\end{bmatrix}, \\
\v{v}_{b,i-1/2,i}  &\coloneqq \left[\pd{e_{b,i-1/2}}{\v{q}_i}\right]^T =
\begin{bmatrix}
- \frac{1}{4 } \frac{q_{3,i-1/2}^2}{  \overline{q}_{1,i-1/2}^2 } + \frac{1}{2 } g_n \left(\d{\widetilde{H}_U}{A_U}\right)_{i} \\
- \frac{1}{4 } \frac{q_{4,i-1/2}^2}{  \overline{q}_{2,i-1/2}^2 } + \frac{1}{2} g_n \left(\d{\widetilde{H}_L}{A_L}\right)_{i} \\
\frac{q_{3,i-1/2}}{\overline{q}_{1,i-1/2}} \\
 \frac{q_{4,i-1/2}}{\overline{q}_{2,i-1/2}} 
\end{bmatrix}.
\end{align*}
The sum of the dot products of $\v{v}_{b,i-1/2,i-1}$ and $\v{v}_{b,i-1/2,i}$ with equation \eqref{eq:semi-discrete/model/finite_volume_scheme} for $\v{q}_{i-1}$ and $\v{q}_{i}$ respectively is 
\begin{multline}
\dotp{ \v{v}_{b,i-1/2,i-1} }{ \d{\v{q}_{i-1} }{t} }  + \dotp{ \v{v}_{b,i-1/2,i} }{\d{\v{q}_i}{t} } + \left< \v{v}_{b,i-1/2,i-1}, \llbracket \v{f}_{i-1} \rrbracket \right> + \left< \v{v}_{b,i-1/2,i}, \llbracket \v{f}_{i} \rrbracket \right>   \\
+ \dotp{ \v{v}_{b,i-1/2,i-1} }{\v{j}_{i-1} } \llbracket p_{i-3/2}  \rrbracket  + \dotp{\v{v}_{b,i-1/2,i} }{ \v{j}_i } \llbracket p_{i-1/2} \rrbracket
 = 0.
  \label{eq:semi-discrete/basic/energy_equation_v1}
 \end{multline}
 In the first two terms, we recognize the time derivative of the energy:
 \begin{equation*}
\dotp{ \v{v}_{b,i-1/2,i-1} }{ \d{\v{q}_{i-1} }{t} }  + \dotp{ \v{v}_{b,i-1/2,i} }{\d{\v{q}_i}{t} } =  \d{e_{b,i-1/2}}{t}.
\end{equation*}

Using the following definitions:
\begin{align*}
\overline{\v{v}}_{b,i,i-1/2} &= \frac{1}{2} \left( \v{v}_{b,i-1/2,i-1} + \v{v}_{b,i+1/2,i} \right), \quad &\quad \overline{\v{v}}_{b,i,i+1/2} &= \frac{1}{2} \left( \v{v}_{b,i-1/2,i} + \v{v}_{b,i+1/2,i+1} \right),  \\
\llbracket \v{v}_{b,i,i-1/2} \rrbracket &= \v{v}_{b,i+1/2,i} - \v{v}_{b,i-1/2,i-1}, \quad &\quad \llbracket \v{v}_{b,i,i+1/2} \rrbracket &= \v{v}_{b,i+1/2,i+1} - \v{v}_{b,i-1/2,i},
\end{align*}
and in addition discrete versions of the geometric relations \eqref{eq:geometric_relations/essential_geometric_relations}\footnote{These are only exactly satisfied by geometries with $\mathrm{d}^2 H_L/\mathrm{d} A_L^2=0$ (for example the 2D channel geometry).},
the volumetric flow constraint \eqref{eq:semi-discrete/model/volumetric_flow_constraint}, 
and substituting our discretization given by \eqref{eq:semi-discrete/model/conservative_numerical_flux_vector}, 
it can be shown that the remaining terms in \eqref{eq:semi-discrete/basic/energy_equation_v1} can be written as the difference of an energy flux  \cite{BuistSanderseDubinkinaEtAl2022}:
\begin{gather*}
  \left< \v{v}_{b,i-1/2,i-1}, \llbracket \v{f}_{i-1} \rrbracket \right>  +  \dotp{\v{v}_{b,i-1/2,i}}{ \llbracket \v{f}_{i} \rrbracket }  = \left\llbracket h_{f,i-1/2} \right\rrbracket, \\
\dotp{ \v{v}_{b,i-1/2,i-1} }{\v{j}_{i-1} } \llbracket p_{i-3/2}  \rrbracket  + \dotp{\v{v}_{b,i-1/2,i} }{ \v{j}_i } \llbracket p_{i-1/2} \rrbracket = \left\llbracket h_{p,i-1/2} \right\rrbracket,
\end{gather*}
with $h_{f,i}$ a discrete version of \eqref{eq:continuous/basic/hf_definition}\footnote{The given expression for $h_{f,i}$ is different from, but equivalent to, the expression given in \cite{BuistSanderseDubinkinaEtAl2023} (under the current assumptions).}:
\begin{equation}
h_{f,i}  =
 g_n  \left( \d{\widetilde{H}_{U}}{A_{U}}\right)_{i} \frac{\overline{q}_{3,i}}{\Delta s} + g_n  \left( \d{\widetilde{H}_{L}}{A_{L}}\right)_{i} \frac{\overline{q}_{4,i}}{\Delta s} 
 +\left( \left[  \overline{\left( \frac{q_{3,i}}{\overline{q}_{1,i}} \right)}  \right]^2
 -\frac{1}{2} \overline{ \left( \frac{q_{3,i}^2}{  \overline{q}_{1,i}^2 } \right) } \right) \frac{\overline{q}_{3,i}}{\Delta s} 
+ \left( \left[  \overline{\left( \frac{q_{4,i}}{\overline{q}_{2,i}} \right)}  \right]^2
 -\frac{1}{2} \overline{ \left( \frac{q_{4,i}^2}{  \overline{q}_{2,i}^2 } \right) } \right) \frac{\overline{q}_{4,i}}{\Delta s},
 \label{eq:semi-discrete/basic/hf_definition_1}
\end{equation}
and $h_{p,i}$ a discrete version of \eqref{eq:continuous/basic/hp_definition}:
\begin{equation}
h_{p,i} = Q(t) p_{i}.
\label{eq:semi-discrete/basic/hp_definition_1}
\end{equation}

Therefore, \eqref{eq:semi-discrete/basic/energy_equation_v1} reduces to the local energy conservation equation
\begin{equation}
\d{e_{b,i-1/2}}{t} + \llbracket h_{b,i-1/2} \rrbracket = 0,
\label{eq:semi-discrete/basic/local_energy_conservation_equation}
\end{equation}
with $h_{b,i} = h_{f,i} + h_{p,i}$.
Like in the continuous case, this equation can be integrated over a closed or periodic domain to yield 
\begin{equation*}
\d{\widehat{E}_{b}}{t} = 0, \quad \text{with} \quad \widehat{E}_{b}(t) = \sum_{i=1}^{N_u} e_{b,i-1/2}(t),
\end{equation*}
which means that the discrete mechanical energy defined by \eqref{eq:semi-discrete/basic/energy_definition} is a secondary conserved quantity of the semi-discrete model described in \autoref{ssec:semi-discrete/model}.

\subsection{Energy equation for the semi-discrete extended model}
\label{ssec:semi-discrete/summary}

Having introduced the energy-conserving discretization of the basic TFM and its energy conservation equation, we will propose discretizations of the three additions to the basic model, that retain the energy properties of their continuous counterparts, which were derived in \autoref{sec:continuous}. 
We will present discretizations of friction and diffusion that are strictly dissipative, and a discretization of surface tension that conserves the discretized energy, extended with a discrete version of the surface energy. 

Additionally, we will present an upwind discretization of the advective terms that can be shown to be strictly dissipative, in contrast to the energy-conserving discretization of the advective terms given by \eqref{eq:semi-discrete/model/conservative_numerical_flux_vector}.
We will then propose to combine this upwind advective flux with the energy-conserving advective flux (using flux limiters), to produce a combined flux that is strictly dissipative, but less dissipative and less diffusive than the purely upwind flux.
The combined flux is energy stable, and adds numerical dissipation only where necessary: near strong gradients and discontinuities. 
This mimics the behavior of weak solutions to the (basic) continuous equations, which instead of the energy equality of \autoref{sec:continuous/basic} (that is only valid for smooth solutions), will satisfy an energy inequality. 

The semi-discrete model, extended with all the additional terms, is given by
\begin{equation}
\boxed{\d{\v{q}_{i}}{t} + \left\llbracket \v{f}_{i} \right\rrbracket + \v{j}_i \left\llbracket p_{i-1/2} \right\rrbracket = \left\llbracket \v{d}_i \right\rrbracket + \v{s}_{i} \Delta s + \v{c}_{g,i} \Delta s + \v{c}_{f,i} \Delta s + \v{c}_{p,i} \Delta s}
\label{eq:semi-discrete/summary/system}
\end{equation}
with $\left\llbracket \v{d}_{i} \right\rrbracket$ representing diffusion, $\v{c}_{f,i}$ representing friction, and $\v{s}_{i}$ representing surface tension.
The expressions for these terms will be given in \eqref{eq:semi-discrete/physical_diffusion/model_addition_d}, \eqref{eq:semi-discrete/friction/model_addition}, and \eqref{eq:semi-discrete/surface_tension/general_discretization}, respectively.
The extended semi-discrete model includes the following contributions from streamwise gravity:
\begin{equation*}
\v{c}_{g,i}^T = 
\begin{bmatrix}
0 &
\quad 0 &
\quad -  g \frac{\left\llbracket y_{i-1/2} \right \rrbracket}{\Delta s} \frac{\overline{q}_{1,i-1/2}}{\Delta s}  &
\quad -  g \frac{\left\llbracket y_{i-1/2} \right \rrbracket}{\Delta s} \frac{\overline{q}_{2,i-1/2}}{\Delta s}
\end{bmatrix}
\quad \text{with} \quad 
 \frac{\left\llbracket y_{i-1/2} \right \rrbracket}{\Delta s} = \left[\sin(\phi)\right]_{i-1/2},
\end{equation*}
and from a constant driving pressure gradient: 
\begin{equation*}
\v{c}_{p,i}^T = 
\begin{bmatrix}
0 &
\quad 0 &
\quad -  \frac{\overline{q}_{1,i-1/2}}{\rho_U \Delta s} \pd{p_0}{s}  &
\quad -  \frac{\overline{q}_{2,i-1/2}}{\rho_L \Delta s} \pd{p_0}{s} 
\end{bmatrix}.
\end{equation*} 

The energy equation that follows from \eqref{eq:semi-discrete/summary/system} reads
\begin{equation}
\boxed{\d{e_{i-1/2}}{t} +  \left\llbracket h_{i-1/2} \right\rrbracket + \left\llbracket h_{n, i-1/2} \right\rrbracket =  - \epsilon_{i-1/2} - \epsilon_{n, i-1/2} + c_{p,i-1/2}}
\label{eq:semi-discrete/summary/energy_equation}
\end{equation}
with
\begin{gather}
e_{i-1/2} = e_{b,i-1/2} + e_{g,i-1/2} + e_{\sigma,i-1/2}, \label{eq:semi-discrete/summary/local_energy} \\
h_{i} = h_{b,i} + h_{g,i} + h_{d,i} + h_{\sigma,i}, \\
\epsilon_{i-1/2} = \epsilon_{d, i-1/2} + \epsilon_{f, i-1/2}.
\end{gather}
The upcoming subsections will give the expressions  \eqref{eq:semi-discrete/physical_diffusion/h_d} for  $h_{d,i}$, \eqref{eq:semi-discrete/physical_diffusion/epsilon_d} for $\epsilon_{d, i-1/2}$, \eqref{eq:semi-discrete/friction/epsilon_f} for $\epsilon_{f, i-1/2}$,  \eqref{eq:semi-discrete/surface_tension/energy_definition_final} for  $e_{\sigma,i-1/2}$, and \eqref{eq:semi-discrete/surface_tension/h_sigma} for $h_{\sigma,i}$.
The contributions from streamwise gravity are given by \cite{BuistSanderseDubinkinaEtAl2023}\footnote{The given expression for $h_{g,i}$ is different from, but equivalent to, the expression given in \cite{BuistSanderseDubinkinaEtAl2023}.}
\begin{equation*}
e_{g,i-1/2} = g  \overline{\left( y_{i-1/2} \left(  q_{1,i-1/2} + q_{2,i-1/2}\right)\right)}, \quad h_{g,i} =  g  y_{i} \overline{\left( \frac{q_{3,i}}{\Delta s} + \frac{q_{4,i}}{\Delta s} \right)},
\end{equation*}
and the contribution of the driving pressure gradient is given by 
\begin{equation*}
c_{p,i-1/2} = - Q_{i-1/2} \pd{p_0}{s} \Delta s,
\end{equation*}
which is positive when the body force is aligned with the mean flow. 
These are all semi-discrete counterparts of the continuous expressions given in \autoref{sec:continuous}.
The terms with subscript $n$ are specific to the semi-discrete setting, and stem from the energy-stable combined advective flux.
The numerical energy flux $h_{n,i}$ will be given by \eqref{eq:semi-discrete/limiters/hl} and the numerical dissipation $\epsilon_{n,i-1/2}$ will be given by \eqref{eq:semi-discrete/limiters/epsilon_l}.

Equation \eqref{eq:semi-discrete/summary/energy_equation} is the second main novel result of this work, as it shows that we have obtained a semi-discrete model with the same energy conservation properties as the continuous model.
The model additions that were conservative in the continuous setting are discretized in such a way that the energy-conserving behavior is retained, and the model additions that were dissipative in the continuous setting are discretized in such a way that the strictly dissipative behavior is retained. 
The local energy equation \eqref{eq:semi-discrete/summary/energy_equation} can be integrated over a periodic domain to yield the global energy equation
\begin{gather}
\boxed{\d{\widehat{E}}{t} = -\widehat{\mathcal{E}} -\widehat{\mathcal{E}}_n + \widehat{C}_p} \quad \text{with} \quad \widehat{E}(t) = \sum_{i=1}^{N_u} e_{i-1/2}(t), \quad \widehat{\mathcal{E}} = \sum_{i=1}^{N_u} \epsilon_{i-1/2}(t), \quad \widehat{\mathcal{E}}_n = \sum_{i=1}^{N_u} \epsilon_{n,i-1/2}(t),
\label{eq:semi-discrete/summary/global_energy_conservation_equation} \\
\widehat{C}_p = \sum_{i=1}^{N_u} c_{p,i-1/2}(t) = -  Q(t) \pd{p_0}{s} L. \nonumber
\end{gather}
This equation determines that the energy of the solution can never increase, except due to an explicitly applied external force (through the source term $\widehat{C}_{p}$). 

Therefore, the novel semi-discrete model is energy stable.
It has physical and numerical dissipation rates ($\widehat{\mathcal{E}}$ and $\widehat{\mathcal{E}}_n$) that can be computed from the solution. 
These dissipation rates can be integrated in time numerically to find the total dissipated energy due to the different contributions. 
The total dissipated energy between two points in time should match the difference in energy between these two points in time, as calculated through an evaluation of $\widehat{E}$ at those two points in time. 

Each term in the semi-discrete model independently results in corresponding terms in the energy equation.
The combined result for the complete extended model, discretized in an energy-consistent manner, was given here.
In the following subsections we will detail the novel contributions separately. 

\subsection{Physical diffusion}

Our first novel contribution in the semi-discrete setting is to propose a discretization of the viscous diffusion terms and prove that it is strictly dissipative, just like its continuous counterpart. 
The diffusion term that can be added to the RHS of \eqref{eq:semi-discrete/model/finite_volume_scheme} is $\left[ \partial \v{d} / \partial s \right]_{i} \Delta s$, 
in which $\left[ \partial \v{d} / \partial s \right]_{i}$ is the discrete version of $\partial \v{d} / \partial s$, with $\v{d}$ given by \eqref{eq:continuous/physical_diffusion/model_addition_d}.
We propose the following straightforward central discretization, which yields the diffusion term in \eqref{eq:semi-discrete/summary/system}:
\begin{equation*}
\left[ \pd{\v{d}}{s} \right]_{i} = \frac{1}{\Delta s} \left\llbracket \v{d}_i \right\rrbracket = \frac{1}{\Delta s} \left(\v{d}_{i+1/2} - \v{d}_{i-1/2}\right),
\end{equation*}
with
\begin{equation}
\v{d}_{i-1/2} \coloneqq
\begin{bmatrix}
d_{1,i-1/2} \\
d_{2,i-1/2} \\
d_{3,i-1}  \\
d_{4,i-1}
\end{bmatrix}
=
\begin{bmatrix}
0 \\
0 \\
\nu_{\mathrm{eff},U} \frac{q_{1,i-1}}{\left(\Delta s\right)^2}  \left\llbracket \frac{q_{3,i-1}}{\overline{q}_{1,i-1}}   \right\rrbracket \\
 \nu_{\mathrm{eff},L} \frac{q_{2,i-1}}{\left(\Delta s\right)^2}  \left\llbracket \frac{q_{4,i-1}}{\overline{q}_{2,i-1}}   \right\rrbracket 
\end{bmatrix}
=
\begin{bmatrix}
0 \\
0 \\
 \rho_U \nu_{\mathrm{eff},U} A_{U,i-1} \frac{\left\llbracket u_{U,i-1}  \right\rrbracket}{\Delta s} \\
 \rho_L \nu_{\mathrm{eff},L}  A_{L,i-1} \frac{\left\llbracket  u_{L,i-1}  \right\rrbracket}{\Delta s} 
\end{bmatrix}.
\label{eq:semi-discrete/physical_diffusion/model_addition_d}
\end{equation}

With diffusion, no extra term is added to the energy.
The steps of the derivation of \autoref{ssec:semi-discrete/basic} can be simply repeated.
With the proposed discretization, the only additional terms in the energy equation (on the LHS) are 
\begin{equation}
\begin{split}
&-\left< \v{v}_{b,i-1/2,i-1}, \llbracket \v{d}_{i-1} \rrbracket \right>  -  \left< \v{v}_{b,i-1/2,i}, \llbracket \v{d}_{i} \rrbracket \right> \\
 & \quad \quad \quad = -\left\llbracket \dotp{ \overline{\v{v}}_{b,i-1/2,i-1} }{ \v{d}_{i-1} } \right\rrbracket  -  \left\llbracket \dotp{ \overline{\v{v}}_{b,i-1/2,i} }{ \v{d}_{i} }  \right\rrbracket  + \overline{ \dotp{\llbracket \v{v}_{b,i-1/2,i-1} \rrbracket }{ \v{d}_{i-1} }} + \overline{  \dotp{ \llbracket \v{v}_{b,i-1/2,i} \rrbracket }{ \v{d}_{i} } } \\
   & \quad \quad \quad \begin{multlined} =
    -\left\llbracket \nu_{\mathrm{eff},U}  \frac{q_{1,i-1/2}}{\left(\Delta s\right)^2} \overline{\left(\frac{q_{3,i-1/2}}{\overline{q}_{1,i-1/2}}\right)} \left\llbracket \frac{q_{3,i-1/2}}{\overline{q}_{1,i-1/2}} \right\rrbracket \right\rrbracket 
 + \overline{\left(  \nu_{\mathrm{eff},U}  \frac{q_{1,i-1/2}}{\left(\Delta s\right)^2} \left\llbracket \frac{q_{3,i-1/2}}{\overline{q}_{1,i-1/2}} \right\rrbracket^2 \right) } \\
-\left\llbracket \nu_{\mathrm{eff},L}  \frac{q_{2,i-1/2}}{\left(\Delta s\right)^2} \overline{\left(\frac{q_{4,i-1/2}}{\overline{q}_{2,i-1/2}}\right)} \left\llbracket \frac{q_{4,i-1/2}}{\overline{q}_{2,i-1/2}} \right\rrbracket \right\rrbracket 
 + \overline{\left(  \nu_{\mathrm{eff},L}  \frac{q_{2,i-1/2}}{\left(\Delta s\right)^2} \left\llbracket \frac{q_{4,i-1/2}}{\overline{q}_{2,i-1/2}} \right\rrbracket^2 \right) }
 \end{multlined} \\
  & \quad \quad \quad = \left\llbracket h_{d,i-1/2} \right\rrbracket + \epsilon_{d,i-1/2},
 \end{split}
 \label{eq:semi-discrete/physical_diffusion/derived_energy_equation_contribution}
\end{equation}
with 
\begin{gather}
h_{d,i} =  - \nu_{\mathrm{eff},U}  \frac{q_{1,i}}{\left(\Delta s\right)^2} \frac{1}{2} \left\llbracket \left(\frac{q_{3,i}}{\overline{q}_{1,i}}\right)^2 \right\rrbracket 
- \nu_{\mathrm{eff},L}  \frac{q_{2,i}}{\left(\Delta s\right)^2} \frac{1}{2} \left\llbracket \left(\frac{q_{4,i}}{\overline{q}_{2,i}}\right)^2 \right\rrbracket, 
\label{eq:semi-discrete/physical_diffusion/h_d} \\
\epsilon_{d,i-1/2} =  \overline{\left(  \nu_{\mathrm{eff},U}  \frac{q_{1,i-1/2}}{\left(\Delta s\right)^2} \left\llbracket \frac{q_{3,i-1/2}}{\overline{q}_{1,i-1/2}} \right\rrbracket^2 \right) } + \overline{\left(  \nu_{\mathrm{eff},L}  \frac{q_{2,i-1/2}}{\left(\Delta s\right)^2} \left\llbracket \frac{q_{4,i-1/2}}{\overline{q}_{2,i-1/2}} \right\rrbracket^2 \right) }.
\label{eq:semi-discrete/physical_diffusion/epsilon_d}
\end{gather}
Here we have used a discrete product rule:
\begin{equation*}
a_{i} \llbracket b_{i} \rrbracket =  \llbracket \overline{a}_{i} b_{i} \rrbracket -  \overline{ \left( \llbracket a_{i} \rrbracket  b_{i} \right) },
\end{equation*}
and a discrete chain rule:
\begin{equation*}
\overline{a}_{i-1/2} \left\llbracket a_{i-1/2} \right\rrbracket = \frac{1}{2} \left\llbracket a_{i-1/2}^2 \right\rrbracket.
\end{equation*}
These can be derived by substituting the definitions \eqref{eq:semi-discrete/model/interpolation_definition} and \eqref{eq:semi-discrete/model/jump_definition}, and applying some algebraic manipulation. 

The result of \eqref{eq:semi-discrete/physical_diffusion/derived_energy_equation_contribution} compares directly to the continuous result, given by \eqref{eq:continuous/physical_diffusion/derived_energy_equation_contribution}.
Like the continuous result, it consists of a conservative part, and a strictly dissipative part.
The latter is due to $\epsilon_{d,i-1/2}$ being strictly positive, and the minus sign that is added when $\epsilon_{d,i-1/2}$ is moved to the RHS. 
Therefore, it has been proven that the proposed discretization of the diffusion terms is strictly dissipative with respect to the energy given by \eqref{eq:semi-discrete/summary/local_energy}. 
Moreover, an explicit expression for the dissipation rate has been obtained, that can be used to measure the dissipation taking place in a numerical simulation.

\subsection{Friction terms}
\label{sec:semi-discrete/friction}

Our second novel contribution in the semi-discrete setting is to show that wall and interface friction result in a strictly dissipative contribution to the semi-discrete energy equation.
In \eqref{eq:semi-discrete/summary/system}, friction is included through the term $\v{c}_{f,i} \Delta s$, 
in which $\v{c}_{f,i} $ is  the discrete version of \eqref{eq:continuous/friction/model_addition_cf}.
With reference to the closure relations given in \autoref{ssec:continous/friction}, we assume the following functional dependencies for the discrete friction terms:
\begin{gather*}
  \tau_{L,i-1/2} \coloneqq   \tau_L(f_{L,i-1/2}, u_{L,i-1/2}), \quad
 \tau_{U,i-1/2} \coloneqq \tau_U(f_{U,i-1/2}, u_{U,i-1/2}), \\
 \tau_{\mathrm{int},i-1/2} \coloneqq  \tau_\mathrm{int}(f_{\mathrm{int},i-1/2}, u_{U,i-1/2}, u_{L,i-1/2}),
 \end{gather*}
with primitive variables given by \eqref{eq:semi-discrete/model/primitive_translation}.
Then we propose the following discretization of the friction source terms: 
\begin{equation}
\begin{split}
\v{c}_{f,i} 
&=
\begin{bmatrix}
c_{f,1,i} \\
c_{f,2,i} \\
c_{f,3,i-1/2}  \\
c_{f,4,i-1/2}
\end{bmatrix}
=
\begin{bmatrix}
0 \\
0 \\
 \tau_{U,i-1/2} P_{U,i-1/2} + \tau_{\mathrm{int},i-1/2} P_{\mathrm{int},i-1/2} \\
  \tau_{L,i-1/2} P_{L,i-1/2} - \tau_{\mathrm{int},i-1/2} P_{\mathrm{int},i-1/2}
\end{bmatrix}.
\end{split}
\label{eq:semi-discrete/friction/model_addition}
\end{equation}

Like with diffusion, adding friction terms to the system does not change the energy definition. 
Since the friction terms do not involve derivatives, the derivation of the contribution to the energy equation is almost the same as in the continuous case.
The only modification to the energy equation is that the following terms are added to the RHS:
\begin{equation*}
\begin{split}
& \dotp{\v{v}_{b,i-1/2,i-1}}{\v{c}_{f,i-1} \Delta s} + \dotp{\v{v}_{b,i-1/2,i}}{\v{c}_{f,i} \Delta s} \\
& \quad 
=\frac{q_{3,i-1/2}}{\overline{q}_{1,i-1/2}}  \left(  \tau_{U,i-1/2} P_{U,i-1/2} + \tau_{\mathrm{int},i-1/2} P_{\mathrm{int},i-1/2} \right) \Delta s+  \frac{q_{4,i-1/2}}{\overline{q}_{2,i-1/2}}  \left(  \tau_{L,i-1/2} P_{L,i-1/2} - \tau_{\mathrm{int},i-1/2} P_{\mathrm{int},i-1/2} \right) \Delta s \\
&  \quad = - \epsilon_{f,i-1/2},
 \end{split}
\end{equation*}
with
\begin{multline}
\epsilon_{f,i-1/2} = 
   \frac{1}{2} f_{U,i-1/2} \rho_U \left( \frac{q_{3,i-1/2}}{\overline{q}_{1,i-1/2}} \right)^2 \abs{ \frac{q_{3,i-1/2}}{\overline{q}_{1,i-1/2}} } P_{U,i-1/2} \Delta s  + \frac{1}{2} f_{L,i-1/2} \rho_L \left( \frac{q_{4,i-1/2}}{\overline{q}_{2,i-1/2}} \right)^2 \abs{ \frac{q_{4,i-1/2}}{\overline{q}_{2,i-1/2}} } P_{L,i-1/2} \Delta s \\
  + \frac{1}{2}  f_{\mathrm{int},i-1/2}  \rho_U  \left( \frac{q_{3,i-1/2}}{\overline{q}_{1,i-1/2}} - \frac{q_{4,i-1/2}}{\overline{q}_{2,i-1/2}} \right)^2 \abs{\frac{q_{3,i-1/2}}{\overline{q}_{1,i-1/2}} - \frac{q_{4,i-1/2}}{\overline{q}_{2,i-1/2}} }P_{\mathrm{int},i-1/2} \Delta s.
  \label{eq:semi-discrete/friction/epsilon_f}
\end{multline}
Since $f_{U,i-1/2}$, $f_{L,i-1/2}$, $f_{\mathrm{int},i-1/2}$, $\rho_U$, $\rho_L$, $P_{U,i-1/2}$, $P_{L,i-1/2}$, and $P_{\mathrm{int},i-1/2}$ must be positive, and the rest of the terms are either quadratic or absolute, $\epsilon_{f,i-1/2}$ must always be positive.
Therefore, $-\epsilon_{f,i-1/2}$ will act as a sink in the energy equation.

This result allows us, for the first time, to compare the dissipation due to wall and interface friction with the dissipation due to axial diffusion. 
Both components of the dissipation rate can be computed from the numerical solution, integrated numerically over time, and compared to one another to determine which has dissipated the most energy. 

\subsection{Surface tension}
\label{ssec:semi-discrete/surface_tension}

Our third novel contribution in the semi-discrete setting is to propose a discretization of the surface tension, and show that it is energy-conserving.
This is key in maintaining the energy-stable nature of the semi-discrete model, while contributing favorably to the linear stability properties of the model; see \autoref{sec:stability}. 

Surface tension is included in \eqref{eq:semi-discrete/summary/system} through the term $\v{s}_{i} \Delta s$, 
in which $\v{s}_{i}$ is  the discrete version of \eqref{eq:continuous/surface_tension/general_model_addition_s}.
A general form of the surface tension, analogous to \eqref{eq:continuous/surface_tension/general_model_addition_s}, is given by
\begin{equation}
\v{s}_{i} =
\begin{bmatrix}
0 \\
0 \\
0 \\
\frac{\overline{q}_{2,i-1/2}}{\rho_L \Delta s}  \frac{1}{\Delta s} \left\llbracket \left[\Delta p \right]_{i-1/2} \right\rrbracket
\end{bmatrix}.
\label{eq:semi-discrete/surface_tension/general_discretization}
\end{equation}
We restrict the analysis to the channel geometry, for which clear results were obtained in the continuous analysis, and propose the following discretization of the pressure jump given in \eqref{eq:continuous/surface_tension/final_results}:
\begin{equation}
\left[\Delta p_\mathrm{int}\right]_{i} = \frac{\sigma}{\Delta s} \left\llbracket \frac{\left\llbracket H_{L,i} \right\rrbracket}{\Delta s} \right\rrbracket.
\label{eq:semi-discrete/surface_tension/pressure_jump_final}
\end{equation}
This constitutes a straightforward discretization of the conventional approximation of the surface tension in the TFM. 

With the proposed discretization, the extra terms on the LHS of the energy equation can be written as
\begin{equation}
\begin{split}
& -\dotp{\v{v}_{b,i-1/2,i-1}}{ \v{s}_{i-1} \Delta s}  -\dotp{\v{v}_{b,i-1/2,i}}{\v{s}_{i} \Delta s}  \\
& \quad \quad \quad \quad
= -\frac{q_{4,i-1/2}}{\rho_L \Delta s}   \left\llbracket \frac{\sigma}{\Delta s} \left\llbracket \frac{ \left\llbracket H_{L,i-1/2} \right\rrbracket }{\Delta s} \right\rrbracket \right\rrbracket \\
& \quad \quad \quad \quad = 
- \frac{\sigma}{\rho_L}  \left\llbracket \overline{\left( \frac{q_{4,i-1/2}}{\Delta s} \right)} \frac{1}{\Delta s} \left\llbracket \frac{ \left\llbracket H_{L,i-1/2} \right\rrbracket}{\Delta s} \right\rrbracket \right\rrbracket
 + \frac{\sigma}{\rho_L} \overline{\left(\left\llbracket \frac{q_{4,i-1/2}}{\Delta s} \right\rrbracket \frac{1}{\Delta s} \left\llbracket \frac{ \left\llbracket H_{L,i-1/2} \right\rrbracket}{\Delta s} \right\rrbracket \right)} \\
& \quad \quad \quad \quad =
 \left\llbracket h_{\sigma,i-1/2}  \right\rrbracket 
+ \frac{\sigma}{\rho_L }  \left\llbracket \frac{1}{\Delta s} \d{q_{2,i-1/2}}{t}  \right\rrbracket \frac{ \left\llbracket H_{L,i-1/2} \right\rrbracket}{\Delta s} \\
& \quad \quad \quad \quad =
 \left\llbracket  h_{\sigma,i-1/2}  \right\rrbracket 
+  \sigma \d{}{t} \left( \frac{\left\llbracket  H_{L,i-1/2} \right\rrbracket}{\Delta s} \right)  \left\llbracket H_{L,i-1/2} \right\rrbracket \\
& \quad \quad \quad \quad=
 \left\llbracket  h_{\sigma,i-1/2}  \right\rrbracket 
+ \sigma \d{e_{\sigma, i-1/2}}{t},
\end{split}
\label{eq:semi-discrete/surface_tension/derivation}
\end{equation}
in which, based on \eqref{eq:continuous/surface_tension/final_results}, we have chosen to define the surface energy as:
\begin{equation}
e_{\sigma, i-1/2} = \sigma \Delta s \left(1 + \frac{1}{2} \left(\frac{\left\llbracket H_{L,i-1/2} \right\rrbracket}{\Delta s}\right)^2\right), 
\label{eq:semi-discrete/surface_tension/energy_definition_final}
\end{equation}
with an energy flux of 
\begin{equation}
h_{\sigma, i} =  \sigma \frac{1}{\Delta s}\left\llbracket \frac{q_{4,i}}{\rho_L \Delta s} \right\rrbracket  \overline{ \left(\frac{\left\llbracket H_{L,i} \right\rrbracket}{\Delta s}  \right)} 
-  \sigma \overline{\left( \frac{q_{4,i}}{\rho_L \Delta s} \right)} \frac{1}{\Delta s}\left\llbracket \frac{\left\llbracket H_{L,i} \right\rrbracket}{\Delta s}  \right\rrbracket.
\label{eq:semi-discrete/surface_tension/h_sigma}
\end{equation}
In this derivation, we have used a discrete version of the product rule that can be derived from \eqref{eq:semi-discrete/model/interpolation_definition} and \eqref{eq:semi-discrete/model/jump_definition}:
\begin{equation*}
 \llbracket \overline{a}_{i-1/2} b_{i-1/2} \rrbracket  = \overline{ \left( \llbracket a_{i-1/2} \rrbracket  b_{i-1/2} \right) } + a_{i-1/2} \llbracket b_{i-1/2} \rrbracket,
 \end{equation*}
 and substituted the semi-discrete mass conservation equation for the lower fluid (the second equation of \eqref{eq:semi-discrete/model/finite_volume_scheme}), specified to the 2D channel.

The derivation \eqref{eq:semi-discrete/surface_tension/derivation} shows that, for the surface tension discretization given by \eqref{eq:semi-discrete/surface_tension/general_discretization} and \eqref{eq:semi-discrete/surface_tension/pressure_jump_final}, the contribution of surface tension to the semi-discrete energy equation can be written in conservative form.
This requires adding an extra term to the energy, which must take a specific form that is tied to this discretization. 
We have succeeded in finding a combination of $\left[\Delta p_\mathrm{int}\right]_{i}$ and $e_{\sigma, i-1/2}$ that is energy-conserving.
These results hold for the 2D channel geometry. 

Therefore, we have found a way to add surface tension to the semi-discrete model, in such a way that it remains energy stable. 
Consistent with the physics of the flow, no dissipation (or production) of energy will result from surface tension.
All dissipation can be attributed to effects that would physically be expected to yield dissipation: diffusion and wall and interface friction.

\subsection{Numerical diffusion}
\label{ssec:semi-discrete/numerical_diffusion}

In addition to the energy-consistent discretizations of the extra terms in the continuous model, a modification is needed to our discretization of the basic model, as given by \eqref{eq:semi-discrete/model/conservative_numerical_flux_vector}. 
In this subsection we will propose an upwind discretization of the advective terms in the momentum equations, that can be shown to add a strictly dissipative term to the energy equation, and is therefore energy stable. 

This is required because our energy-conserving central advective flux, included in \eqref{eq:semi-discrete/model/conservative_numerical_flux_vector}, is sensitive to discontinuities. 
At discontinuities, the proofs of energy conservation for the continuous equations no longer hold, and the continuous equations should dissipate energy \cite{Jameson2008a}.
However, our energy-conserving flux expressly forbids this.
As a result, when discontinuities appear in the solution, numerical oscillations are generated.
Adding physical diffusion mitigates the problem, but it acts at small scales, and in order to incorporate its full effect, a high grid refinement is required.

Therefore, it is necessary to introduce some form of (strictly dissipative) numerical diffusion. 
The adjective `numerical' indicates that it should be grid-dependent: it is primarily needed at coarse resolutions.
Such diffusion can be provided by an upwind discretization of the advective flux.

Different upwind discretizations can be conceived, by taking different interpolations and by upwinding different parts of the numerical flux, see e.g.\ \cite{IssaKempf2003, KrasnopolskyLukyanov2018, LiaoMeiKlausner2008}. 
Here we present a new upwind discretization that is based on the conservative variables, and closely resembles the energy-conserving flux of \eqref{eq:semi-discrete/model/conservative_numerical_flux_vector}, with the exception that for the advecting velocities we use an upwind interpolation, instead of a central interpolation:
\begin{subequations}
\begin{align}
f_{3,a,i-1,u} &=  \left( \frac{q_{3}}{\overline{q}_{1}} \right)_{\mathrm{up},i-1} \frac{\overline{q}_{3,i-1}}{\Delta s} = \rho_U u_{U,\mathrm{up},i-1}  \overline{ \left(   \overline{A}_{U,i-1} u_{U,i-1} \right)},  \\
f_{4,a,i-1,u} &=  \left( \frac{q_{4}}{\overline{q}_{2}} \right)_{\mathrm{up},i-1} \frac{\overline{q}_{4,i-1}}{\Delta s} = \rho_L u_{L,\mathrm{up},i-1}  \overline{ \left(   \overline{A}_{L,i-1} u_{L,i-1} \right)},
\end{align}
\label{eq:semi-discrete/numerical_diffusion/upwind_fluxes}
\end{subequations}
with
\begin{equation*}
u_{U,\mathrm{up},i-1} =
\begin{cases}
u_{U,i-3/2}, & \text{if}\  \overline{q}_{3,i-1}>0 \\
u_{U,i-1/2}, & \text{otherwise}
\end{cases}
\quad \quad
u_{L,\mathrm{up},i-1} =
\begin{cases}
u_{L,i-3/2}, & \text{if}\  \overline{q}_{4,i-1}>0  \\
u_{L,i-1/2}, & \text{otherwise}
\end{cases}    
\end{equation*}
This upwind flux is atypical in its choice to have $u_U$ and $u_L$ as the upwinded variables instead of $q_3$ and $q_4$, and in its choice to base the upwind directions on $q_3$ and $q_4$ instead of $u_U$ and $u_L$. 
These choices are needed to prove the strictly dissipative property.

We only apply an upwind discretization to the advective terms of the momentum equations, and not to those of the mass equations.
The mass equations are left unchanged in order to not interfere with the coupling between the mass and momentum equations, the connection between the volume and volumetric flow constraints, and the energy analysis which at a few points requires substitution of the mass conservation equations.  

We now show that the contribution to the energy equation of these upwind fluxes can be divided into a conservative part and a non-conservative part.
To this end, the analysis of \autoref{ssec:semi-discrete/basic} is repeated, this time with \eqref{eq:semi-discrete/numerical_diffusion/upwind_fluxes} taking the place of the momentum advection part of the fluxes given in \eqref{eq:semi-discrete/model/conservative_numerical_flux_vector}. 
The contribution of the flux terms to \eqref{eq:semi-discrete/basic/energy_equation_v1} can then be written as
\begin{equation*}
\begin{split}
 & \left< \v{v}_{b,i-1/2,i-1}, \llbracket \v{f}_{i-1} \rrbracket \right> + \left< \v{v}_{b,i-1/2,i}, \llbracket \v{f}_{i} \rrbracket \right> \\
 & \quad \quad \quad = \left\llbracket \dotp{ \overline{\v{v}}_{b,i-1/2,i-1} }{ \v{f}_{i-1} } \right\rrbracket  +  \left\llbracket \dotp{ \overline{\v{v}}_{b,i-1/2,i} }{ \v{f}_{i} }  \right\rrbracket  - \overline{ \dotp{\llbracket \v{v}_{b,i-1/2,i-1} \rrbracket }{ \v{f}_{i-1} }} - \overline{  \dotp{ \llbracket \v{v}_{b,i-1/2,i} \rrbracket }{ \v{f}_{i} } } \\
  & \quad \quad \quad =
\left\llbracket h_{f,i-1/2} \right\rrbracket + \left\llbracket h_{u,i-1/2} \right\rrbracket + \epsilon_{u,i-1/2},
 \end{split}
\end{equation*}
with $h_{f,i}$ given by \eqref{eq:semi-discrete/basic/hf_definition_1}, $h_{u,i}$ given by
\begin{equation}
h_{u,i} 
= - \overline{\left( \frac{q_{3,i}}{\overline{q}_{1,i}} \right)} \left(  \overline{\left( \frac{q_{3,i}}{\overline{q}_{1,i}} \right)}  -  \left( \frac{q_3}{  \overline{q}_1 } \right)_{\mathrm{up},i}   \right) \frac{\overline{q}_{3,i}}{\Delta s}  
 - \overline{\left( \frac{q_{4,i}}{\overline{q}_{2,i}} \right)} \left(  \overline{\left( \frac{q_{4,i}}{\overline{q}_{2,i}} \right)} -  \left( \frac{q_4}{  \overline{q}_2 } \right)_{\mathrm{up},i}   \right) \frac{\overline{q}_{4,i}}{\Delta s}, 
 \label{eq:semi-discrete/upwind/hu}
\end{equation}
and $\epsilon_{u,i-1/2}$ given by
\begin{multline}
\epsilon_{u,i-1/2} = \epsilon_{u,U,i-1/2} + \epsilon_{u,L,i-1/2} = 
 \overline{ \left( \left\llbracket \frac{q_{3,i-1/2}}{  \overline{q}_{1,i-1/2} } \right\rrbracket \left( \overline{\left( \frac{q_{3,i-1/2}}{\overline{q}_{1,i-1/2}} \right)} -  \left( \frac{q_3}{  \overline{q}_1 } \right)_{\mathrm{up},i-1/2} \right) \frac{\overline{q}_{3,i-1/2}}{\Delta s} \right)  } \\
+  \overline{ \left( \left\llbracket \frac{q_{4,i-1/2}}{  \overline{q}_{2,i-1/2} } \right\rrbracket \left( \overline{\left( \frac{q_{4,i-1/2}}{\overline{q}_{2,i-1/2}} \right)} -  \left( \frac{q_4}{  \overline{q}_2 } \right)_{\mathrm{up},i-1/2} \right) \frac{\overline{q}_{4,i-1/2}}{\Delta s} \right)  }.
\label{eq:semi-discrete/upwind/epsilon_u}
\end{multline}
We have split $\epsilon_{u,i-1/2}$ into a part pertaining to the upper fluid ($\epsilon_{u,U,i-1/2}$, first line) and a part pertaining to the lower fluid ($\epsilon_{u,L,i-1/2}$, second line).

This shows that the contribution of the upwind flux can be written as the contribution $\left\llbracket h_{f,i-1/2} \right\rrbracket$ of the energy-conserving flux, plus some extra terms.
Some of these terms are completely between double brackets, meaning they are energy-conserving, and can be included in the energy flux $h_{u,i}$.
The remaining terms will become source terms in the energy equation. 

We will now show that these source terms are strictly dissipative. 
It is sufficient to only examine the terms pertaining to the upper fluid, since the terms pertaining to the lower fluid have the same structure, so their analysis will yield similar results. 
The source terms for the upper fluid can be rewritten in the following manner:
 \begin{equation*}
 \begin{split}
 \epsilon_{u,U,i-1/2} 
& = \frac{1}{2} \left\llbracket \frac{q_{3,i-1}}{  \overline{q}_{1,i-1} } \right\rrbracket \frac{\overline{q}_{3,i-1}}{\Delta s}  \left[ \overline{\left(  \frac{q_{3,i-1}}{  \overline{q}_{1,i-1} } \right)} - \left( \frac{q_3}{  \overline{q}_1 } \right)_{\mathrm{up},i-1}  \right] + \frac{1}{2} \left\llbracket \frac{q_{3,i}}{  \overline{q}_{1,i} } \right\rrbracket  \frac{\overline{q}_{3,i}}{\Delta s} \left[ \overline{\left(  \frac{q_{3,i}}{  \overline{q}_{1,i} } \right)} - \left( \frac{q_3}{  \overline{q}_1 } \right)_{\mathrm{up},i} \right].
\end{split}
\end{equation*}
Now, we consider the case that $\overline{q}_{3,i-1}>0$ and $\overline{q}_{3,i}>0$, so that $u_{U,\mathrm{up},i-1} = u_{U,i-3/2}$ and $u_{U,\mathrm{up},i} = u_{U,i-1/2}$:
 \begin{equation*}
 \begin{split}
 \epsilon_{u,U,i-1/2} 
& = \frac{1}{2} \left\llbracket \frac{q_{3,i-1}}{  \overline{q}_{1,i-1} } \right\rrbracket \frac{\overline{q}_{3,i-1}}{\Delta s}  \left[ \overline{\left(  \frac{q_{3,i-1}}{  \overline{q}_{1,i-1} } \right)} - \frac{q_{3,i-3/2}}{  \overline{q}_{1,i-3/2} }  \right] + \frac{1}{2} \left\llbracket \frac{q_{3,i}}{  \overline{q}_{1,i} } \right\rrbracket  \frac{\overline{q}_{3,i}}{\Delta s} \left[ \overline{\left(  \frac{q_{3,i}}{  \overline{q}_{1,i} } \right)} - \frac{q_{3,i-1/2}}{  \overline{q}_{1,i-1/2} }  \right] \\
& =
\frac{1}{4} \left\llbracket \frac{q_{3,i-1}}{  \overline{q}_{1,i-1} } \right\rrbracket^2 \frac{\overline{q}_{3,i-1}}{\Delta s} + \frac{1}{4} \left\llbracket \frac{q_{3,i}}{  \overline{q}_{1,i} } \right\rrbracket^2  \frac{\overline{q}_{3,i}}{\Delta s}.
\end{split}
\end{equation*}
Clearly, these terms must be positive, given that we have specified that $\overline{q}_{3,i-1}>0$ and $\overline{q}_{3,i}>0$.

Similarly, we consider the case that $\overline{q}_{3,i-1}<0$ and $\overline{q}_{3,i}<0$, so that $u_{U,\mathrm{up},i-1} = u_{U,i-1/2}$ and $u_{U,\mathrm{up},i} = u_{U,i+1/2}$:
 \begin{equation*}
 \begin{split}
  \epsilon_{u,U,i-1/2} 
& = \frac{1}{2} \left\llbracket \frac{q_{3,i-1}}{  \overline{q}_{1,i-1} } \right\rrbracket \frac{\overline{q}_{3,i-1}}{\Delta s}  \left[ \overline{\left(  \frac{q_{3,i-1}}{  \overline{q}_{1,i-1} } \right)} - \frac{q_{3,i-1/2}}{  \overline{q}_{1,i-1/2} }  \right] + \frac{1}{2} \left\llbracket \frac{q_{3,i}}{  \overline{q}_{1,i} } \right\rrbracket  \frac{\overline{q}_{3,i}}{\Delta s} \left[ \overline{\left(  \frac{q_{3,i}}{  \overline{q}_{1,i} } \right)} - \frac{q_{3,i+1/2}}{  \overline{q}_{1,i+1/2} }  \right] \\
& =
-\frac{1}{4} \left\llbracket \frac{q_{3,i-1}}{  \overline{q}_{1,i-1} } \right\rrbracket^2 \frac{\overline{q}_{3,i-1}}{\Delta s} - \frac{1}{4} \left\llbracket \frac{q_{3,i}}{  \overline{q}_{1,i} } \right\rrbracket^2  \frac{\overline{q}_{3,i}}{\Delta s}.
\end{split}
\end{equation*}
Again, these terms must be positive, given that we have specified that $\overline{q}_{3,i-1}<0$ and $\overline{q}_{3,i}<0$.
The third and fourth options (with differing signs between $\overline{q}_{3,i-1}$ and $\overline{q}_{3,i}$) are just simple recombinations of these two results, so they too will be strictly positive. 
Move these terms to the RHS, and they become strictly negative source terms.
This means that our proposed upwind discretization adds a strictly negative source term to the energy equation, which acts to dissipate the energy given by \eqref{eq:semi-discrete/summary/local_energy}. 

Comparing this numerical dissipation term to the physical dissipation term given by \eqref{eq:semi-discrete/physical_diffusion/epsilon_d}, we see that, among other differences, the numerical dissipation has an additional factor $\Delta s$. 
It is proportional to the cell size and will decrease at a first order rate with increasing grid resolution. 

Note that alternative upwind fluxes, such as those used by \cite{KrasnopolskyLukyanov2018, LiaoMeiKlausner2008}, do not yield contributions to the energy equation that can be written as the sum of a conservative term and a strictly negative dissipation term. 
In contrast, our new upwind advective numerical flux does possess the property of energy stability.
However, it may be more dissipative than necessary, and for this reason we will combine it with the energy-conserving flux, in an energy-stable manner. 

\subsection{Energy-stable combined advective flux}
\label{ssec:semi-discrete/flux_limiters}

Our fifth and key novel contribution in the semi-discrete setting is that we combine the strictly dissipative upwind advective flux with the energy-conserving central advective flux, in such a way that the resulting advective flux is energy stable, but less dissipative than a purely upwind discretization.
The proposed combination possesses the best properties of both schemes.

Our  energy-conserving advective fluxes were defined in \eqref{eq:semi-discrete/model/conservative_numerical_flux_vector} as
\begin{align*}
f_{3,a,i-1,\mathrm{ec}} &=  \overline{ \left( \frac{q_{3,i-1}}{  \overline{q}_{1,i-1} } \right) }  \frac{\overline{q}_{3,i-1}}{\Delta s} = \rho_U \overline{u}_{U,i-1} \overline{ \left(   \overline{A}_{U,i-1} u_{U,i-1} \right)}, \\
f_{4,a,i-1,\mathrm{ec}} &=   \overline{ \left( \frac{q_{4,i-1}}{  \overline{q}_{2,i-1} } \right) }  \frac{\overline{q}_{4,i-1}}{\Delta s} = \rho_L \overline{u}_{L,i-1} \overline{ \left(   \overline{A}_{L,i-1} u_{L,i-1} \right)}. 
\end{align*}
Following the conventional manner of combining low-order and higher-order fluxes \cite{Toro1999}, we propose the following combination of the energy-conserving fluxes and the upwind fluxes given by \eqref{eq:semi-discrete/numerical_diffusion/upwind_fluxes}, using flux limiters:
\begin{subequations}
\begin{align}
f_{3,a,i-1} &= \left( 1 - \phi\left(r_{U,i-1}\right) \right) f_{3,a,i-1,u} + \phi\left(r_{U,i-1}\right) f_{3,a,i-1,\mathrm{ec}},  \\
f_{4,a,i-1} &= \left( 1 - \phi\left(r_{L,i-1}\right) \right) f_{4,a,i-1,u} + \phi\left(r_{L,i-1}\right) f_{4,a,i-1,\mathrm{ec}}, 
\end{align}
\label{eq:semi-discrete/limiters/fluxes}
\end{subequations}
with $\phi(r_{U,i-1})$ and $\phi(r_{L,i-1})$ the limiter functions which determine the weighting between the upwind flux and the energy-conserving flux. 
Here, the upwind flux is a low-order flux, and the energy-conserving flux is a higher-order flux.
The limiting coefficients depend on the slope of the solution:
\begin{equation*}
r_{U,i-1} =
\begin{cases}
\frac{q_{3,i-5/2}-q_{3,i-3/2}}{q_{3,i-3/2}-q_{3,i-1/2}}, & \text{if}\  \overline{u}_{U,i-1}>0 \\
1, & \text{if}\ \overline{u}_{U,i-1} = 0 \\
\frac{q_{3,i-1/2}-q_{3,i+1/2}}{q_{3,i-3/2}-q_{3,i-1/2}}, & \text{if}\  \overline{u}_{U,i-1}<0
\end{cases}
\quad \quad
r_{L,i-1} =
\begin{cases}
\frac{q_{4,i-5/2}-q_{4,i-3/2}}{q_{4,i-3/2}-q_{4,i-1/2}}, & \text{if}\  \overline{u}_{L,i-1}>0 \\
1, & \text{if}\ \overline{u}_{L,i-1} = 0 \\
\frac{q_{4,i-1/2}-q_{4,i+1/2}}{q_{4,i-3/2}-q_{4,i-1/2}}, & \text{if}\  \overline{u}_{L,i-1}<0
\end{cases}
\end{equation*}

These coefficients are fed to the limiter functions, for which many options exist.
Here we choose the minmod function:
\begin{equation*}
\phi(r) = \mathrm{max}\left[0,\mathrm{min}\left(r,1\right) \right].
\end{equation*}
The minmod function will always yield a value between 0 and 1, which is an important property that we need to show energy stability of the combined scheme.
When the solution is smooth, $\phi(r)$ will be close to 1, and the energy-conserving flux will be used. 
When the solution is less smooth, the upwind flux will be weighted more heavily. 

The energy analysis of \autoref{ssec:semi-discrete/numerical_diffusion} can be repeated for the fluxes given by \eqref{eq:semi-discrete/limiters/fluxes}, with similar results.
The contribution of the flux terms to \eqref{eq:semi-discrete/basic/energy_equation_v1} can be written as
\begin{equation*}
\begin{split}
 & \left< \v{v}_{b,i-1/2,i-1}, \llbracket \v{f}_{i-1} \rrbracket \right> + \left< \v{v}_{b,i-1/2,i}, \llbracket \v{f}_{i} \rrbracket \right> \\
 & \quad \quad \quad = \left\llbracket \dotp{ \overline{\v{v}}_{b,i-1/2,i-1} }{ \v{f}_{i-1} } \right\rrbracket  +  \left\llbracket \dotp{ \overline{\v{v}}_{b,i-1/2,i} }{ \v{f}_{i} }  \right\rrbracket  - \overline{ \dotp{\llbracket \v{v}_{b,i-1/2,i-1} \rrbracket }{ \v{f}_{i-1} }} - \overline{  \dotp{ \llbracket \v{v}_{b,i-1/2,i} \rrbracket }{ \v{f}_{i} } } \\
  & \quad \quad \quad =
\left\llbracket h_{f,i-1/2} \right\rrbracket + \left\llbracket h_{n,i-1/2} \right\rrbracket + \epsilon_{n,i-1/2},
 \end{split}
\end{equation*}
with $h_{f,i}$ given by \eqref{eq:semi-discrete/basic/hf_definition_1}, $h_{n,i}$ given by
\begin{equation}
h_{n,i} 
= - \left( 1 - \phi\left(r_{U,i}\right) \right) \overline{\left( \frac{q_{3,i}}{\overline{q}_{1,i}} \right)} \left(  \overline{\left( \frac{q_{3,i}}{\overline{q}_{1,i}} \right)}  -  \left( \frac{q_3}{  \overline{q}_1 } \right)_{\mathrm{up},i}   \right) \frac{\overline{q}_{3,i}}{\Delta s}  
 - \left( 1 - \phi\left(r_{L,i}\right) \right) \overline{\left( \frac{q_{4,i}}{\overline{q}_{2,i}} \right)} \left(  \overline{\left( \frac{q_{4,i}}{\overline{q}_{2,i}} \right)} -  \left( \frac{q_4}{  \overline{q}_2 } \right)_{\mathrm{up},i}   \right) \frac{\overline{q}_{4,i}}{\Delta s}, 
 \label{eq:semi-discrete/limiters/hl}
\end{equation}
and $\epsilon_{n,i-1/2}$ given by
\begin{multline}
\epsilon_{n,i-1/2} = 
 \overline{ \left( \left( 1 - \phi\left(r_{U,i-1/2}\right) \right) \left\llbracket \frac{q_{3,i-1/2}}{  \overline{q}_{1,i-1/2} } \right\rrbracket \left( \overline{\left( \frac{q_{3,i-1/2}}{\overline{q}_{1,i-1/2}} \right)} -  \left( \frac{q_3}{  \overline{q}_1 } \right)_{\mathrm{up},i-1/2} \right) \frac{\overline{q}_{3,i-1/2}}{\Delta s} \right)  } \\
+  \overline{ \left( \left( 1 - \phi\left(r_{L,i-1/2}\right) \right) \left\llbracket \frac{q_{4,i-1/2}}{  \overline{q}_{2,i-1/2} } \right\rrbracket \left( \overline{\left( \frac{q_{4,i-1/2}}{\overline{q}_{2,i-1/2}} \right)} -  \left( \frac{q_4}{  \overline{q}_2 } \right)_{\mathrm{up},i-1/2} \right) \frac{\overline{q}_{4,i-1/2}}{\Delta s} \right)  }.
\label{eq:semi-discrete/limiters/epsilon_l}
\end{multline}
Since with the minmod limiter function, the factors $\left( 1 - \phi\left(r_{U,i}\right) \right)$ and $\left( 1 - \phi\left(r_{L,i}\right) \right)$ have values between 0 and 1, this dissipation term has the same positivity property as the upwind dissipation term given by \eqref{eq:semi-discrete/upwind/epsilon_u}. 

Therefore, our novel advective numerical flux, formed by combining our upwind flux and our energy-conserving flux, is energy stable. 
The new energy-stable flux will be less dissipative than the pure upwind flux, since it uses the energy-conserving flux where possible. 
Where the solution is smooth, the continuous equations conserve energy, and our energy-stable flux replicates this property.
Where the solution is discontinuous, the energy-stable flux effectively adds numerical diffusion which dissipates energy.

\section{Stability}
\label{sec:stability} 

In this section, we discuss the stability of the basic and extended TFM. 
The stability of the TFM is a topic that has received much attention since the discovery of the ill-posedness issue by \cite{LyczkowskiGidaspowSolbrigEtAl1978}.
Here we focus on providing a detailed motivation for our proposed model additions of physical diffusion and surface tension.
Together these effects produce a reliable model that yields convergent solutions under flow conditions where the basic two-fluid model fails.
Friction plays a less important role in the model's stability, but is an important physical effect that an accurate model must include, and is included in the stability analysis given here.

The basic two-fluid model, as described in \autoref{ssec:continuous/model}, is known to be conditionally hyperbolic \cite{Montini2011}.
In the region of state space where the velocity difference is below the inviscid Kelvin-Helmholtz (IKH) limit, the eigenvalues of the model are real, but outside this region the eigenvalues of the model are complex \cite{LiaoMeiKlausner2008}. 
Linear stability analysis (see \autoref{sec:linear_stability_analysis}) confirms the issue put forward by the characteristic analysis: within the hyperbolic region the model is stable, but in the non-hyperbolic region the (linear) growth rates for small wavelength perturbations tend towards infinity. 
Therefore the model is said to be (linearly) ill-posed: the common view is that this precludes meaningful solutions to the continuous model, and prevents convergence of numerical solutions \cite{IssaKempf2003}.
This prevents the use of the basic model in its non-hyperbolic region. 

\begin{figure}[htbp] 
\centering
\includegraphics[width=0.45\linewidth]{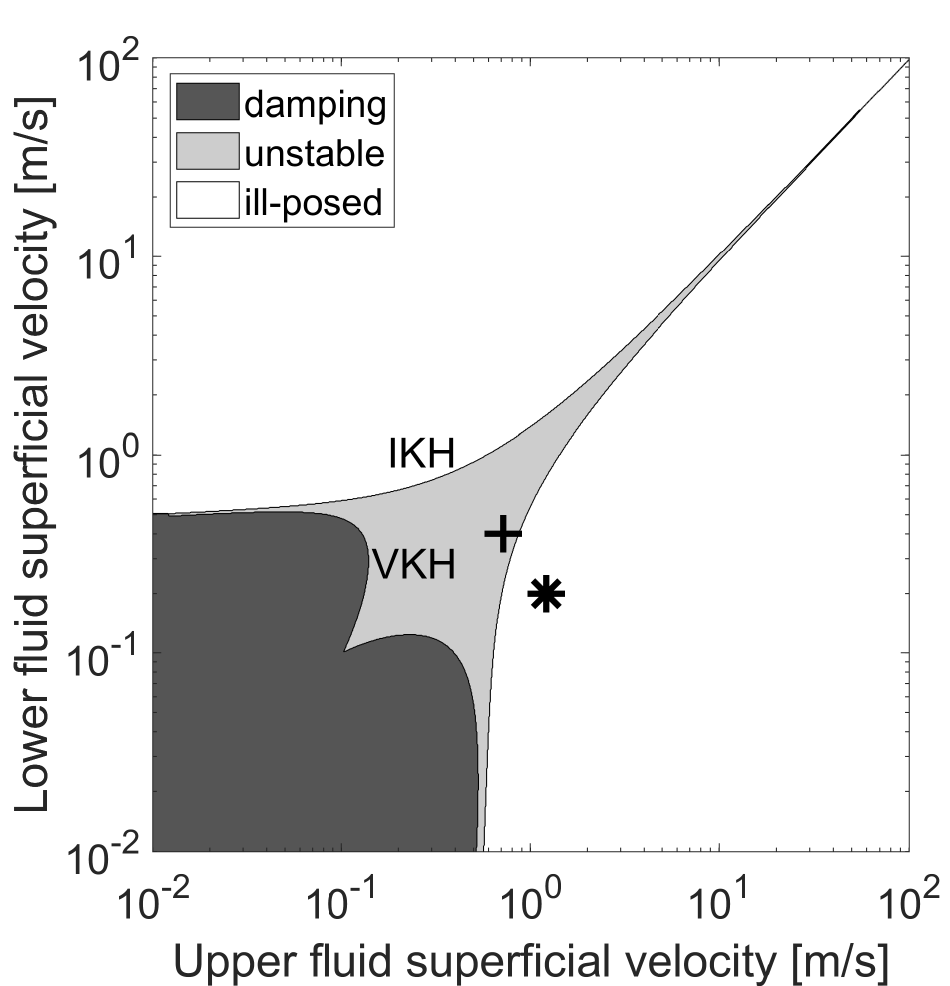} 
\caption{A map of the linear stability of perturbations to steady states of the TFM with wall and interface friction, using the basic model without diffusion and surface tension.}
\label{fig:stability/basic_stability_map}
\end{figure}

\autoref{fig:stability/basic_stability_map} is a stability map similar to those of \cite{BarneaTaitel1993, BarneaTaitel1994}.
It maps the stability of steady states of the TFM, where friction is balanced by a constant driving pressure gradient (acting as a body force). 
The parameters are given in \autoref{tab:Thorpe_parameters_LSA} and the geometry is that of a 2D channel.
Given the lower fluid superficial velocity $u_L \alpha_L$ and the upper fluid superficial velocity $u_U (1-\alpha_L)$, the hold-up $\alpha_L=A_L/A$ and driving pressure gradient $\partial p_0 /\partial s$ follow from the demand for a fully developed steady state (derivatives to $s$ and $t$ must be zero). 
For these steady states, the two dispersion relations $\omega(\lambda)$ can be calculated according to \autoref{sec:linear_stability_analysis}.
For a given perturbation wavelength $\lambda$, we consider the dispersion relation for which the imaginary component of $\omega$ is largest: this is the most unstable mode for this wavelength.
If for this most unstable mode $\mathrm{Im}(\omega(\lambda)) < 0$, the steady state is stable and damping (to perturbations of wavelength $\lambda$), if  $\mathrm{Im}(\omega(\lambda)) = 0$ it is neutrally stable, and if  $\mathrm{Im}(\omega(\lambda)) > 0$ it is unstable.
As long as $\mathrm{Im}(\omega(\lambda))$ is bounded as $\lambda \rightarrow 0$, the state is well-posed.
If $\mathrm{Im}(\omega(\lambda)) \rightarrow \infty$ for $\lambda \rightarrow 0$, the state is labeled `ill-posed'. 

\begin{figure}[htbp] 
\centering
\includegraphics[width=0.45\linewidth]{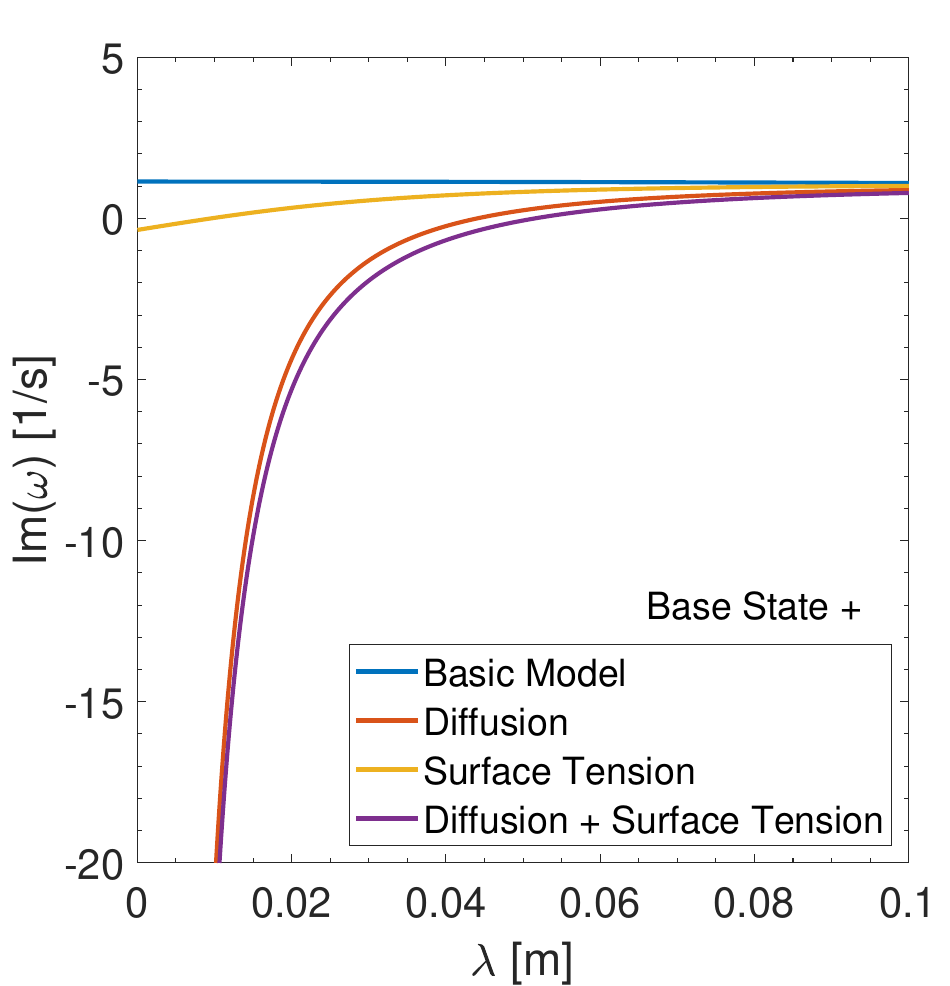} 
\includegraphics[width=0.45\linewidth]{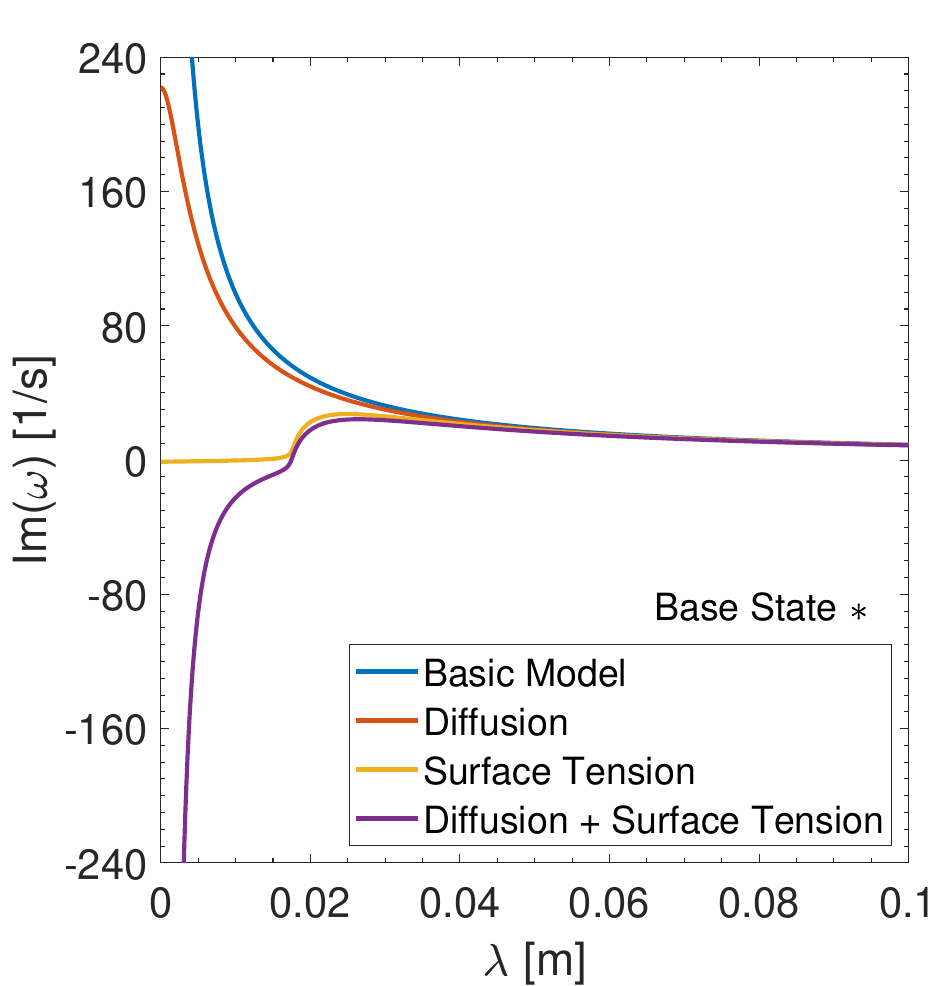} 
\caption{Imaginary component of the angular frequency $\omega$ for the most unstable mode, for the set of parameters given in \autoref{tab:Thorpe_parameters_LSA}, plotted as a function of wavelength $\lambda=2\pi/k$. All models include wall and interface friction. Left: base state given by \autoref{tab:LSA_base_state_cross}, marked by a cross in the stability maps. Right: base state given by \autoref{tab:Thorpe_base_state_asterisk}, marked by an asterisk in the stability maps.}
\label{fig:stability/dispersion_comparison}
\end{figure}

\begin{table}[htb]
\small
\centering
\caption{Parameters used for the linear stability analysis. These resemble the parameters of the Thorpe experiment \cite{Thorpe1969} as described by \cite{FullmerRansomLopezdeBertodano2014}. } 
\begin{tabular}{llll}
\toprule
Parameter & Symbol & Value & Units\\
\midrule
Lower fluid density & $\rho_L$ & $1000$ & $  \mathrm{kg} \, \mathrm{m}^{-3}$ \\
Upper fluid density & $\rho_U$ & $780$ & $ \mathrm{kg} \, \mathrm{m}^{-3}$ \\
Acceleration of gravity & $g$ & $9.81$ & $\mathrm{m} \, \mathrm{s}^{-2}$ \\
Channel inclination & $\phi$ & $0$ & degrees  \\
Lower fluid material viscosity & $\nu_{\mathrm{m},L}$ & $1.0 \cdot 10^{-6}$ & $ \mathrm{m}^{2} \, \mathrm{s}^{-1}$ \\
Upper fluid material viscosity & $\nu_{\mathrm{m},U}$ & $1.9 \cdot 10^{-6}$ & $ \mathrm{m}^{2} \, \mathrm{s}^{-1}$ \\
Lower fluid turbulent viscosity & $\nu_{\mathrm{t},L}$ & $1.3 \cdot 10^{-5}$ & $ \mathrm{m}^{2} \, \mathrm{s}^{-1}$ \\
Upper fluid turbulent viscosity & $\nu_{\mathrm{t},U}$ & $1.3 \cdot 10^{-5}$ & $ \mathrm{m}^{2} \, \mathrm{s}^{-1}$ \\
Effective viscosity adjustment factor & $ C_\epsilon $ & $8.1$ & $-$ \\
Surface tension & $\sigma$ & $0.04$ & $ \mathrm{kg} \, \mathrm{m} \, \mathrm{s}^{-2}$ \\
\bottomrule
\end{tabular}
\label{tab:Thorpe_parameters_LSA}
\end{table}

\begin{table}[htb]
\small
\centering
\caption{Base state used for the linear stability analysis, corresponding to the left plot in \autoref{fig:stability/dispersion_comparison}, and marked with a cross in the stability maps.}
\begin{tabular}{llll}
\toprule
Variable & Symbol & Value & Units\\
\midrule
Initial lower fluid hold-up & $\alpha_{L,0}$ & $0.4$ & $-$ \\
Initial lower fluid velocity & $u_{L,0}$ & $1$ & $\mathrm{m} \, \mathrm{s}^{-1}$ \\
Initial upper fluid velocity & $u_{U,0}$ & $1.198$ & $\mathrm{m} \, \mathrm{s}^{-1}$ \\
Driving pressure gradient & $\partial p_0/\partial s$ & $-204.2$ & $\mathrm{kg} \, \mathrm{m}^{-2} \, \mathrm{s}^{-2}$ \\
\bottomrule
\end{tabular}
\label{tab:LSA_base_state_cross}
\end{table}

\begin{table}[htb]
\small
\centering
\caption{Base state used for the linear stability analysis, corresponding to the right plot in \autoref{fig:stability/dispersion_comparison}, and marked with an asterisk in the stability maps.}
\begin{tabular}{llll}
\toprule
Variable & Symbol & Value & Units\\
\midrule
Initial lower fluid hold-up & $\alpha_{L,0}$ & $0.2$ & $-$ \\
Initial lower fluid velocity & $u_{L,0}$ & $1$ & $\mathrm{m} \, \mathrm{s}^{-1}$ \\
Initial upper fluid velocity & $u_{U,0}$ & $1.515$ & $\mathrm{m} \, \mathrm{s}^{-1}$ \\
Driving pressure gradient & $\partial p_0/\partial s$ & $-268.4$ & $\mathrm{kg} \, \mathrm{m}^{-2} \, \mathrm{s}^{-2}$ \\
\bottomrule
\end{tabular}
\label{tab:Thorpe_base_state_asterisk}
\end{table}

For the basic model without diffusion and surface tension, stability is independent of wavelength  \cite{JonesProsperetti1985}: if long wavelengths are unstable then short wavelengths are also unstable (though with different growth rates). 
This can be seen in \autoref{fig:stability/dispersion_comparison} (similar to figure 3.2 in \cite{LopezdeBertodanoFullmerClausseEtAl2017}), which shows the growth rate of the most unstable mode, for two different steady states, and for different versions of the TFM. 
Therefore the stability map of the basic model in \autoref{fig:stability/basic_stability_map} is independent of wavelength.
It is divided into an ill-posed region and a well-posed region.
Without friction the whole well-posed region would be neutrally stable. 
Friction divides the well-posed region of \autoref{fig:stability/basic_stability_map} into a region with damping and an unstable region (but with bounded growth rates).
The stability boundary with friction is referred to as the viscous Kelvin-Helmholtz (VKH) boundary, while the ill-posedness boundary is referred to as the inviscid Kelvin-Helmholtz (IKH) boundary \cite{BarneaTaitel1993, BarneaTaitel1994}.

\begin{figure}[htbp] 
\centering
\includegraphics[width=0.32\linewidth]{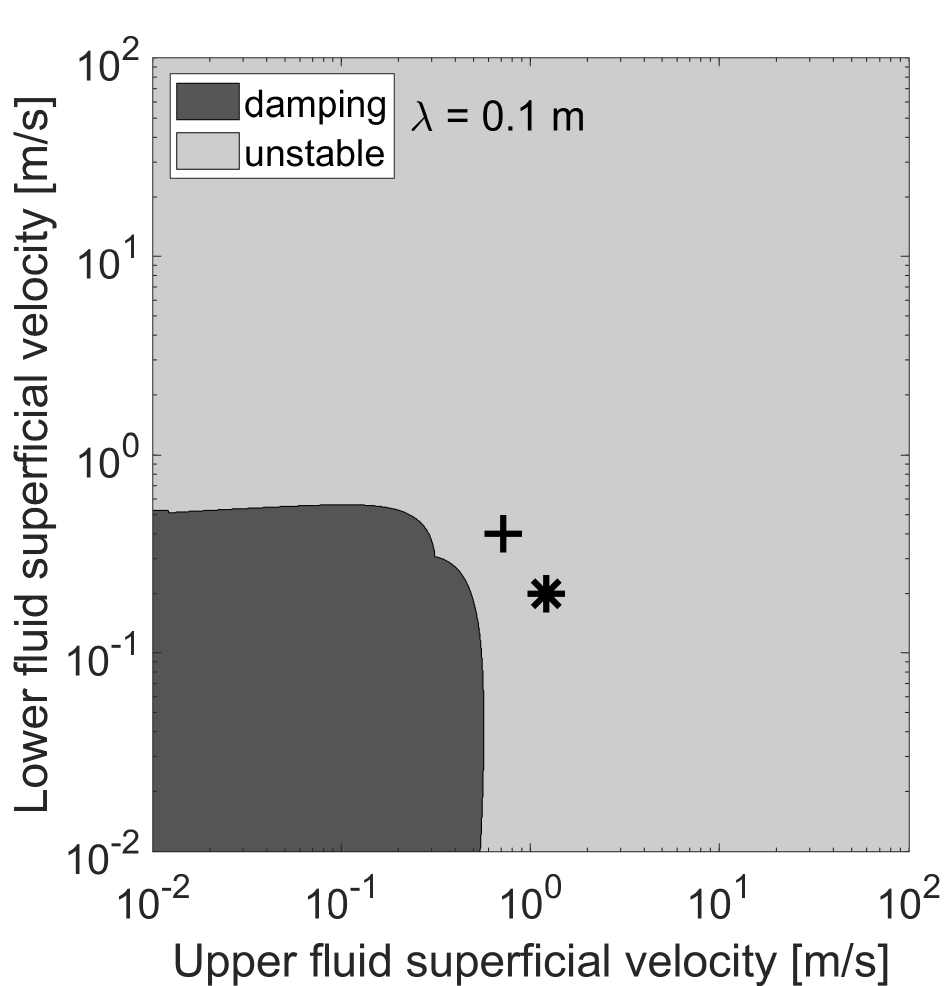} 
\includegraphics[width=0.32\linewidth]{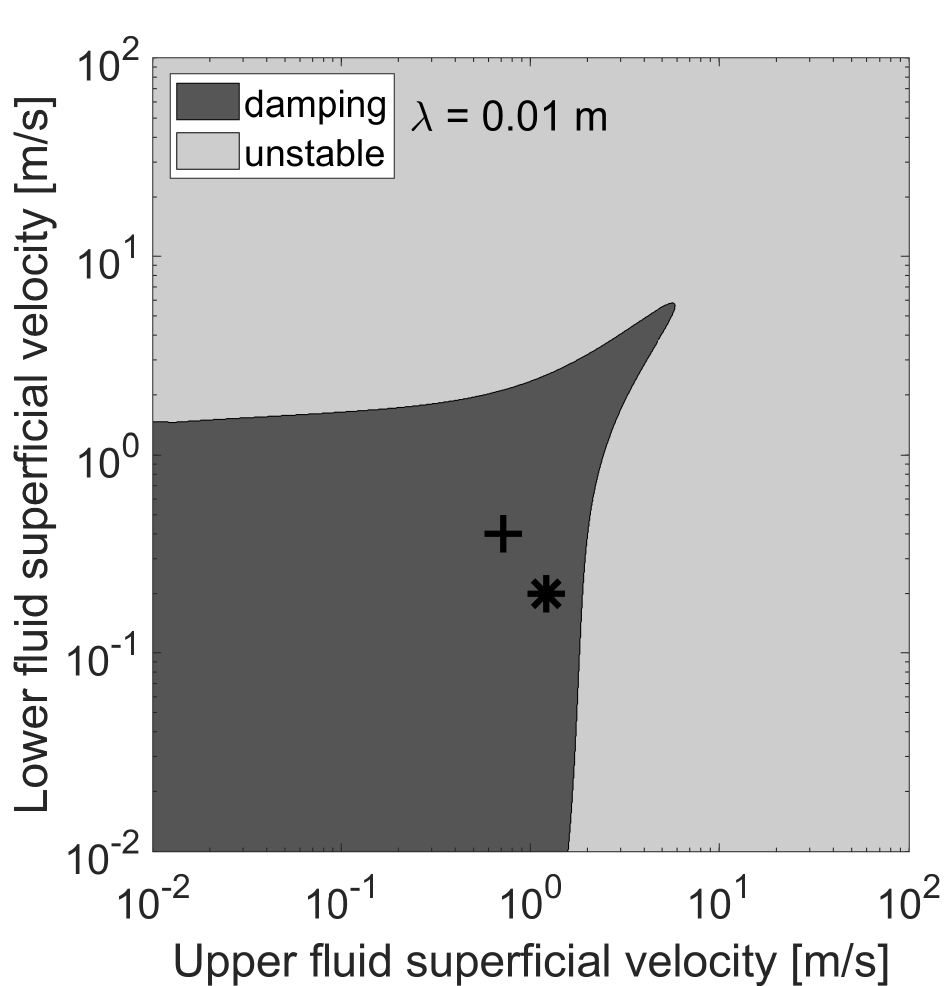}
\includegraphics[width=0.32\linewidth]{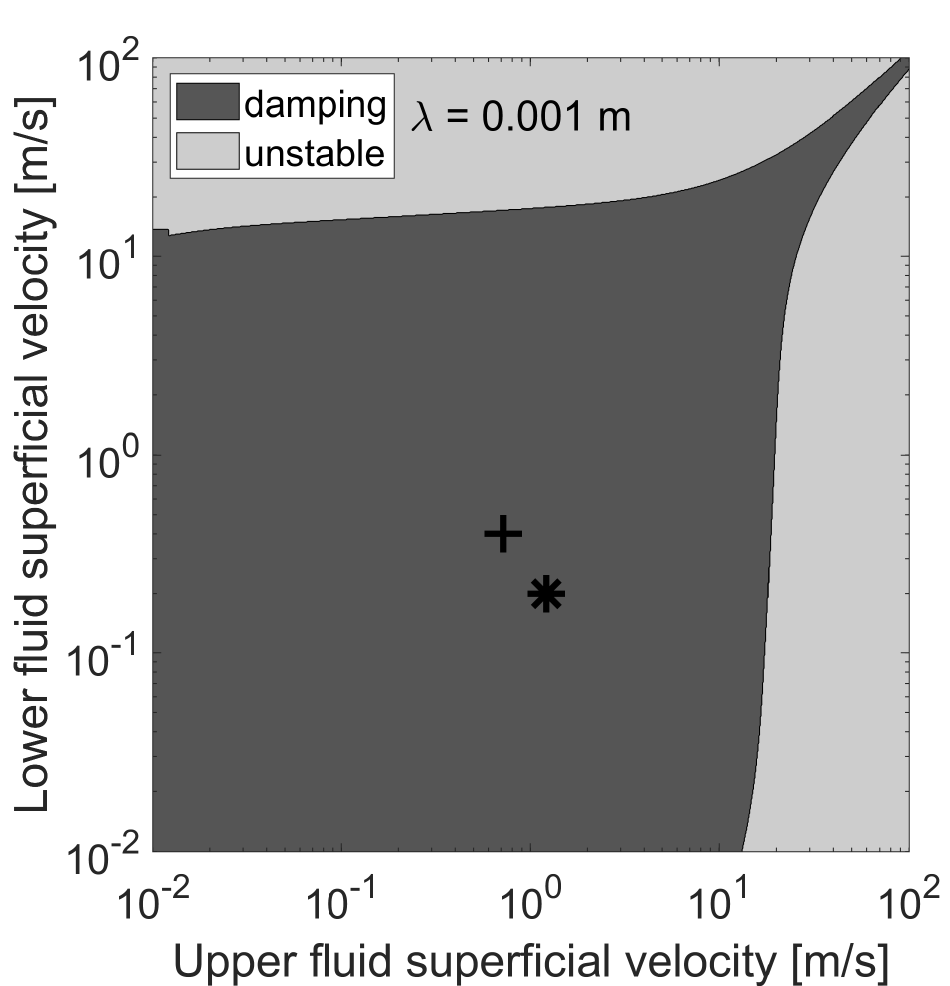} 
\caption{Maps of the linear stability of short wavelength perturbations to steady states of the TFM, with wall and interface friction, diffusion, and surface tension. The stability of perturbations with a specific wavelength is shown. Left: $\lambda = 0.1 \, \mathrm{m}$. Middle: $\lambda = 0.01 \, \mathrm{m}$. Right: $\lambda = 0.001 \, \mathrm{m}$.}
\label{fig:stability/short_wavelength_stability_map}
\end{figure}

Adding physical diffusion and surface tension changes the dispersion relations, as can be seen in \autoref{fig:stability/dispersion_comparison}. 
Short wavelength perturbations are stabilized, while at long wavelengths, the dispersion relations are unchanged.
The growth rate no longer tends to infinity for short wavelengths, removing the ill-posedness issue.
In case only diffusion is added, the growth rate is bounded, but its value still increases rapidly at short wavelengths, which causes short wavelength perturbations to dominate the solution.
In a numerical model, upon refining the grid, increasingly unstable scales are resolved, making it impossible to reach convergence  \cite{HolmasSiraNordsveenEtAl2008}.

When both physical diffusion and surface tension are added to the model, as suggested by \cite{FullmerRansomLopezdeBertodano2014}, a cut-off wavelength is introduced below which perturbations are damped, as can be seen in \autoref{fig:stability/dispersion_comparison} at approximately $\lambda_c =0.0174 \, \mathrm{m}$. 
This removes the unphysical short wavelength instabilities.
Meanwhile, the long wavelength instabilities, which are physical instabilities that are an integral part of the model, can still be resolved dynamically.
The removal of the severe short wavelength instabilities means that the impediment to grid convergence is removed, and implies that theoretically all dynamics could be resolved without refining past $\Delta s = \lambda_c/2$.
The combination of physical diffusion and surface tension is crucial to achieving the damping effect: with only surface tension the short wavelengths are nearly neutrally stable, with only the shortest wavelengths being very weakly damped, due to the influence of friction.

The cut-off wavelength depends on the state and model parameters.
As the difference between the velocities of the two fluids is increased further into the region of instability beyond the IKH limit, the cut-off wavelength is decreased.
This is apparent from comparing the stability maps for different wavelengths in \autoref{fig:stability/short_wavelength_stability_map}.
The marked states in the maps are unstable (with bounded growth rates) to long wavelength perturbations, but stable for short wavelengths, as shown in \autoref{fig:stability/dispersion_comparison}. 
For each possible state there will always be a cut-off wavelength below which damping takes place.
Therefore, the extended model is unconditionally well-posed.

This method of regularization leaves intermediate scale perturbations intact that might lie outside of the range of validity of the model dictated by the long-wavelength assumption.
An alternative option is to also damp these scales, by using artificial diffusion, added to both the mass and momentum equations \cite{BonzaniniPicchiPoesio2017, FullmerLeeLopezdeBertodano2014, HolmasSiraNordsveenEtAl2008}. 
However, the fact that these scales are not modeled to complete accuracy does not mean that it is more accurate to artificially eliminate these perturbations.
Our approach is to leave these perturbations intact, for as far as they are not stabilized by physically motivated model components.

Depending on the state, the cut-off wavelength may become quite low and the instability quite severe. 
For such an unstable state, the solution will be dominated by intermediate scale instabilities, that lie above the cut-off wavelength, but are still of very short wavelength. 
Practically, the grid resolution required to reach the cut-off wavelength may be prohibitive.
Therefore, in engineering applications, it may not be possible to resolve all the dynamics of the model in a numerical simulation. 
This does not affect the linear well-posedness: if the small scales are not resolved, they will cause no harm, and if they are resolved, they will be regularized by diffusion and surface tension.

The longer scale instabilities that remain present in our extended model have bounded growth rates, but would grow indefinitely, according to the linear stability analysis.
In reality, when the perturbations grow large, the assumptions made in the linearization of the model become invalid, and the behavior of the full model will depart from the behavior of the linearized model.
At this point, the nonlinear stability of the model must be considered \cite{LopezdeBertodanoFullmerClausseEtAl2017}.
Nonlinear effects can bound perturbations that grow initially due to linear instability.
Typically, the unstable perturbations develop into shocks, which must be bounded by a dissipative mechanism.
In \autoref{ssec:numerical_experiments/double_wave} we show numerically that physical diffusion plays a crucial role in the nonlinear damping of linear instabilities that develop into shocks.
At coarse grid resolutions, this role is taken over by numerical diffusion.

An analytical indication of a form of nonlinear stability, which the model should satisfy even when it is linearly unstable, is given by the energy conservation property.
The extended local energy, given by \eqref{eq:continuous/summary/extended_energy}, can be written in the following primitive form for the 2D channel geometry:
\begin{multline}
e_{\mathrm{ch}} = \frac{1}{2} \rho_U g_n H_U^2 + \rho_U g_n  H_U H_L +  \frac{1}{2} \rho_L g_n H_L^2  + \frac{1}{2}  \rho_U  u_U^2 H_U + \frac{1}{2}\rho_L u_L^2 H_L \\
+ \rho_U g y H_U +  \rho_L g y H_L  +  \sigma \left( 1 + \frac{1}{2} \left( \pd{H_L}{s}\right)^2 \right).
\end{multline}
Since $H_U$ and $H_L$ must be positive, each term in this expression is positive.
Therefore the global energy equation \eqref{eq:continuous/summary/global_energy_equation}  implies a bound on the velocities, the heights, and their spatial derivatives. 
A numerical model that conserves (or strictly dissipates) this energy can be expected to have solutions that are bounded in this way.

In conclusion, the extended model linearly damps short wavelength perturbations, nonlinearly damps unstable long wavelength perturbations when they grow large, and possesses an energy bound.
These properties are achieved by adding only physically derived terms to the basic model (physical diffusion, friction, and surface tension).
The end result is a reliable model that can be expected to handle difficult flow states, while still resolving physical instabilities. 

\section{Numerical experiments} 
\label{sec:numerical_experiments}

\subsection{Introduction}

In this section, the energy stability and well-posedness properties of our new framework are demonstrated through three different numerical experiments. 
These were conducted using a code based on the spatial discretizations given in \autoref{sec:semi-discrete}.
The numerical experiments consider a 2D channel geometry, for which the original advective flux described in \autoref{ssec:semi-discrete/model} and the surface tension discretization described in \autoref{ssec:semi-discrete/surface_tension} are exactly energy-conserving.

For the time integration we use the fourth order constraint-consistent Runge-Kutta method described in \cite{SanderseVeldman2019}. 
This method is explicit for the mass and momentum equations and implicit for the pressure, requiring the solution of a pressure Poisson equation.
It requires the mass and momentum equations to be coupled as has been done by setting the fluxes in the mass equations according to \eqref{eq:semi-discrete/model/conservative_numerical_flux_vector}.

First, in \autoref{ssec:numerical_experiments/traveling_wave} we consider a traveling wave solution to the basic TFM without diffusion or surface tension.
We show that our novel energy-stable advective flux (described in \autoref{ssec:semi-discrete/flux_limiters}) yields smooth solutions without excessive numerical diffusion or numerical oscillations. 
We compare this flux to our original energy-conserving flux (described in \autoref{ssec:semi-discrete/model}) and to our strictly dissipative upwind flux (described in \autoref{ssec:semi-discrete/numerical_diffusion}).
Additionally, we compare these to a naive central scheme which is neither energy-conserving nor strictly dissipative.

In \autoref{ssec:numerical_experiments/traveling_wave_surf} we repeat the traveling wave case, but with surface tension added according to \autoref{ssec:semi-discrete/surface_tension}.
We show that this addition is energy-conserving, as predicted by the analysis. 

Last, in \autoref{ssec:numerical_experiments/double_wave} we consider an unstable perturbation to a shear flow base state that would be ill-posed for the basic TFM. 
We demonstrate that the complete new framework with friction, diffusion, surface tension, and the energy-stable advective flux, is able to obtain solutions that converge with increasing grid resolution, for this challenging test case.
We quantify the contributions of numerical diffusion, physical diffusion and friction to the nonlinear damping by computing the dissipation, using our derived expressions for the dissipation rates.

\subsection{Traveling wave with the basic model and different advective fluxes}
\label{ssec:numerical_experiments/traveling_wave}

We conduct a test case with a traveling wave, induced as a perturbation upon a uniform base state, for the basic model without diffusion, friction, or surface tension. 
The base state is given in \autoref{tab:traveling_wave_parameters}, and the flow parameters are those of \autoref{tab:Thorpe_parameters_LSA}.
The perturbation is defined according to the analysis in \autoref{sec:linear_stability_analysis}.
It is the initial condition for the exact solution to the linearized system, for one of the two modes $\omega(k)$.
We set the wavelength of the perturbation to $\lambda= 0.1 \, \mathrm{m}$, and select the wave mode $\omega = 39.89 \, \mathrm{s}^{-1}$ (the other option is $\omega = 22.94\, \mathrm{s}^{-1}$ and would yield a slower wave).
The perturbation is then limited to the range between $s_\mathrm{pert}-\lambda/2$ and $s_\mathrm{pert}+\lambda/2$, with $s_\mathrm{pert}= 2\lambda + \lambda/4$.
Outside of this range the base state is kept.
The computational domain has length $L=0.5$ and has periodic boundaries.
A pressure projection step is performed on the complete initial condition, adjusting the velocities to ensure that the volumetric flow constraint is satisfied (see \cite{SanderseVeldman2019}).

\begin{table}[htb]
\small
\centering
\caption{Base state for the traveling wave case.}
\begin{tabular}{llll}
\toprule
Variable & Symbol & Value & Units\\
\midrule
Initial lower fluid hold-up & $\alpha_{L,0}$ & $0.5$ & $-$ \\
Initial lower fluid velocity & $u_{L,0}$ & $0.5$ & $\mathrm{m} \, \mathrm{s}^{-1}$ \\
Initial upper fluid velocity & $u_{U,0}$ & $0.5$ & $\mathrm{m} \, \mathrm{s}^{-1}$ \\
\bottomrule
\end{tabular}
\label{tab:traveling_wave_parameters}
\end{table}

\begin{figure}[htbp] 
\centering
\includegraphics[width=0.45\linewidth]{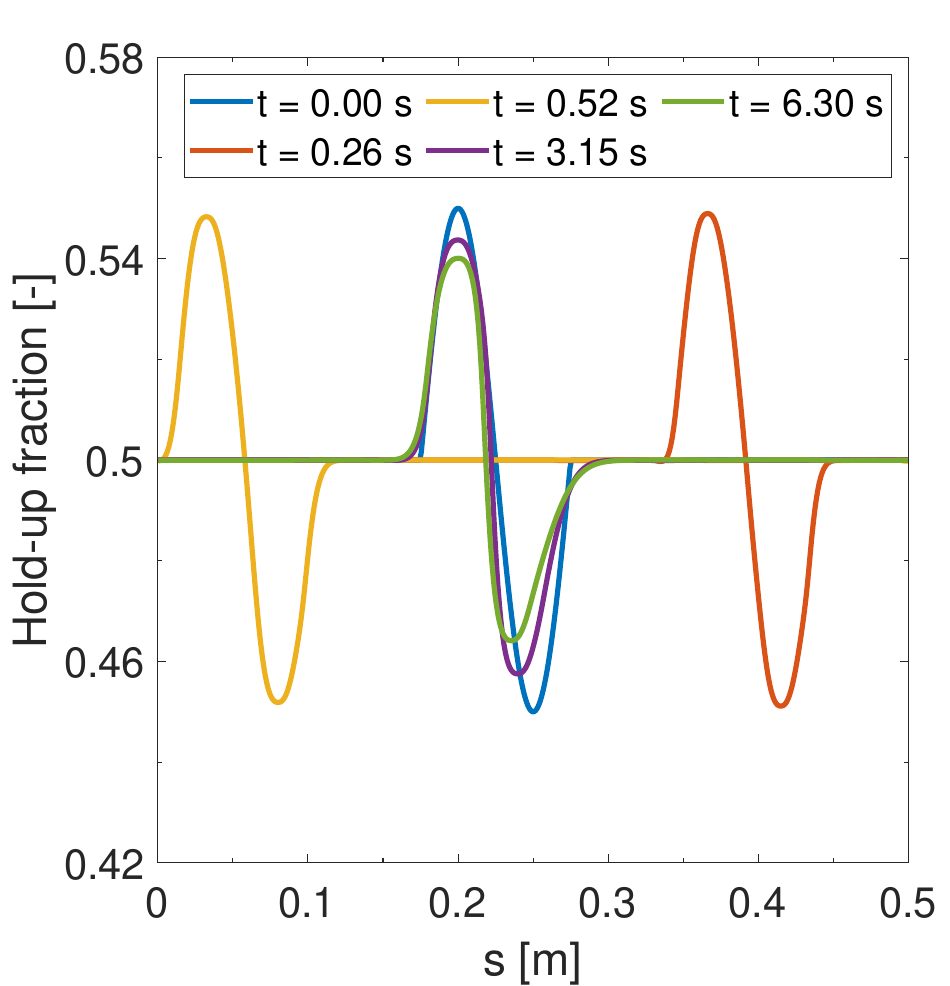} 
\includegraphics[width=0.45\linewidth]{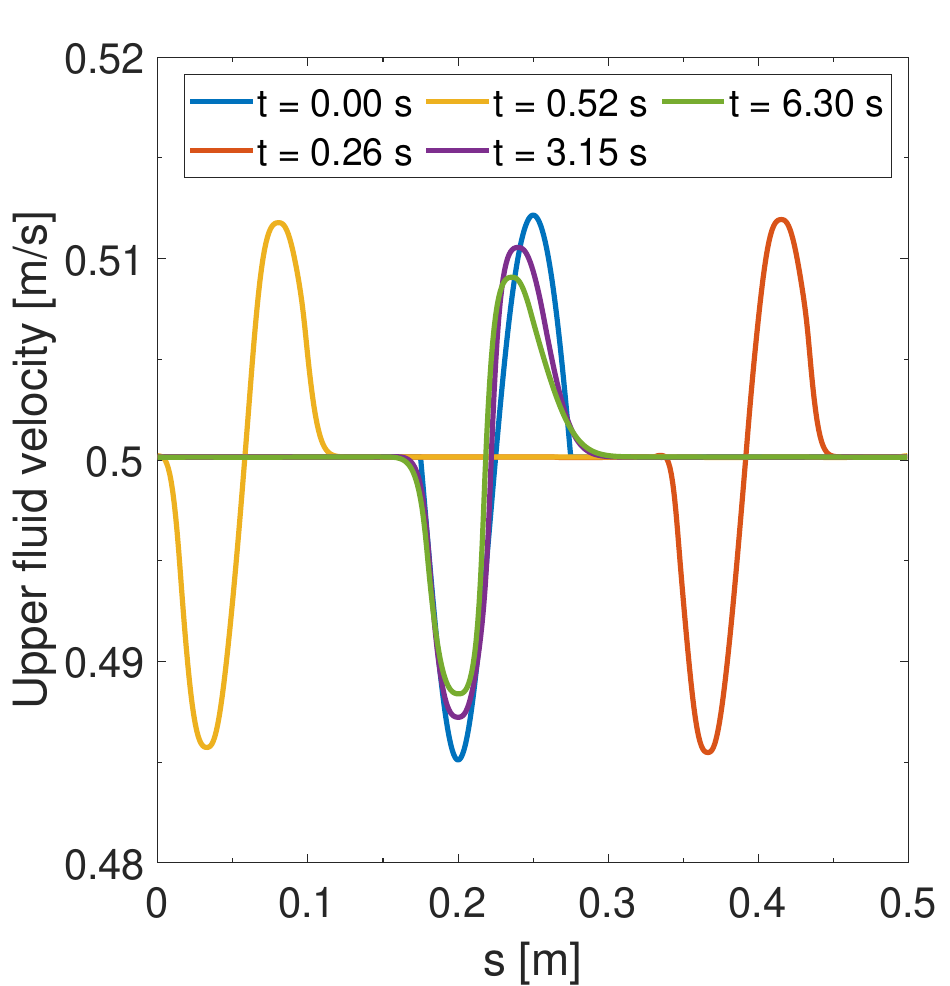} 
\caption{Two components of the solution to the traveling wave case for the basic model, using the energy-stable advective flux described in \autoref{ssec:semi-discrete/flux_limiters}, with $\Delta s = 10^{-3} \, \mathrm{m}$ and $\Delta t = 10^{-4} \, \mathrm{s}$.}
\label{fig:numerical_experiments/trav_wave/ec_lim_snapshots}
\end{figure}

The exact solution of the linearized model is a wave traveling to the right without deformation.
The solution to the full nonlinear model, using the energy-stable advective flux described in \autoref{ssec:semi-discrete/flux_limiters}, is shown in \autoref{fig:numerical_experiments/trav_wave/ec_lim_snapshots}.
We show the hold-up $\alpha_L=A_L/A$ and the upper fluid velocity $u_U$. 
The wave travels to the right at constant velocity, with little deformation.
At $t=3.15\, \mathrm{s}$ and $t=6.30 \, \mathrm{s}$, the wave has traveled through the domain an integer number of times (4 and 8 times respectively), and the wave can be compared to the initial perturbation.
In the middle of the wave, a slight steepening has taken place, tending towards the formation of a discontinuity.
At the edges, the wave has diffused slightly. 

\begin{figure}[htbp] 
\centering
\includegraphics[width=0.45\linewidth]{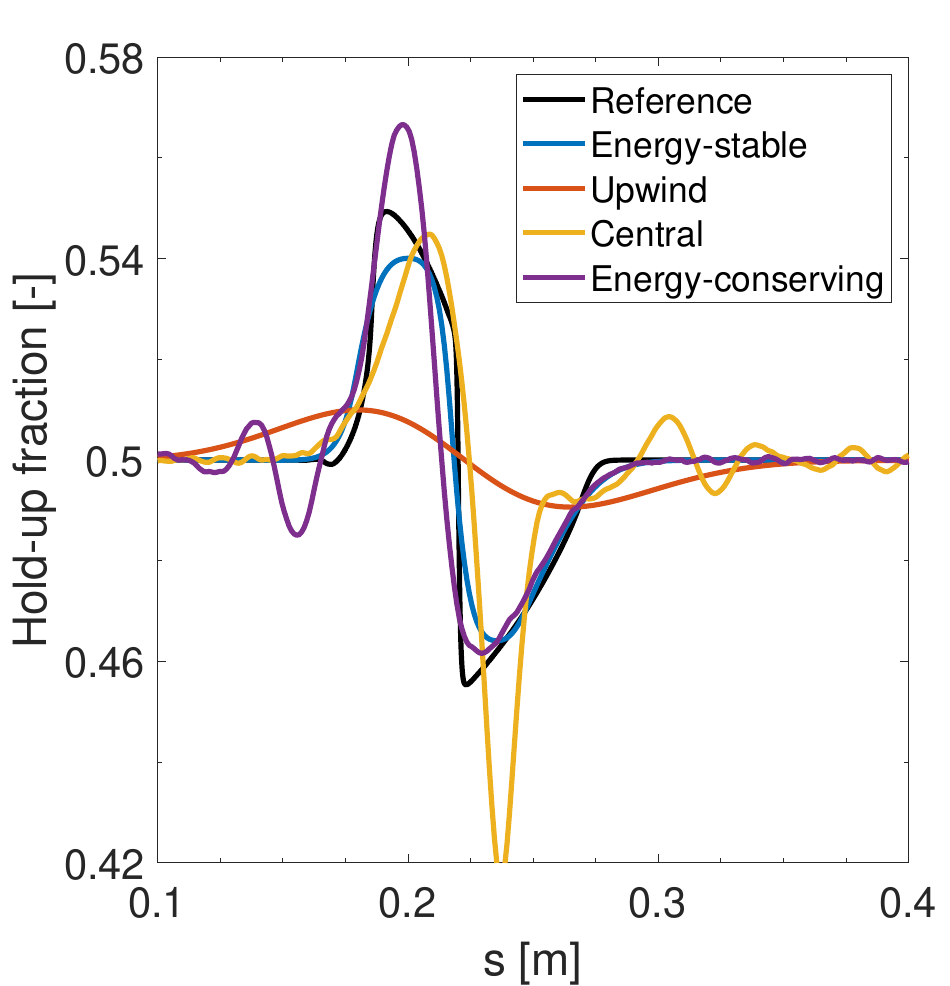} 
\includegraphics[width=0.45\linewidth]{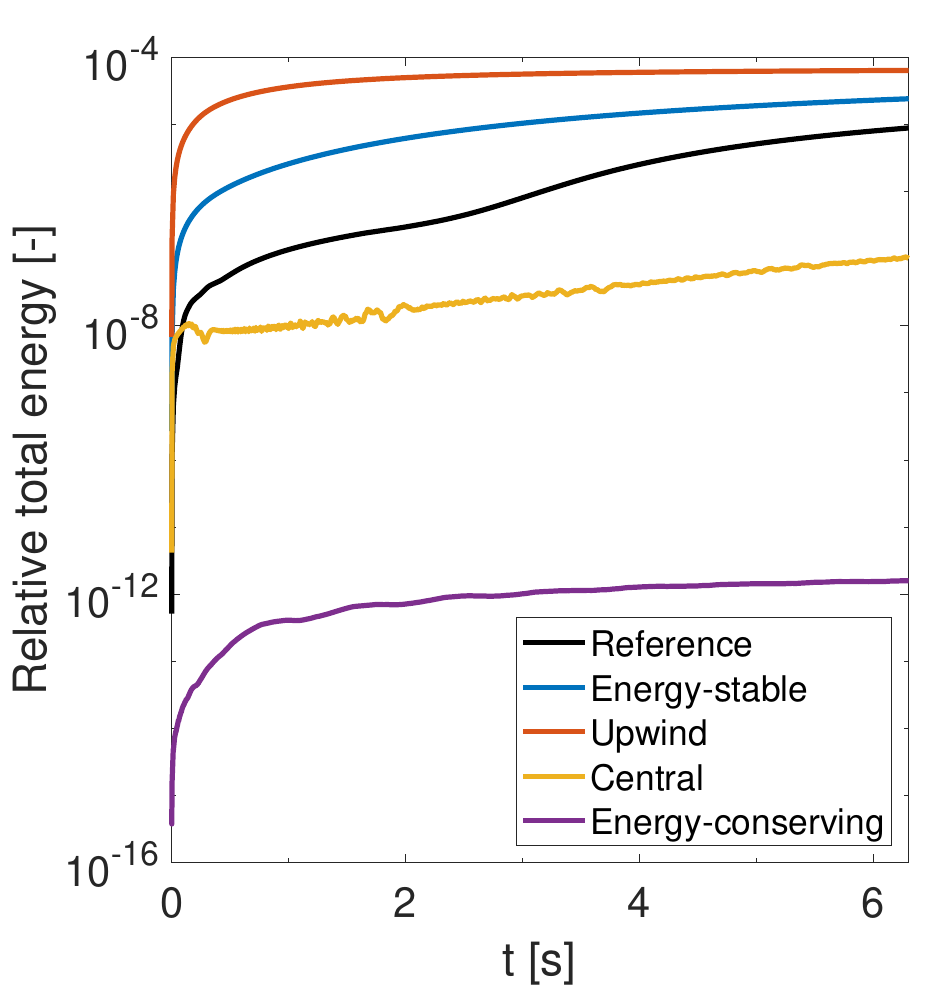} 
\caption{Comparison of results of the traveling case for the basic model, using different advective flux discretizations. In all cases, $\Delta s = 10^{-3} \, \mathrm{m}$ and $\Delta t = 10^{-4} \, \mathrm{s}$. The reference is a high-resolution solution ($\Delta s = 1.25\cdot10^{-4} \, \mathrm{m}$ and $\Delta t = 1.25\cdot10^{-5} \, \mathrm{s}$) obtained using the energy-stable scheme. Left: the solution for the hold-up at time $t=6.3 \, \mathrm{s}$. Right: the absolute difference between the energy as a function of time and the initial energy, divided by the initial energy.}
\label{fig:numerical_experiments/trav_wave/discretization_comparison}
\end{figure}

In \autoref{fig:numerical_experiments/trav_wave/discretization_comparison}, we compare results for this test case, using different discretizations of the advective flux in the momentum equations. 
We compare the energy-stable scheme given by \eqref{eq:semi-discrete/limiters/fluxes}, the upwind scheme given by \eqref{eq:semi-discrete/numerical_diffusion/upwind_fluxes}, the energy-conserving scheme given in \eqref{eq:semi-discrete/model/conservative_numerical_flux_vector}, and a naive central interpolation scheme in which the momentum advection fluxes are given by 
\begin{align*}
f_{3,a,i-1,\mathrm{cen}} &= \left[ \overline{ \left( \frac{q_{3,i-1}}{  \overline{q}_{1,i-1} } \right) } \right]^2 \frac{q_{1,i-1}}{\Delta s} = \rho_U \left(\overline{u}_{U,i-1}\right)^2 A_{U,i-1}, \\
f_{4,a,i-1,\mathrm{cen}} &=  \left[ \overline{ \left( \frac{q_{4,i-1}}{  \overline{q}_{2,i-1} } \right) } \right]^2  \frac{q_{2,i-1}}{\Delta s} = \rho_L \left(\overline{u}_{L,i-1}\right)^2 A_{L,i-1}. 
\end{align*}
A high-resolution solution ($\Delta s = 1.25\cdot10^{-4} \, \mathrm{m}$ and $\Delta t = 1.25\cdot10^{-5} \, \mathrm{s}$), obtained using the energy-stable scheme, is used as a reference.
The results show that the central and energy-conserving schemes produce numerical oscillations in the presence of strong gradients, while the upwind scheme is excessively diffusive. 
The proposed energy-stable scheme yields the most accurate solution, without numerical oscillations and with much less diffusion than the upwind scheme.

This behavior can be understood from the perspective of the (global) energy, of which \autoref{fig:numerical_experiments/trav_wave/discretization_comparison} shows the absolute and nondimensional difference with respect to the initial condition. 
The energy-conserving scheme conserves energy up to a very small time integration error.
The energy-stable and upwind schemes lose energy with respect to the initial condition, while the central scheme gains energy (this is not visible since only the absolute difference is plotted).
The energy-stable scheme is less dissipative than the upwind scheme, which is reflected in its less diffused solution.
Some dissipation is physically necessary near strong gradients or discontinuities, and the lack of this in the central and energy-conserving schemes can be understood to lead to numerical oscillations.

\subsection{Traveling wave with surface tension}
\label{ssec:numerical_experiments/traveling_wave_surf}

Adding surface tension to the basic model yields a model that is still energy-conserving, but for the modified energy $e = e_b + e_s$. 
We repeat the previous case with this model.
We test the surface tension implementation described in \autoref{ssec:semi-discrete/surface_tension} with the energy-conserving advective flux to show that the addition is energy-conserving.
We also test the surface tension implementation with the energy-stable flux and show that practically this yields the best results.

\begin{figure}[htbp] 
\centering
\includegraphics[width=0.45\linewidth]{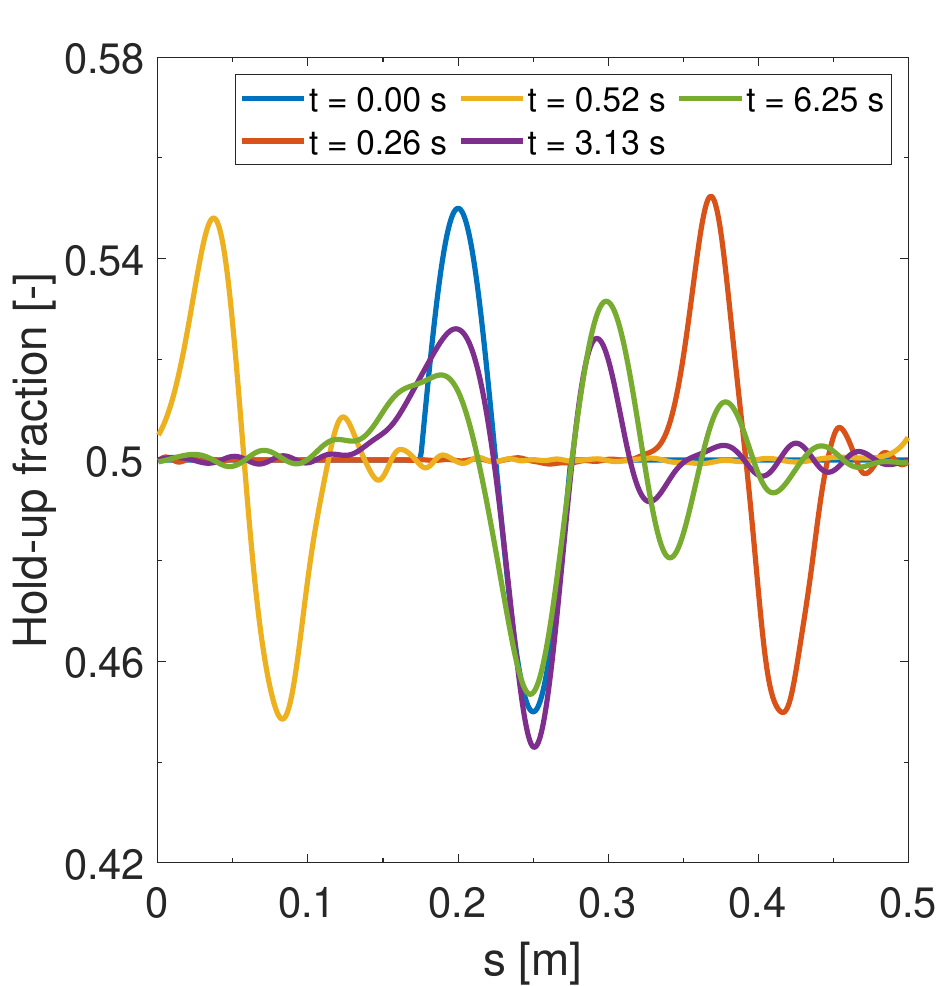} 
\includegraphics[width=0.45\linewidth]{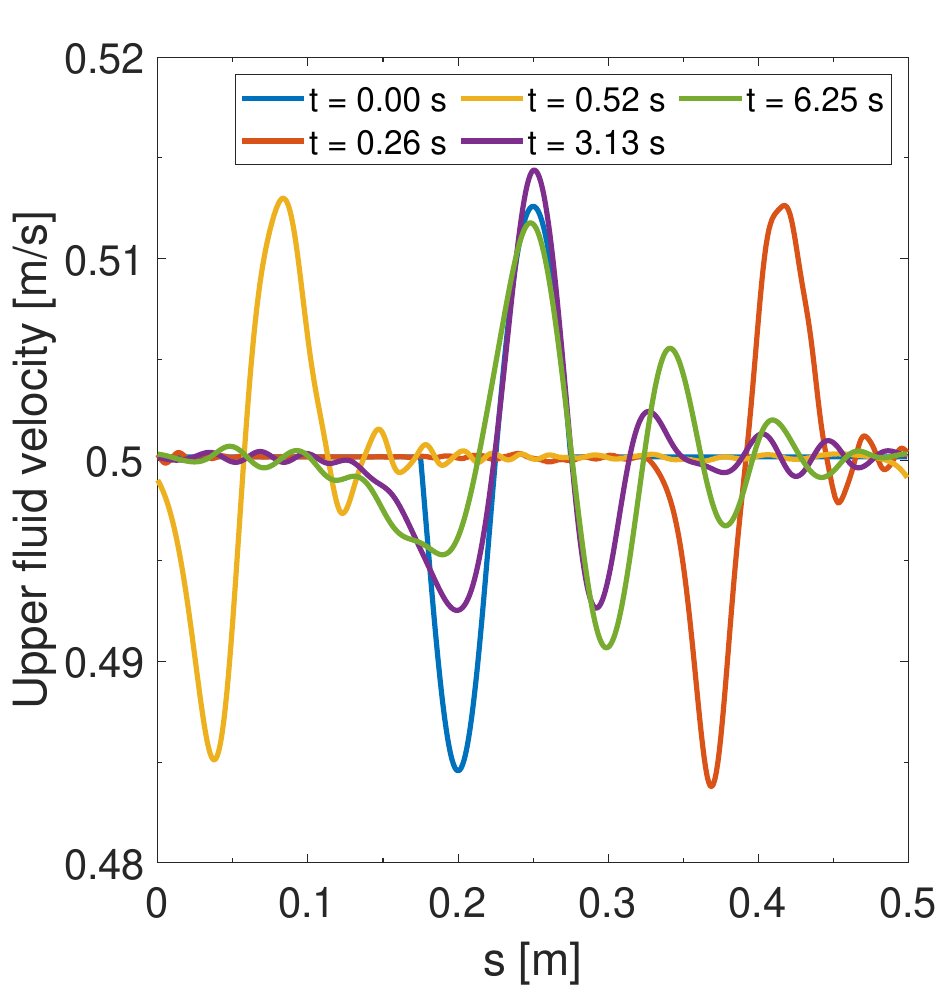} 
\caption{Two components of the solution to the traveling wave case for the basic model plus surface tension, using the energy-stable advective flux, with $\Delta s = 10^{-3} \, \mathrm{m}$ and $\Delta t = 10^{-4} \, \mathrm{s}$.}
\label{fig:numerical_experiments/trav_wave_surf/ec_lim_snapshots}
\end{figure}

The addition of surface tension results in a different angular frequency of $\omega = 40.19 \, \mathrm{s}^{-1}$, for $\lambda = 0.1 \, \mathrm{m}$ (the other mode is $\omega = 22.64 \, \mathrm{s}^{-1}$).
The solution at various points in time is shown in \autoref{fig:numerical_experiments/trav_wave_surf/ec_lim_snapshots}.
Due to the slightly increased wave speed, the snapshots at $t = 3.13 \, \mathrm{s}$ and $6.25 \, \mathrm{s}$ are now the points at which the wave has traveled through the domain 4 and 8 times respectively.
The addition of surface tension has a dispersive effect: the traveling wave spreads out into smaller oscillations, which are not of numerical origin. 
This can be determined from the fact that they do not vanish upon grid refinement. 

\begin{figure}[htbp] 
\centering
\includegraphics[width=0.45\linewidth]{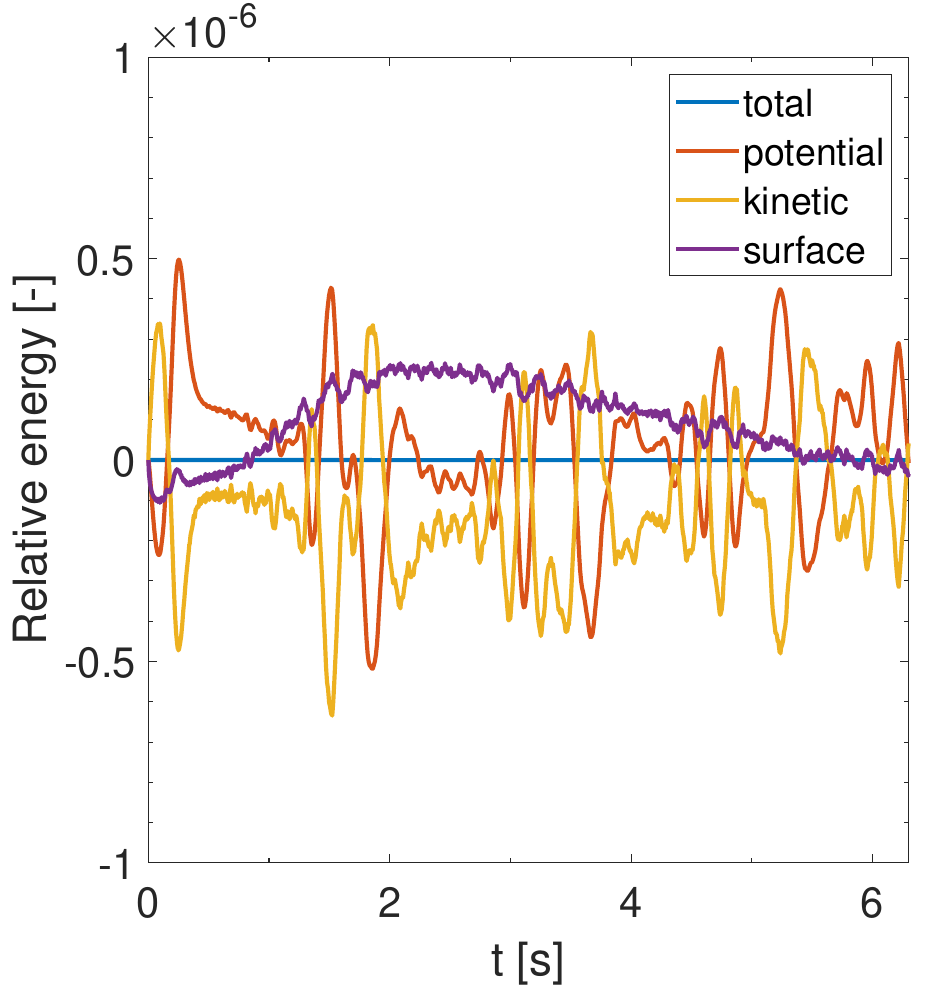} 
\includegraphics[width=0.45\linewidth]{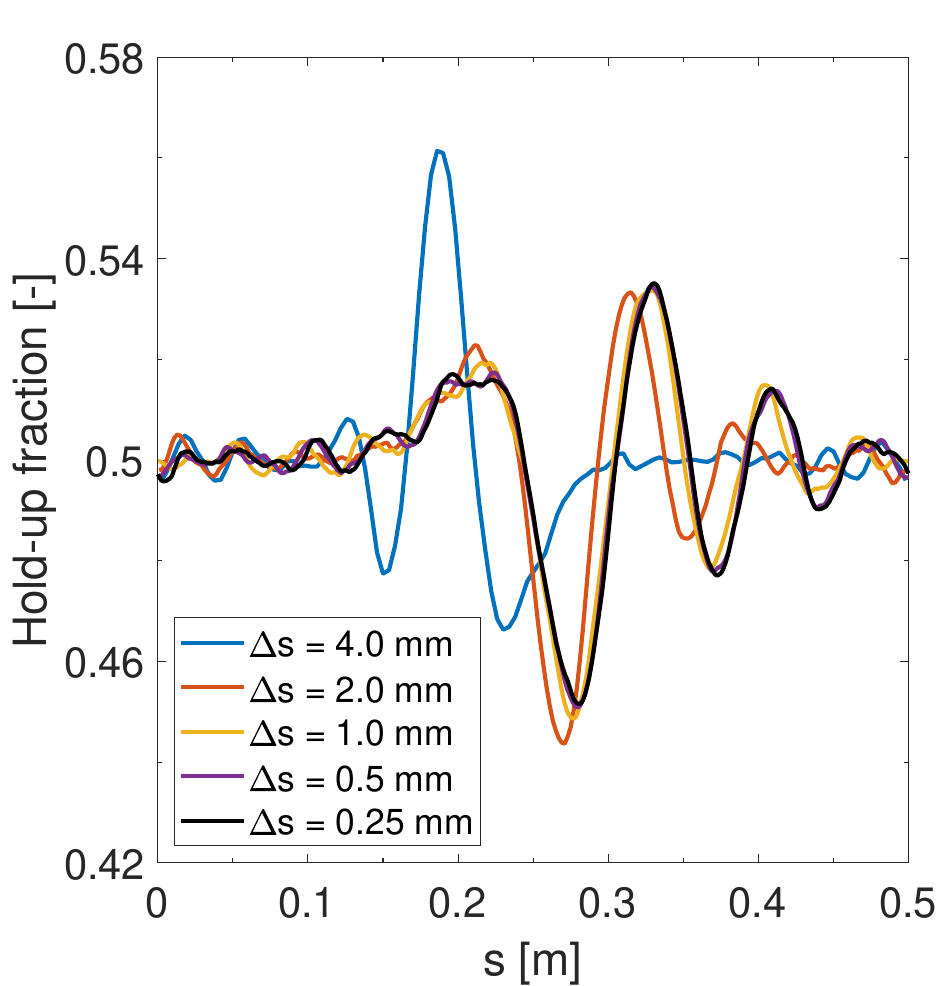} 
\caption{Results for the traveling wave case with surface tension, using the energy-conserving advective flux. Left: components of the energy of the solution, with $\Delta s = 10^{-3} \, \mathrm{m}$ and $\Delta t = 10^{-4} \, \mathrm{s}$. Right: convergence of the hold-up at time $t=6.3 \, \mathrm{s}$, with a constant ratio $\Delta t/\Delta s = 0.1  \, \mathrm{s}/\mathrm{m}$. }
\label{fig:numerical_experiments/trav_wave_surf/ec_convergence_and_energy}
\end{figure}

\autoref{fig:numerical_experiments/trav_wave_surf/ec_convergence_and_energy} shows energy and convergence results using the energy-conserving advective flux. 
Using the energy-conserving advective flux makes it possible to isolate the effect of the surface tension implementation on the (global) energy. 
The figure shows how the total energy remains constant in time. 
This confirms our theoretical analysis: the surface tension implementation is indeed energy-conserving.
The different components of the energy (potential, kinetic, and surface energy) are free to increase or decrease, exchanging with one another. 
The magnitude of the exchange is small. 

\begin{figure}[htbp]
\centering
\includegraphics[width=0.45\linewidth]{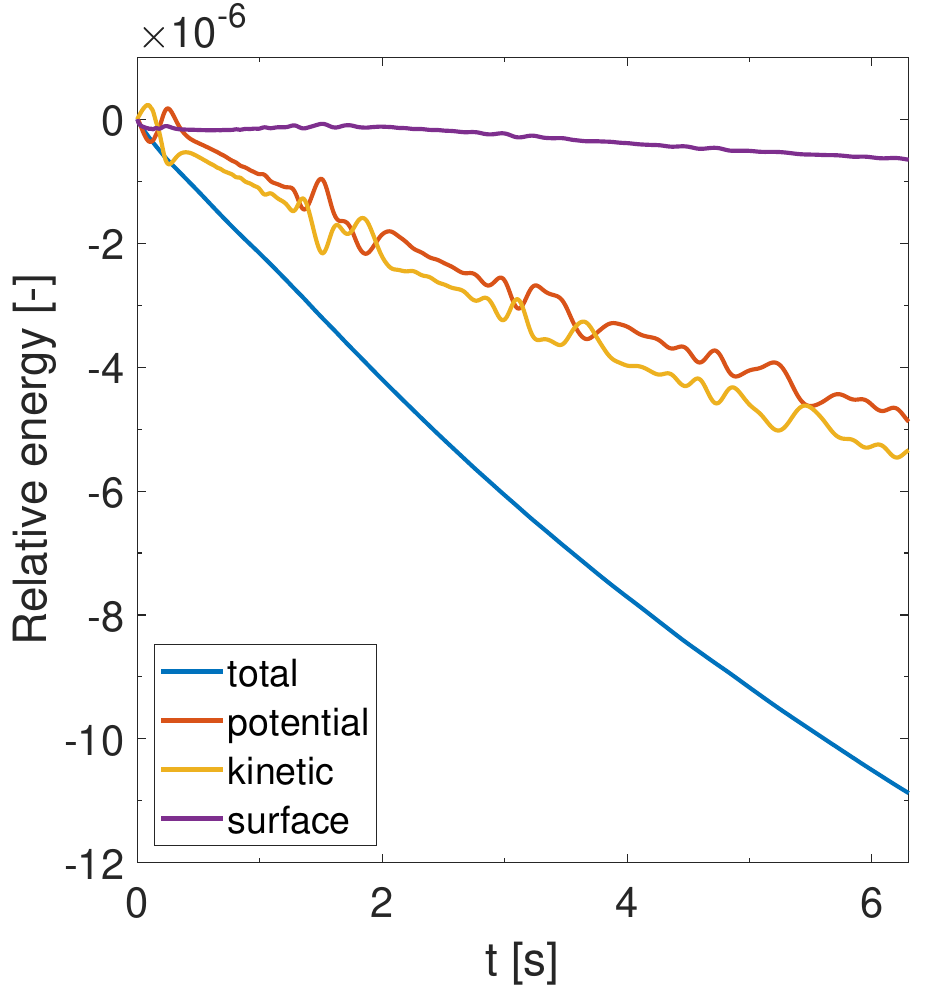} 
\includegraphics[width=0.45\linewidth]{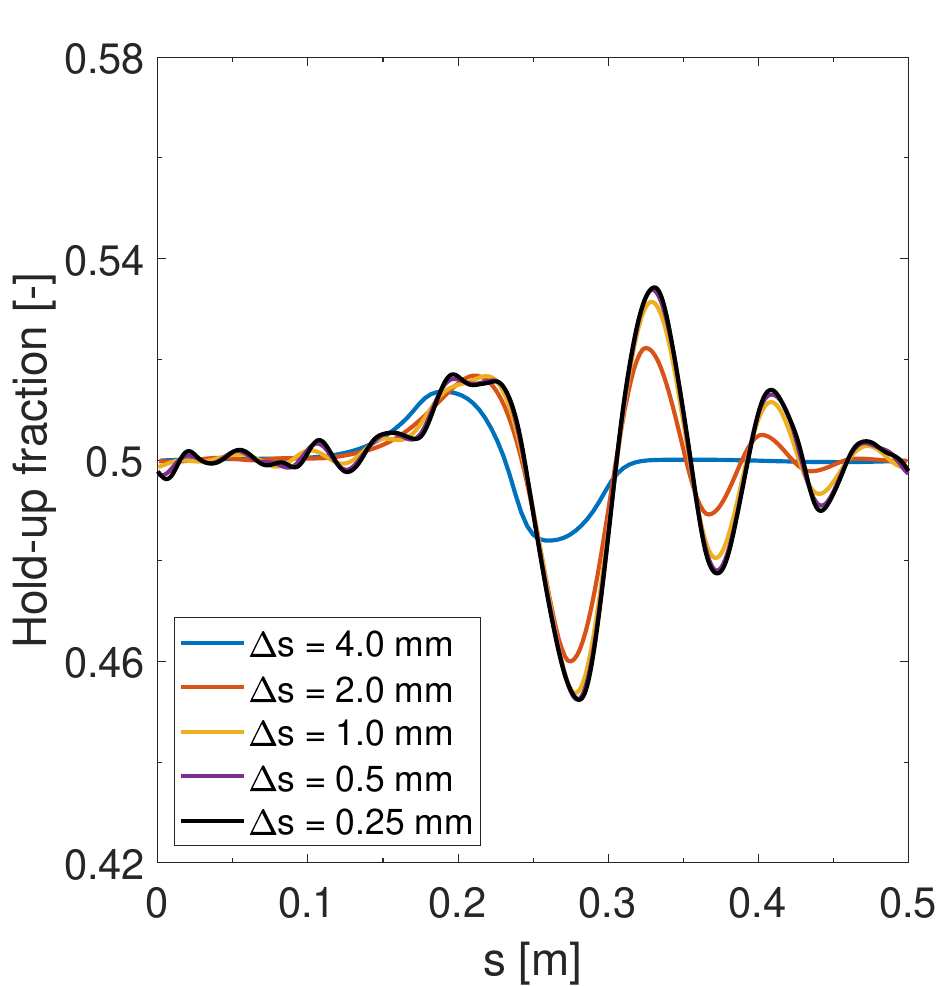} 
\caption{Results for the traveling wave case with surface tension, using the energy-stable advective flux. Left: components of the energy of the solution, with $\Delta s = 10^{-3} \, \mathrm{m}$ and $\Delta t = 10^{-4} \, \mathrm{s}$. Right: convergence of the hold-up at time $t=6.3 \, \mathrm{s}$, with a constant ratio $\Delta t/\Delta s = 0.1  \, \mathrm{s}/\mathrm{m}$.}
\label{fig:numerical_experiments/trav_wave_surf/ec_lim_convergence_and_energy}
\end{figure}

\autoref{fig:numerical_experiments/trav_wave_surf/ec_lim_convergence_and_energy} shows energy and convergence results using the energy-stable advective flux. 
Using this flux, the total energy is not conserved, but decreases monotonically, as discussed in \autoref{ssec:numerical_experiments/traveling_wave}. 
A comparison between the right plots of \autoref{fig:numerical_experiments/trav_wave_surf/ec_convergence_and_energy} and \autoref{fig:numerical_experiments/trav_wave_surf/ec_lim_convergence_and_energy} shows that this comes with the advantages of smoother convergence and absence of numerical oscillations.
Therefore the energy-stable flux is favored, in combination with the energy-conserving surface tension discretization. 

\FloatBarrier

\subsection{Shock formation and dissipation in unstable region}
\label{ssec:numerical_experiments/double_wave}

In this test case we test our complete proposed framework, with all physical effects and the energy-stable advective flux, on a challenging case involving the rapid growth of a perturbation and development into a shock.
The flow is in the region of state space where the basic model is ill-posed: it is marked with an asterisk in the stability maps of \autoref{sec:stability}. 
However, with our extended model we are able to obtain good convergence and a well-resolved shock. 

\begin{figure}[htbp] 
\centering
\includegraphics[width=0.45\linewidth]{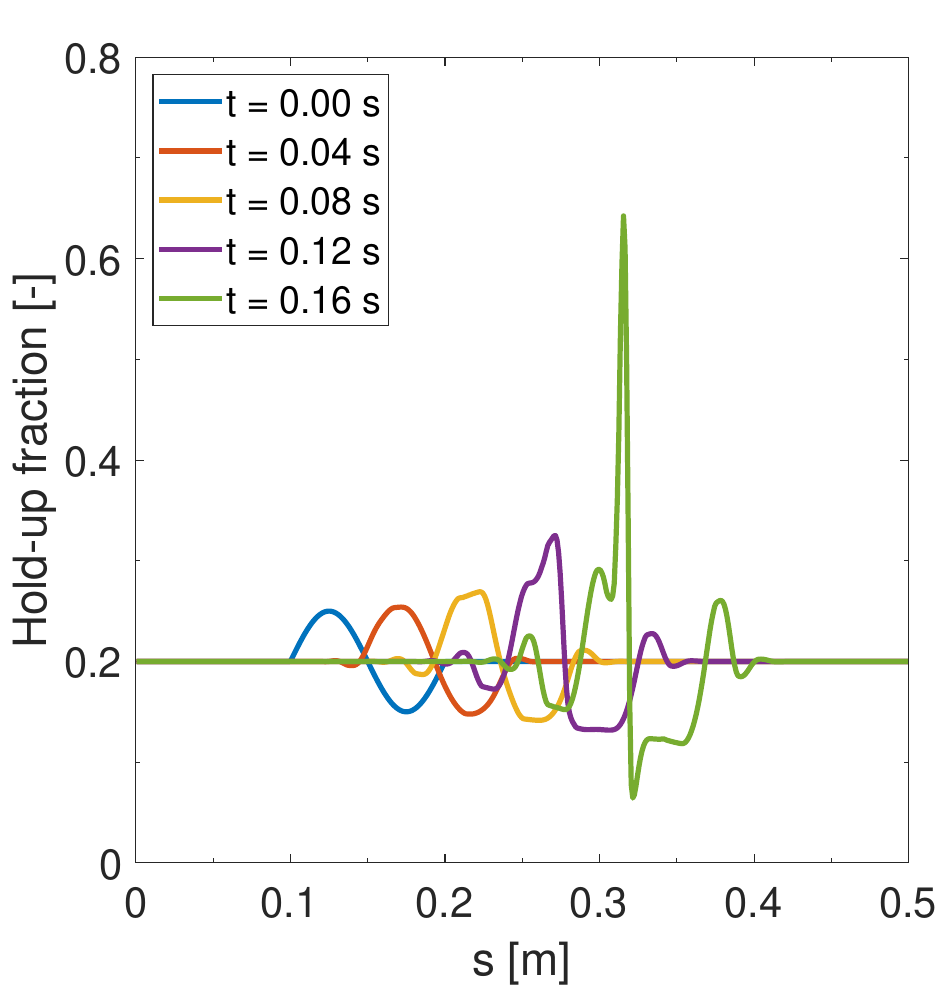} 
\includegraphics[width=0.45\linewidth]{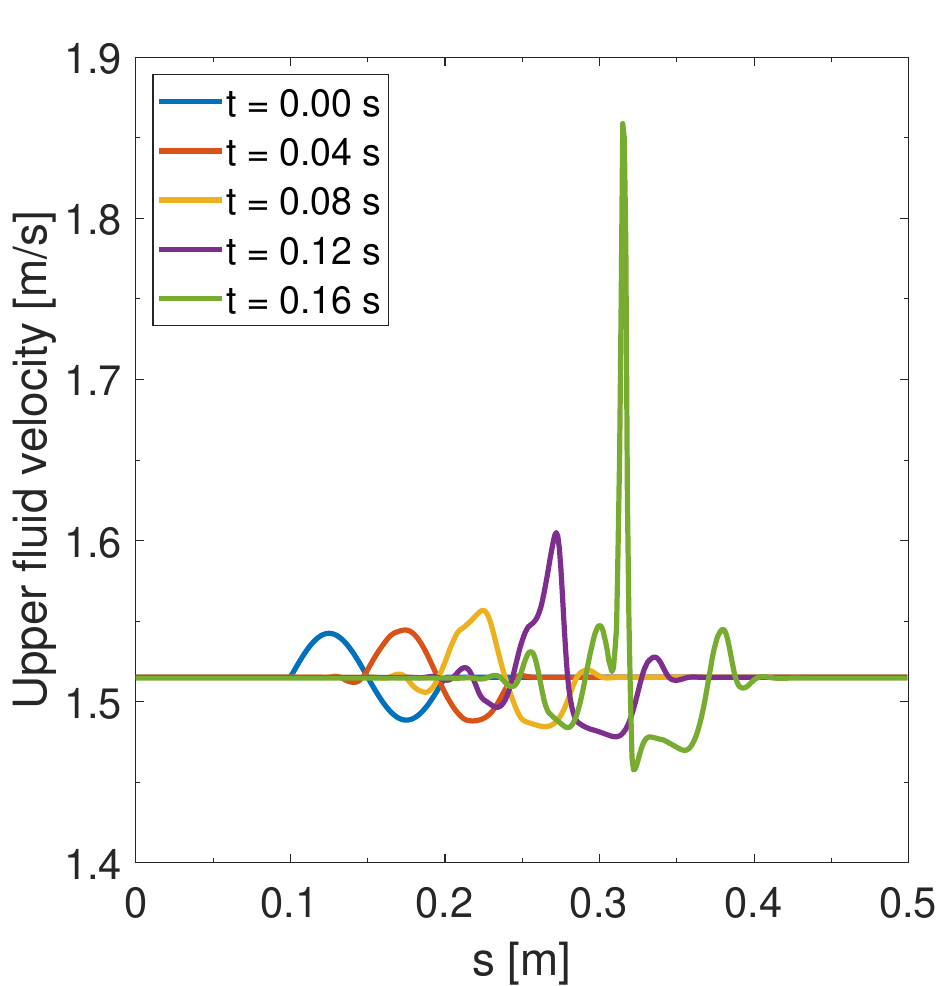} 
\caption{Two components of the solution to the unstable shock formation case, using the energy-stable advective flux, with $\Delta s = 10^{-3} \, \mathrm{m}$ and $\Delta t = 10^{-4} \, \mathrm{s}$.}
\label{fig:numerical_experiments/double_wave/ec_lim_snapshots}
\end{figure}

This test case is inspired by a case from \cite{FullmerRansomLopezdeBertodano2014}, which is in turn derived from \cite{HolmasSiraNordsveenEtAl2008}.
The boundaries are periodic, the flow parameters are given by \autoref{tab:Thorpe_parameters_LSA}, and the base state is given by \autoref{tab:Thorpe_base_state_asterisk}.
This base state is perturbed by a single wave (with wavelength $\lambda = 0.1 \, \mathrm{m}$) in the hold-up, while the other components of the solution are kept constant. 
Then, a projection step is performed, adjusting the velocities such that the volumetric flow constraint is satisfied. 
Two components of the resulting initial condition can be seen in \autoref{fig:numerical_experiments/double_wave/ec_lim_snapshots}, along with the evolution of the wave in time.
The main difference between this case and the case from \cite{FullmerRansomLopezdeBertodano2014} is that we add wall and interface friction, so that all sources of dissipation are included in the numerical experiment.
The friction is balanced by a driving pressure gradient, so that the base state is a steady state (see \autoref{sec:stability}).
Without the external forcing provided by a driving pressure gradient, the initial perturbation would quickly die out. 

While the basic model possesses an unbounded short wave growth rate for these flow conditions, the extended model damps short wavelength perturbations (see \autoref{fig:stability/dispersion_comparison}).
The wavelength of the perturbation considered here is still unstable, and indeed the perturbation is observed to grow rapidly in \autoref{fig:numerical_experiments/double_wave/ec_lim_snapshots}.
Fortunately, as predicted by the linear stability analysis, it grows at a finite rate, and is not dominated by extreme short wavelength instabilities.
Finally, after developing into a shock its growth is stopped by nonlinear effects. 
After this point, secondary perturbations will start to grow, eventually developing into a full wave train consisting of several shocks, but all with limited amplitude. 

\begin{figure}[htbp] 
\centering
\includegraphics[width=0.45\linewidth]{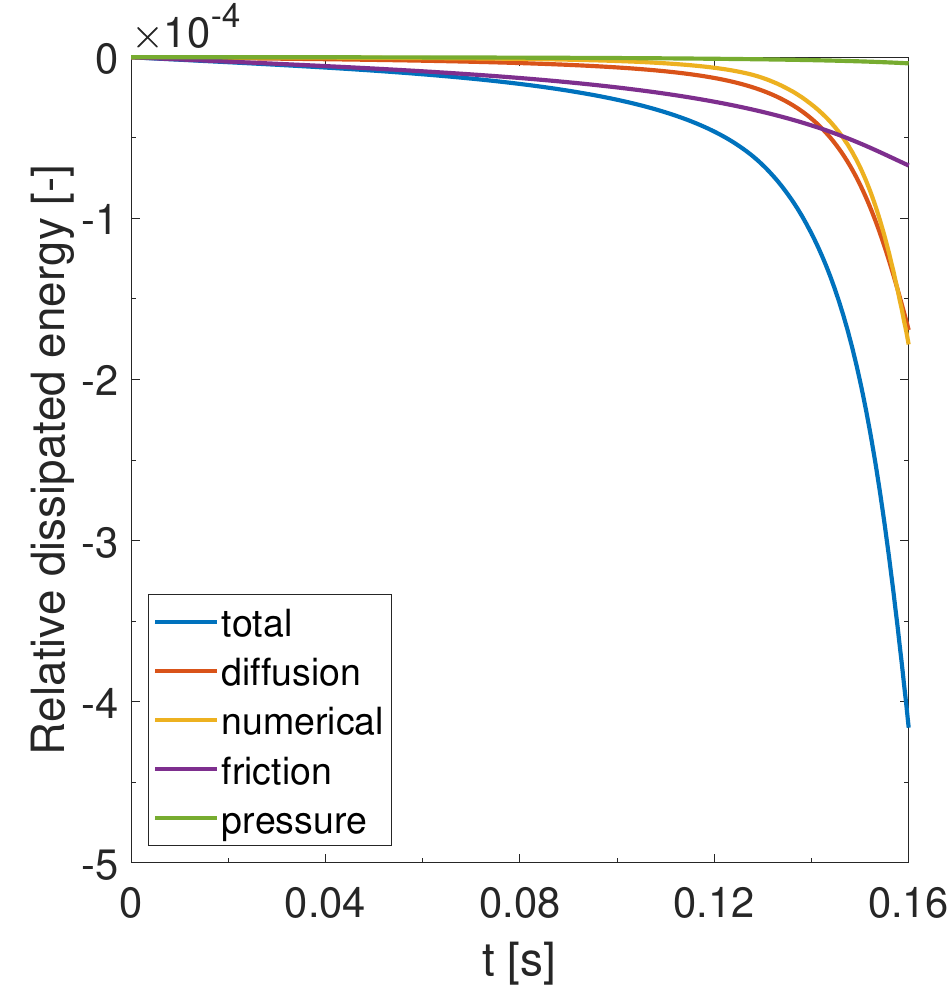} 
\includegraphics[width=0.45\linewidth]{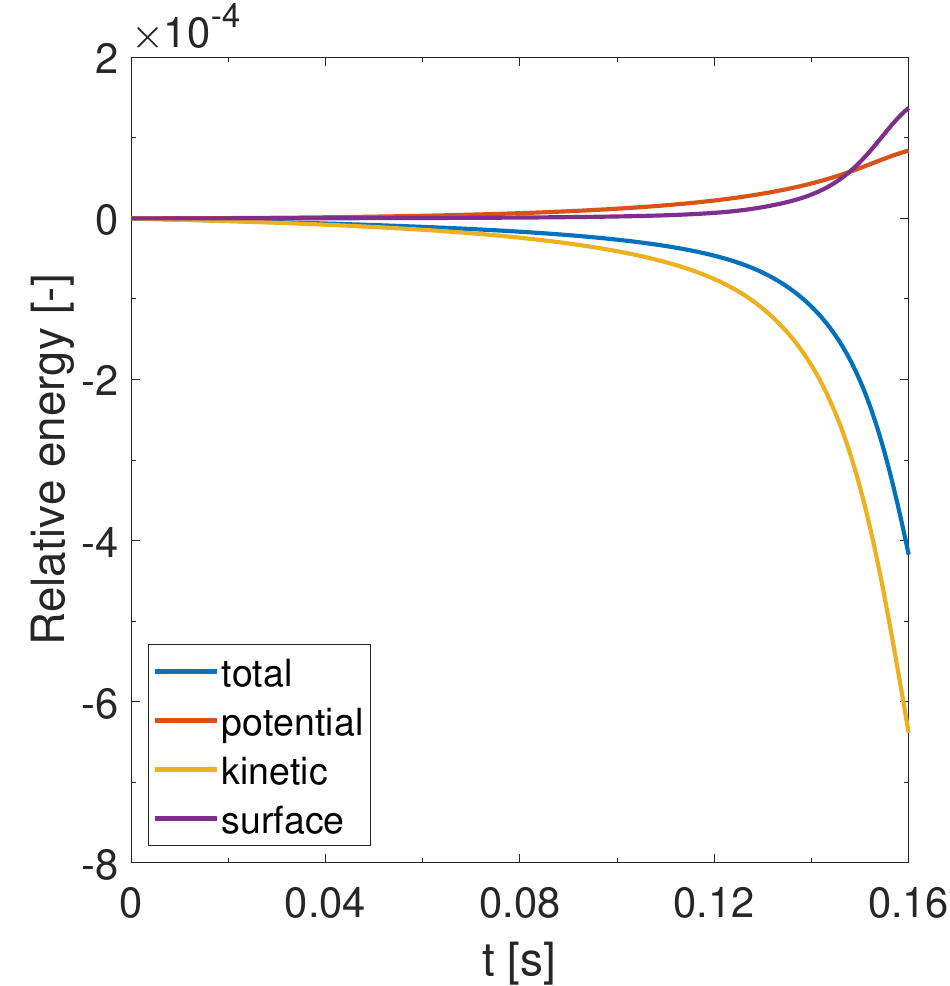} 
\caption{For the unstable shock formation case, this figure shows the dissipated energy (left) and the instantaneous energy (right), made non-dimensional by the total energy of the initial condition. Using the energy-stable advective flux, with $\Delta s = 10^{-3} \, \mathrm{m}$ and $\Delta t = 10^{-4} \, \mathrm{s}$. The total dissipation is divided into contributions from physical diffusion, numerical diffusion, wall and interface friction, and a production term due to an externally applied driving pressure gradient.}
\label{fig:numerical_experiments/double_wave/dissipation}
\end{figure}

The global dissipation as a function of time is shown in \autoref{fig:numerical_experiments/double_wave/dissipation}.
Here the dissipated energy is calculated using the expressions for the local dissipation - $\epsilon_{d,i-1/2}$, $\epsilon_{f,i-1/2}$, $\epsilon_{n,i-1/2}$ - and the expression for energy production due to a driving pressure gradient - $c_{p,i-1/2}$.
These expressions are summed over the domain and integrated in time (numerically) according to \eqref{eq:semi-discrete/summary/global_energy_conservation_equation}, and their sum yields the total dissipated energy. 
Since the initial base state is uniform, it has no (physical or numerical) diffusion, but it does have high dissipation due to friction which is balanced by an energy input from the driving pressure gradient.
These base state dissipation and production terms have been subtracted so that friction and the driving pressure gradient do not dominate the plot.

In the second plot of \autoref{fig:numerical_experiments/double_wave/dissipation}, the instantaneous energy is calculated using the expression for $e_{i-1/2}$, summed over the domain to yield the global energy. 
The left and right plots of \autoref{fig:numerical_experiments/double_wave/dissipation} show the same decrease in total energy, confirming that the two methods of calculation are consistent.  

\autoref{fig:numerical_experiments/double_wave/dissipation} reveals exactly how nonlinear effects bound the amplitude of the shock. 
The respective contributions of the physical and numerical diffusion to the nonlinear damping can now be quantified, by examining their effect on the energy of the solution.
The figure shows that as the shock develops, the physical and numerical diffusion and their resulting dissipation grow large, and decrease the energy of the solution.  
They act to decrease the kinetic energy of the solution, allowing the potential and surface energy to grow slightly. 
We note that a calculation of the local dissipation shows the dissipation to be localized around the shock.

The dissipation due to friction also grows with time, but less dramatically, since it is proportional to the size of the wave, not to its steepness.
It has a smaller stabilizing effect.
Regarding the driving pressure gradient, the energy input remains roughly constant, since it is not dependent on the perturbation but only on the volumetric flow rate, which is a property of the complete flow. 
Its negative value at $t=0.16 \, \mathrm{s}$ means that the energy input is slightly lower than it was for the initial base state, since the volumetric flow rate has decreased slightly, indicating that the flow has been slowed down slightly.

\begin{figure}[htbp] 
\centering
\includegraphics[width=0.45\linewidth]{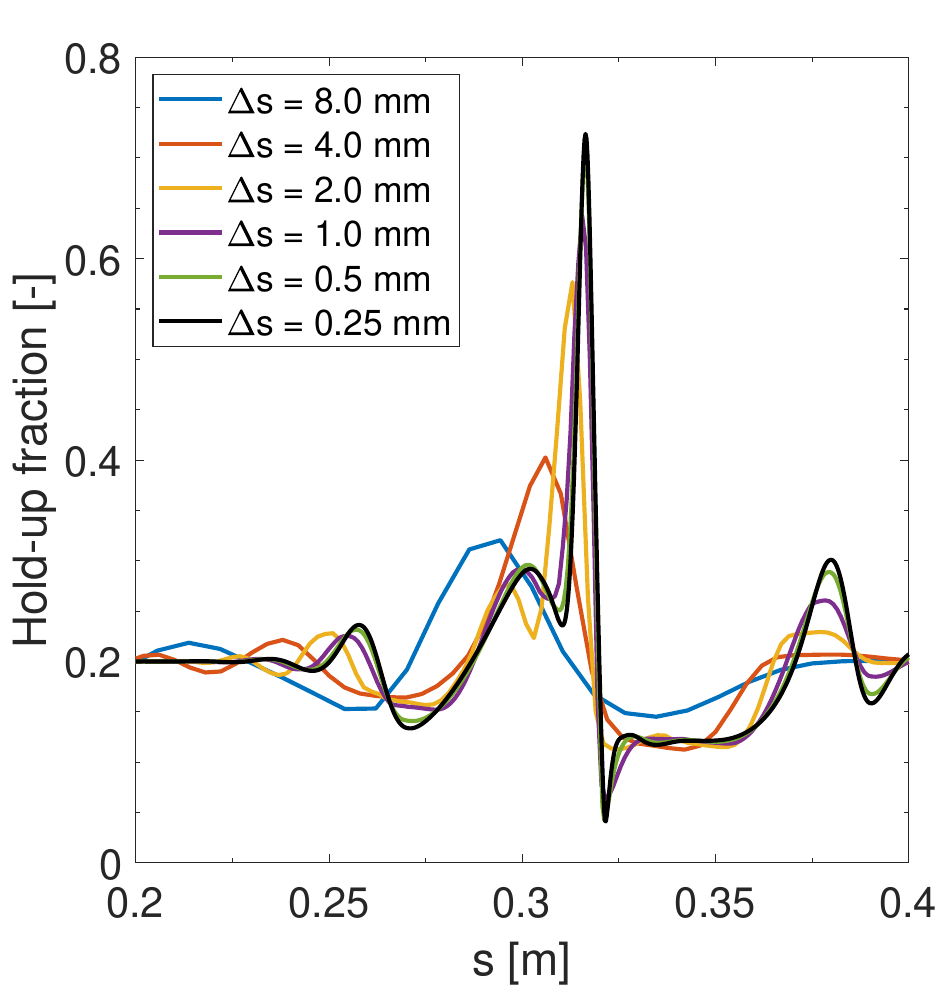} 
\includegraphics[width=0.45\linewidth]{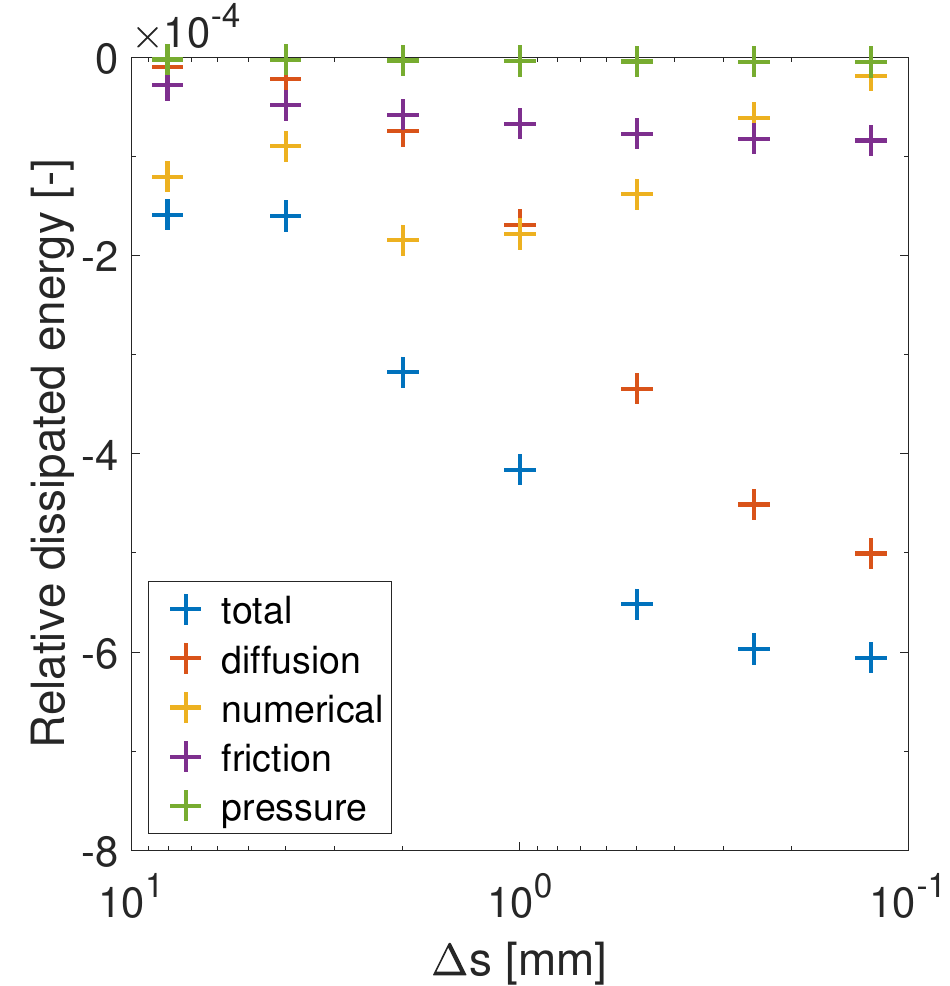} 
\caption{For the unstable shock formation case, this figure shows the convergence with grid resolution, of the solution at time $t=0.16 \, \mathrm{s}$. Using the energy-stable advective flux, with a constant ratio $\Delta t/\Delta s = 0.1  \, \mathrm{s}/\mathrm{m}$. Left: hold-up fraction $\alpha_L=A_L/A$. Right: dissipated energy, divided by the total energy of the initial condition.}
\label{fig:numerical_experiments/double_wave/convergence}
\end{figure}

\autoref{fig:numerical_experiments/double_wave/convergence} shows how the solution converges with grid resolution, confirming that the extended model is well-posed, as discussed in \autoref{sec:stability}.
Also shown in \autoref{fig:numerical_experiments/double_wave/convergence} is the convergence of the dissipation, divided into its different components.
As the grid is refined, the small scales at which the physical diffusion acts are better resolved, allowing the corresponding dissipation to grow and converge to its full physical effect.
In contrast, beyond a certain resolution before which the solution is relatively smooth, the numerical dissipation decreases with grid resolution. 
Only at coarse resolutions, numerical dissipation is needed to compensate for the lack of physical dissipation.
Dissipation due to friction only varies slightly with grid resolution, since it is not a small scale phenomenon.

Previous work has described how the linearly unstable wave is bounded by (nonlinear) dissipation in the shock, due to numerical and physical diffusion \cite{FullmerRansomLopezdeBertodano2014}.
However, up to now, dissipation has remained an abstract concept for the TFM.
Here we provide definitions for the various components of the dissipation, and specify their effect on a well-defined energy.
Therefore, dissipation has become a concrete quantity that can be measured.
This provides a stronger basis for discussions of the nonlinear damping of unstable waves.
We confirm the conclusions of \cite{FullmerRansomLopezdeBertodano2014}, who observe that with only numerical diffusion, the solution fails to converge (with oscillations appearing at high resolutions), due to the lack of numerical diffusion at high grid resolutions.
We similarly observe (results not shown here) that physical diffusion without numerical diffusion leads to a less clear convergence, with coarse grid solutions being insufficiently diffused.
With \autoref{fig:numerical_experiments/double_wave/convergence} we have made concrete that with a combination of numerical and physical diffusion, the dissipation in the shock smoothly converges to a finite value, realizing a grid-independent bound on the amplitude of the shock. 

\FloatBarrier

 \section{Conclusions} \label{sec:conclusion}
 
This paper has proposed a complete energy-stable framework -- including diffusion, friction, surface tension, and an energy-stable advective flux scheme -- for reliable simulations with the one-dimensional two-fluid model (TFM).
The paper builds on our earlier work on the energy-conserving basic TFM, which we have extended in an energy-consistent manner. 
We have shown that for the channel geometry, surface tension can be added to the model in an energy-conserving manner.
The additions of friction and momentum diffusion have been shown to be strictly dissipative. 
Therefore, these extensions yield an energy-stable model.

Besides their implications for energy stability, the additions to the model also solve the basic model's issue of unbounded linear instability at short wavelengths: diffusion and surface tension introduce a cut-off wavelength below which perturbations are damped. 
The cut-off wavelength is shown to depend on the state: it decreases with increasing velocity difference between the phases.
Nevertheless, there exists a cut-off wavelength for any state, rendering the model unconditionally well-posed.
These cut-off wavelengths may be shorter than the scales at which the TFM is usually employed, but it is these scales at which the ill-posedness issue of the one-dimensional model resides and it is these scales which need to be stabilized.
Diffusion and surface tension offer a clearly physically motivated way to do so.
Precisely because diffusion and surface tension are physically motivated, they fit well into our energy-stable framework. 

The energy conservation and dissipation properties of the model have been proven to carry over to the semi-discrete model, when the model and its energy are discretized in a specific manner. 
The semi-discrete model with surface tension is exactly energy-conserving, if the surface energy is added to the basic energy as an extra term. 
Diffusion and friction add strictly dissipative terms to the local energy conservation equation, with expressions for the dissipation rates that can be evaluated as functions of the local instantaneous solution.

However, the key highlight of our semi-discrete model is a new discretization of the advective terms, which combines a previously developed energy-conserving discretization and a strictly dissipative upwind discretization.
These discretizations are combined using flux limiters.
The novel combined advective flux is energy stable, and comes with an explicit expression for the numerical dissipation rate, that can be evaluated during a numerical simulation.
It is designed to be energy-conserving where the solution is smooth, and dissipative where the solution has strong gradients.
This dissipation is motivated by the fact that the energy conservation property of the continuous model does not hold for discontinuous solutions: dissipation is required in this case.
The novel energy-stable flux retains the advantages in stability and physical fidelity of the original energy-conserving flux, without the latter's tendency to generate numerical oscillations near discontinuities.

In numerical experiments, spatially exact energy conservation is demonstrated for the basic model extended with surface tension, using the original energy-conserving flux. 
The upwind and energy-stable advective fluxes are demonstrated to be strictly dissipative, as opposed to a naive central discretization that is neither conservative nor strictly dissipative.
The energy-stable scheme is shown to yield smooth solutions without numerical oscillations.
It is much less diffusive than a first-order upwind scheme, and this is reflected in the dissipation, which is lower for the energy-stable scheme than for the upwind scheme.

A challenging test of our complete framework is provided by the simulation of an unstable wave in a region of state space where the basic model is linearly ill-posed.
Our proposed framework yields a convergent solution, confirming that it is well-posed. 
The unstable perturbation develops into a shock, which is bounded by nonlinear dissipation.
The analytical results of this work enable a precise analysis of the dissipation and better insight into the nonlinear damping taking place. 
The dissipation due to numerical and physical diffusion are observed to grow as the wave steepens, with numerical dissipation dominating at coarse resolutions and physical dissipation dominating at fine resolutions.
Together, numerical and physical diffusion yield a smoothly converging total dissipation, and a smoothly converging solution.

In order to see the full effect of the physical diffusion and surface tension and reach convergence, the grid needs to be refined to high resolutions that may be impractical in engineering applications. 
This is not a problem from the perspective of stability: it only means that not all short scale dynamics of the model can be resolved in practice.
At coarser resolutions, our energy-stable framework provides solutions that are similar to the converged solutions, except that sharp perturbations are diffused.
The convergence plots show a monotonic steepening of the waves, without spurious oscillations. 

\section*{CRediT}
\textbf{J.F.H. Buist}: Conceptualization, Formal Analysis, Investigation, Software, Writing - Original Draft; \textbf{B. Sanderse}: Conceptualization, Software, Writing - Review \& Editing; \textbf{S. Dubinkina}: Writing - Review \& Editing; \textbf{C.W. Oosterlee}: Writing - Review \& Editing; \textbf{R.A.W.M. Henkes}: Writing - Review \& Editing.

\section*{Funding}

This work was supported by the research program Shell-NWO/FOM Computational Sciences for Energy Research (CSER), project number 15CSER017, which is partly financed by the Netherlands Organization for Scientific Research (NWO).

\numberwithin{equation}{section}
\numberwithin{figure}{section}
\begin{appendices}

\section{Geometric relations} 
\label{sec:geometric_relations} 

The equations are written in terms of the cross-sectional areas occupied by each fluid, which in general can be defined to be related to the interface height $H_L$ via
\begin{align}
A_L &= \int_{0}^{H_L} w(h) \, \mathrm{d} h,  &
A_U &= \int_{H_L}^{H} w(h) \, \mathrm{d} h,
\label{eq:geometric_relations/cross-section}
\end{align}
with $w(h)$ the local duct width. Note that $w(H_L)=P_\mathrm{int}$, where $P_\mathrm{int}$ is the (generalized) interface perimeter which is shown for a circular pipe geometry in \autoref{fig:continuous/model/two-fluid_schematic}. 
For the 2D channel geometry, $w(h)=1$, and the cross-sections $A_L$ and $A_U$ are identical to the fluid heights $H_L$ and $H_U$.

The geometric quantities $\widehat{H}_L$ and $\widehat{H}_U$
which appear in the governing equations of the two-fluid model, also have general definitions:
\begin{align}
\widehat{H}_L &\coloneqq   \int_{0}^{H_L} (h-H_L) w(h) \, \mathrm{d} h, &
\widehat{H}_U &\coloneqq  \int_{H_L}^H (h-H_L) w(h) \, \mathrm{d} h.
\label{eq:geometric_relations/general_level_gradient_integral}
\end{align}
For a 2D channel geometry the integrals evaluate to 
\begin{align*}
\widehat{H}_L &=  -\frac{1}{2} H_L^2, & 
\widehat{H}_U &=  \frac{1}{2}  H_U^2,
\end{align*}
where we have substituted $H_L=H-H_U$. For the pipe geometry, the results of the integrals are given by \cite{SanderseSmithHendrix2017}:
\begin{align*}
\widehat{H}_L &=  \left[ (R-H_L)A_L - \frac{1}{12} P_\mathrm{int}^3 \right], & 
\widehat{H}_U &= - \left[(R-H_U) A_U - \frac{1}{12} P_\mathrm{int}^3 \right].
\end{align*}

Besides $\widehat{H}_L$ and $\widehat{H}_{U}$, the following geometric quantities are used in the energy definition:
\begin{align}
\widetilde{H}_L &\coloneqq   \int_{0}^{H_L} h w(h) \, \mathrm{d} h = \widehat{H}_L + H_L A_L,  &
\widetilde{H}_U &\coloneqq  \int_{H_L}^H h w(h) \, \mathrm{d} h = \widehat{H}_U + (H-H_U) A_U. 
\label{eq:geometric_relations/general_potential_energy_integral}
\end{align}

We can apply Leibniz' rule to \eqref{eq:geometric_relations/cross-section} (see \cite{BuistSanderseDubinkinaEtAl2022}) to obtain
\begin{align*}
   \d{A_L}{H_L} &= P_\mathrm{int},  &   \d{A_U}{H_U} &= P_\mathrm{int},  
\end{align*}
where we have used $H_L + H_U = H$. 
Applying the same technique to \eqref{eq:geometric_relations/general_level_gradient_integral} and \eqref{eq:geometric_relations/general_potential_energy_integral} yields
\begin{align*}
  \d{\widehat{H}_L}{H_L} &= - A_L, &  \d{\widehat{H}_U}{H_U} &= A_U, \\
  \d{\widetilde{H}_L}{H_L} &=  H_L P_\mathrm{int}, &  \d{\widetilde{H}_U}{H_U} &=  (H-H_U) P_\mathrm{int}, 
\end{align*}
which leads to
\begin{align}
\d{\widehat{H}_L}{A_L} &= \d{\widehat{H}_L}{H_L} \left[\d{A_L}{H_L} \right]^{-1} = -\frac{A_L}{P_\mathrm{int}},  & \d{\widehat{H}_U}{A_U} &= \d{\widehat{H}_U}{H_U} \left[\d{A_U}{H_U} \right]^{-1} = \frac{A_U}{P_\mathrm{int}}, \\
\d{\widetilde{H}_L}{A_L} &= \d{\widetilde{H}_L}{H_L} \left[\d{A_L}{H_L} \right]^{-1} = H_L,  & \d{\widetilde{H}_U}{A_U} &= \d{\widetilde{H}_U}{H_U} \left[\d{A_U}{H_U} \right]^{-1} = H-H_U.
\label{eq:geometric_relations/general_potential_energy_integral/derivative_to_A}
\end{align}
A similar calculation, in which we assume that $H$ is constant, yields
\begin{align*}
  \pd{\widehat{H}_L}{s} &= -A_L \pd{H_L}{s}, & \pd{\widehat{H}_U}{s} &= -A_U \pd{(H-H_U)}{s}.
\end{align*}
Finally, comparison to \eqref{eq:geometric_relations/general_potential_energy_integral/derivative_to_A} yields
\begin{align}
  \pd{\widehat{H}_L}{s} &= -A_L \pd{}{s} \left( \d{\widetilde{H}_L}{A_L}  \right), & \pd{\widehat{H}_U}{s} &= -A_U \pd{}{s} \left( \d{\widetilde{H}_U}{A_U} \right),
  \label{eq:geometric_relations/essential_geometric_relations}
\end{align}
and these geometric relations are critical to deriving the local energy conservation equation.

\section{Friction closure relations}
\label{sec:friction_closure_relations} 

The wall and interface stresses of the two-fluid model are typically modeled in the following manner \cite{TaitelDukler1976}:
\begin{equation*}
\tau_{L} = -\frac{1}{2} f_L \rho_L u_L |u_L|, \quad \tau_{U} = -\frac{1}{2} f_U \rho_U u_U |u_U|, \quad  \tau_{\mathrm{int}} = -\frac{1}{2} f_{\mathrm{int}} \rho_U \left(u_U - u_L \right) |u_U - u_L|,
\end{equation*}
in which $f_L$, $f_U$, and $f_{\mathrm{int}}$ are the Fanning friction factors, which require further closure relations.
The friction factors depend on the Reynolds numbers
\begin{align*}
\mathrm{Re}_L &= \frac{|u_L| D_L}{\nu_{m,L}}, & \mathrm{Re}_U &= \frac{ |u_U| D_U}{\nu_{m,U}},
\end{align*}
with hydraulic diameters 
\begin{align*}
D_L &= \frac{4 A_L}{P_L}, & D_U &= \frac{4 A_U}{P_U + P_{\mathrm{int}}}.
\end{align*}
In this work we use the Taitel and Dukler friction model \cite{BarneaTaitel1993, TaitelDukler1976}
\begin{equation*}
f_L = \frac{C}{\mathrm{Re}_L^n}, \quad \quad f_U = \frac{C}{\mathrm{Re}_U^{n}}, \quad \quad f_{\mathrm{int}} = \mathrm{max} \left(f_U, 0.014 \right),
\end{equation*}
with coefficients $C = 0.046$ and $n = 0.2$ (valid for turbulent flow).

\section{Linear stability analysis}
\label{sec:linear_stability_analysis} 

We conduct a linear stability analysis of the (continuous) model, following \cite{FullmerRansomLopezdeBertodano2014, LiaoMeiKlausner2008, LopezdeBertodanoFullmerClausseEtAl2017}.
The analysis starts by writing \eqref{eq:continuous/summary/extended_governing_equations} in quasilinear matrix form, which can be done by substituting the volume constraint and assuming the solution is smooth:
\begin{equation}
\v{A}(\v{w}) \pd{\v{w}}{t}  + \v{B}(\v{w}) \pd{\v{w}}{s}  + \v{E}(\v{w}) \pdd{\v{w}}{s}  + \v{G}(\v{w}) \pddd{\v{w}}{s}  = \v{c}(\v{w}),
\label{eq:LSA/quasilinear_system}
\end{equation}
with
\begin{equation*}
\v{w} 
=
\begin{bmatrix}
w_1 \\
w_2 \\
w_3 \\
w_4
\end{bmatrix}
=
\begin{bmatrix}
A_L \\
u_L \\
u_U \\
p
\end{bmatrix},
\quad \quad 
\v{A} = 
\begin{bmatrix}
1	&	0			&	0			&	0	\\
-1		&	0			&	0			&	0	\\
0 & 1 & 0 &	0	\\
0	& 0		& 1	&	0	
\end{bmatrix}, 
\end{equation*}
\begin{equation*}
\v{B} = 
\begin{bmatrix}
w_2	&	 w_1	&	0				& 0	\\
- w_3 	&	0				&	A - w_1	& 0	\\
-  \frac{g_n}{w_1} \d{\widehat{H}_L}{A_L} &  w_2 -\frac{1}{w_1} \pd{}{s} \left(\nu_{\mathrm{eff},L} w_1 \right)	&	0				& 1/\rho_L 	\\
  \frac{g_n}{A-w_1} \d{\widehat{H}_U}{A_U}  & 0 &    w_3 -\frac{1}{A-w_1} \pd{}{s} \left( \nu_{\mathrm{eff},U} (A-w_1) \right)& 1/\rho_U 
\end{bmatrix}, 
\end{equation*}
\begin{equation*}
\v{c} = 
\begin{bmatrix}
0 \\
0 \\
-\frac{1}{\rho_L} \pd{p_0}{s}  + \frac{\tau_L P_L}{\rho_L w_1} - \frac{\tau_\mathrm{int} P_\mathrm{int}}{\rho_L w_1} -   g_s   \\
 -\frac{1}{\rho_U} \pd{p_0}{s} + \frac{\tau_U P_U}{\rho_U (A-w_1)} + \frac{\tau_\mathrm{int} P_\mathrm{int}}{\rho_U (A-w_1)} - g_s  
\end{bmatrix}, 
\end{equation*}
\begin{equation*}
\v{E} =
\begin{bmatrix}
0 & 0 & 0 & 0\\
0 & 0 & 0 & 0\\
\frac{\sigma}{\rho_L} \frac{1}{P_\mathrm{int}^2} \d{P_\mathrm{int}}{A_L} \pd{w_1}{s} & -\nu_{\mathrm{eff},L}   & 0 & 0\\
0 & 0 &  -\nu_{\mathrm{eff},U}  & 0
\end{bmatrix},
\quad \quad
\v{G} =
\begin{bmatrix}
0 & 0 & 0 & 0\\
0 & 0 & 0 & 0\\
 -  \frac{\sigma}{\rho_L} \frac{1}{P_\mathrm{int}}  & 0  & 0 & 0\\
0 & 0 & 0  & 0
\end{bmatrix}.
\end{equation*}
Here we have used the second expression in \eqref{eq:continuous/surface_tension/dp_approximation} for the surface tension, which can be applied to both the 2D channel and the circular pipe geometries.

A general method for the linearization of systems of quasilinear partial differential equations is given by \cite{ProsperettiTryggvason2007}. 
The general solution is decomposed into $\v{w} = \v{\bar{w}} + \Delta \v{w}$ with $\Delta \v{w}$ a small disturbance ($\Delta \v{w} \ll \v{\bar{w}}$), and $\v{\bar{w}}$ a base state that is itself also a solution to the equations.
Additionally, we assume that the base state is a uniform steady state (its derivatives to $s$ and to $t$ are zero). 
Then, neglecting terms that are higher order in $\Delta \v{w}$, and subtracting the equation for the base state (which is satisfied by definition), a system of the form \eqref{eq:LSA/quasilinear_system} can be approximated by
\begin{equation}
\v{A}(\v{\bar{w}}) \pd{\Delta \v{w}}{t}  + \v{B}(\v{\bar{w}})  \pd{\Delta \v{w}}{s}  + \v{E}(\v{\bar{w}} )\pdd{\Delta \v{w} }{s}  + \v{G}(\v{\bar{w}} )\pddd{ \Delta \v{w}}{s}  = \v{D}_C(\v{\bar{w}}) \Delta \v{w},
\label{eq:LSA/general_linearized_equation/third-order}
\end{equation}
with
\begin{equation*}
\v{D}_C(\v{\bar{w}} ) = \pd{\v{c}(\v{\bar{w}})}{\v{\bar{w}}}.
\end{equation*}
Here, $\v{D}_C$ is a Jacobian matrix. 
Due to the assumption of the uniform base state, and the neglecting of higher order terms, the terms in $\v{B}(\v{\bar{w}})$ and $\v{E}(\v{\bar{w}})$ involving partial derivatives to $s$ drop out.

We write $\Delta  \v{w}$ as a Fourier series and substitute an arbitrary Fourier mode 
\begin{equation*}
\Delta \v{w} =\Delta \widehat{\v{w}} \exp \left[i \left(k s - \omega t \right)\right], 
\label{eq:LSA/w_perturbation_Fourier_mode}
\end{equation*}
with $\Delta \hat{\v{w}}$ the amplitude, $k$ the wavenumber, and $\omega$ the angular frequency, into \eqref{eq:LSA/general_linearized_equation/third-order}. 
This yields the following linear system \cite{FullmerRansomLopezdeBertodano2014}:
\begin{equation}
\left[-  \omega \v{A}(\v{\bar{w}}) + k \v{B}(\v{\bar{w}}) + i \v{D}(\v{\bar{w}}) + ik^2 \v{E}(\v{\bar{w}})  - k^3 \v{G}(\v{\bar{w}}) \right] \Delta \widehat{\v{w}} = \v{0}.
\label{eq:LSA/linear_system/third-order}
\end{equation}
For nontrivial solutions to exist, the determinant of the term between brackets must be zero, and solving for this yields two dispersion relations $\omega(k)$. 

The perturbation amplitudes $\Delta \widehat{\v{w}}$ corresponding to the found dispersion relations can be found by substituting these in \eqref{eq:LSA/linear_system/third-order} and solving for $\Delta \widehat{\v{w}}$. 
This can be understood as, for each dispersion relation, finding the null space of the term between brackets in \eqref{eq:LSA/linear_system/third-order}, which will consist of one vector.
The associated phase angles can be calculated component-wise:
\begin{equation*}
\gv{\theta} = \arctan{[\mathrm{Im}(\Delta \widehat{\v{w}})/\mathrm{Re}(\Delta \widehat{\v{w}})]},
\end{equation*}
where each component of $\gv{\theta}$ has a range $\left[-\pi, \pi \right]$ (use the four-quadrant inverse tangent).

This makes it possible to write the evolution in time of a perturbation as 
\begin{equation}
\Delta \v{w} = \sum_j \left| \Delta \widehat{\v{w}}_j \right| \mathrm{e}^{\mathrm{Im}\left\{\omega_j \right\} t} \cos \left( k s - \mathrm{Re}\left\{\omega_j \right\} t + \gv{\theta}_j \right),
\label{eq:LSA/analytical_solution_in_time}
\end{equation}
where we take the sum over the different solutions for $\omega$ for a given $k$, and the associated amplitude vectors.
If a sinusoidal perturbation is initialized with a given wavenumber $k$, and an amplitude vector exactly corresponding to one of the two angular frequencies $\omega(k)$, then the sum in \eqref{eq:LSA/analytical_solution_in_time} can be left out and the perturbation will propagate as a single wave with speed $\mathrm{Re}\left\{\omega \right\}$ and growth rate $\mathrm{Im}\left\{\omega \right\}$. 
This holds exactly for the linearized system, but solutions to the full nonlinear system will deviate from this solution over time. 

\end{appendices}


\bibliography{library_bibtex}      
\bibliographystyle{abbrv} 

\end{document}